\documentclass[a4paper,fleqn,usenatbib]{mnras}

\usepackage[T1]{fontenc}
\usepackage{ae,aecompl}
\usepackage{placeins}
\usepackage{capt-of}

\usepackage{graphicx}	% Including figure files
\usepackage{amsmath}	% Advanced maths commands
\usepackage{multirow}
\usepackage{morefloats}
\usepackage{float}
\usepackage[dvipsnames]{xcolor}
\title[{\it Kepler} B-type stars]{Variability of newly identified B-type stars observed by  {\it Kepler}}

\author[Szewczuk, Walczak \& Daszy{\'n}ska-Daszkiewicz]{
Wojciech Szewczuk\thanks{E-mail: wojciech.szewczuk@uwr.edu.pl (WS)},
Przemys{\l}aw Walczak\thanks{E-mail: przemyslaw.walczak@uwr.edu.pl (PW)},
Jadwiga Daszy{\'n}ska-Daszkiewicz\thanks{E-mail:jadwiga.daszynska-daszkiewicz@uwr.edu.pl (JDD)}
\\
Astronomical Institute, University of Wroc{\l}aw, Kopernika 11, PL-51-622 Wroc{\l}aw, Poland
}

\date{Accepted XXX. Received YYY; in original form ZZZ}

\pubyear{2015}

\begin{document}
\label{firstpage}
\pagerange{\pageref{firstpage}--\pageref{lastpage}}
\maketitle

\begin{abstract}
Recent re-determination of stellar atmospheric parameters for a sample of stars observed during the {\it Kepler} mission
allowed to enlarge the number of {\it Kepler} B-type stars. We present the detailed frequency analysis for all these objects.
All stars exhibit pulsational variability with maximum amplitudes at frequencies corresponding to high-order g modes. 
Peaks that could be identified with low-order p/g modes are also extracted for a few stars.
We identified some patters in the oscillation spectra that can be associated with the  period spacings that can result from the asymptotic nature of the detected pulsational modes.
We  also tentatively confront  the observed oscillation characteristics with predictions
from linear nonadiabatic computations of stellar pulsations.  For high-order g modes the traditional approximation 
was employed to include the effects of rotation on the frequency values  and mode instability. 
\end{abstract}


%%%%%%%%%%%%%%%%%%%%%%%%%%%%%%%%%%%%%%%%%%%%%%%%%%

%%%%%%%%%%%%%%%%% BODY OF PAPER %%%%%%%%%%%%%%%%%%

\section{Introduction}

%Although B-type stars are relatively rare objects, compared to less massive stars, they are very important ingredients 
%of stellar populations. 
B-type stars, compared to the less massive stars, are relatively rare objects. Nevertheless, they are very important ingredients 
of stellar populations.
They enrich interstellar medium in heavy elements as well as nitrogen and carbon and, due to their high masses, determine
the structure of stellar systems. Unfortunately, some aspects of their
evolution, especially element mixing processes and angular momentum transfer,  are still poorly understood.

In the beginning of previous century, it has been found that some B-type stars  exhibit photometric
and spectroscopic variability \citep{1902ApJ....15..340F,1913AN....195..265G}
that was associated with radial and non-radial pulsations \citep{1951ApJ...114..373L}.
These pulsations  are driven on the local opacity increase caused by iron group elements,
\citep[so  called Z-bump,][]{1992A&A...256L...5M,1992ApJ...393..272C,1993MNRAS.262..204D,1993MNRAS.265..588D,1993MNRAS.262..213G}.

B-type stars that pulsate in low-overtone p and g modes are called $\beta$ Cephei \citep[e.g.\,][]{2005ApJS..158..193S}, 
whereas B-type stars pulsating in high-overtone g modes are called Slowly Pulsating B-type stars \citep[SPB,][]{1991A&A...246..453W}, and those exhibiting both types of pulsations are called $\beta$ Cep/SPB hybrid pulsators
\citep[e.g.\,][]{2005MNRAS.360..619J, 2006MNRAS.365..327H, 2006A&A...448..697C, 2011MNRAS.413.2403B}.

Apart from uncertainties mentioned in the first paragraph, asteroseismic studies of B-type pulsators e.g.\,\citet{2017sbcs.conf..173W, 2017EPJWC.15206005W, 2017sbcs.conf..138D,2017MNRAS.466.2284D, 
2018MNRAS.478.2243S, 2019MNRAS.485.3544W} suggest that stellar opacities still need some revision.
In particular, a huge increase of opacity is necessary 
to excite high-order g modes in early  B-type stars \citep{2012MNRAS.422.3460S,2017MNRAS.466.2284D}.
Theoretical work by \citet{2018A&A...610L..15H}
showed, that including diffusion of heavy elements does cause the accumulation of iron and nickel 
in the vicinity of driving zone in massive stars, which significantly increases the opacity in these layers.

In the recent decade, the space born missions: {\it MOST}, {\it CoRoT} {\it Kepler, K2},  {\it BRITE}, {\it TESS} have provided 
high precision photometric data. It has became obvious that variability is common
among B-type stars \citep[e.g.][]{2011MNRAS.413.2403B, 2012AJ....143..101M},
although there are known, up to the level of available photometric precision,
constant stars \citep[eg. KIC\,11817929,][]{2011MNRAS.413.2403B}.

\begin{table*}
\label{Star_sample}
	\centering
	\caption{Atmospheric parameters and available  {\it Kepler} data for our sample of stars.
	The columns contain sequentially: our ID number, {\it Kepler} number, effective temperature, $T_\mathrm{eff}$, gravity,
	$\log g$, projected rotational velocity, $V\sin i$ and
	available {\it Kepler} quarters. The stellar parameters are taken from H19.}
	\label{tab:param}
	\begin{tabular}{rrrrrr} % four columns, alignment for each
		\hline
\#  & KIC ID   &    $T_\mathrm{eff}$ (K) & $\log g$ (dex)  & $V\sin i$ (km\,s$^{-1}$) & {\it Kepler} data \\
\hline
 1 &  1430353 & 17000(2000)  &  3.30(0.27)   &    210(26)    &Q0-Q17\\
 2 &  3459297 & 13100(600)   &  3.61(0.14)   &    123(12)    & Q0-Q5; Q7-Q9; Q11-Q13; Q15-Q17\\
 3 &  3839930 & 16950(650)   &  4.18(0.13)   &     51(6)     & Q0-Q2; Q4-Q5; Q7-Q9; Q11-Q13; Q15-Q17\\
 4 &  3862353 & 14000(1000)  &  3.71(0.18)   &     64(15)    & Q2-Q3; Q7-Q9\\
 5 &  4077252 & 12100(900)   &  3.51(0.22)   &     65(23)    & Q0-Q5; Q7-Q9; Q11-Q13; Q15-Q17\\
 6 &  4936089 & 12250(600)   &  4.05(0.18)   &     48(7)     & Q1-Q5; Q7-Q9; Q11-Q13; Q15-Q17\\
 7 &  4939281 & 17500(1000)  &  3.87(0.14)   &    115(20)    & Q0-Q5; Q7-Q9; Q11-Q13; Q15-Q17\\
 8 &  5477601 & 11950(350)   &  4.10(0.14)   &     88(11)    & Q0-Q17\\
 9 &  7630417 & 19200(1900)  &  3.73(0.24)   &    135(16)    & Q1-Q17\\
10&  8167938 & 13400(1000)  &  4.00(0.22)   &     70(3)     & Q2-Q4; Q6-Q8; Q10-Q12; Q14-Q16\\
11 &  8264293 & 13150(500)   &  4.08(0.14)   &    284(13)    & Q0-Q17\\
12 &  8381949 & 20200(2000)  &  3.72(0.27)   &    245(21)    & Q0-Q17\\
13 &  8714886 & 18200(900)   &  4.10(0.14)   &     52(9)     & Q0-Q17\\
14 &  9227988 & 14000(1200)  &  3.34(0.18)   &     50(3)     & Q0-Q17\\
15 &  9278405 & 11100(200)   &  4.14(0.11)   &    110(17)    & Q0-Q17\\
16 &  9468611 & 10750(550)   &  3.70(0.22)   &    263(30)    & Q0-Q17\\
17 &  9715425 & 15750(1500)  &  3.52(0.27)   &    122(12)    & Q0-Q17\\
18 &  9910544 & 12100(300)   &  4.10(0.14)   &     72(5)     & Q2-Q6; Q8-Q10; Q12-Q14; Q16-Q17\\
19 &  9964614 & 20100(850)   &  3.84(0.14)   &     77(14)    & Q0-Q6; Q8-Q10; Q12-Q14; Q16-Q17\\
20 & 10118750 & 10950(500)   &  4.27(0.22)   &    271(16)    & Q1-Q17\\
21 & 10526294 & 11500(500)   &  4.18(0.14)   &     36(10)    & Q1-Q17\\
22 & 10790075 & 11850(400)   &  3.63(0.14)   &     71(5)     & Q0-Q17\\
23 & 11293898 & 16400(1000)  &  3.79(0.18)   &    355(35)    & Q0-Q17\\
24 & 11360704 & 16500(1050)  &  3.72(0.14)   &    303(12)    & Q0-Q17\\
25 & 11671923 & 11800(200)   &  4.05(0.11)   &     91(8)     & Q0-Q17\\
\hline
	\end{tabular}
\end{table*}

Satellite missions dedicated to perform  long and continues
photometric time series opened a new window in asteroseismic studies of B-type stars.
These observations  have allowed for the detection of a much larger number of frequencies than 
was accessible from ground-based observations \citep[e.g.\,][]{2019ApJ...872L...9P}.
Moreover, it turned out that a vast majority of B-type pulsators, if not all,  are the $\beta$ Cep/SPB hybrid pulsators.
This fact has increased an opportunity of probing stellar interiors using asteroseismic tools with unprecedented precision.
Unfortunately, it quickly turned out that we were dealing with a kind of failure of richness  \citep[e.g.\,][]{2014IAUS..301..109S}.
We have hundreds of frequencies, but we do not know what theoretical modes correspond to them.
Thus, we still have to face with the old standing problem of mode identification, i.e. determination of
the mode degree $\ell$,  azimuthal order $m$ and radial order $n$.

In the case of one-band photometric data the only way to determine $\ell$ and $m$
is to look for regularities in the   spectrum of oscillations.
One can expect some characteristic features, that are multiplets caused by rotational splitting of modes with the same $\ell$
and radial order $n$. In the case of low frequencies corresponding to  high-order g modes, the asymptotic theory of oscillation predicts a regular period spacing between consecutive radial orders for a given $\ell$ and $m$ \citep[e.g.\,][]{1980ApJS...43..469T, 1993MNRAS.265..588D, 2013MNRAS.429.2500B}. 

Up to now {\it Kepler} satellite give the best, in terms of quality and time span, photometric time series of B-type stars. 
The general view of oscillations of B-type stars observed by Kepler was published by \cite{2011MNRAS.413.2403B}
and \cite{2012AJ....143..101M}, but only the first quarters of {\it Kepler} observations  were available at that time.

The patterns of regular period spacings in {\it Kepler}'s B stars were found by
\citet{2014A&A...570A...8P, 2015ApJ...803L..25P, 2017A&A...598A..74P}, \citet{2018MNRAS.478.2243S}
and \citet{2018ApJ...854..168Z}. Three stars, namely KIC\,10526294 \citep{2015A&A...580A..27M, 2015ApJ...810...16T},
KIC\,7760680 \citep{2016ApJ...823..130M, 2017EPJWC.16003012S}
and KIC\,3240411 \citep{2018MNRAS.478.2243S}, were subjects of detailed seismic analysis.

Because B-type stars in {\it Kepler} field  are not numerous, any additional enlarging sample is desirable. Since effective temperatures and gravities, especially for hot stars in KIC ({\it Kepler} Input Catalogue)
are not accurate and often underestimated, some B-type stars can be misclassified as cool stars
\citep[e.g.\,][]{2011AJ....142..112B, 2012ApJS..199...30P, 2018ApJ...854..168Z}. Recently \citet{2019AJ....157..129H}
redetermined stellar parameters of 25 stars observed by {\it Kepler}.
In most cases,  \citet{2019AJ....157..129H} found that their actual effective temperatures correspond to B-type stars,
whereas they were previously considered as cool stars. 

The aim of this paper is to characterize oscillation properties of B-type stars listed in \citet{2019AJ....157..129H}.
In particular, the ultimate goal is to find regular patterns in the spectra of oscillations.

In Section \ref{Star_sample}, we give a short description of observational data and periodogram analysis.
Section \ref{individual_stars} contains results for individual stars. In Section \ref{red_noise}
we describe red noise that was found in the data. Conclusions are summarized in Section \ref{conclusions}.
Some additional material is available in the appendix \ref{appendix:A}.

\section{Star sample}

\begin{figure*}
	\includegraphics[angle=0, width=2\columnwidth]{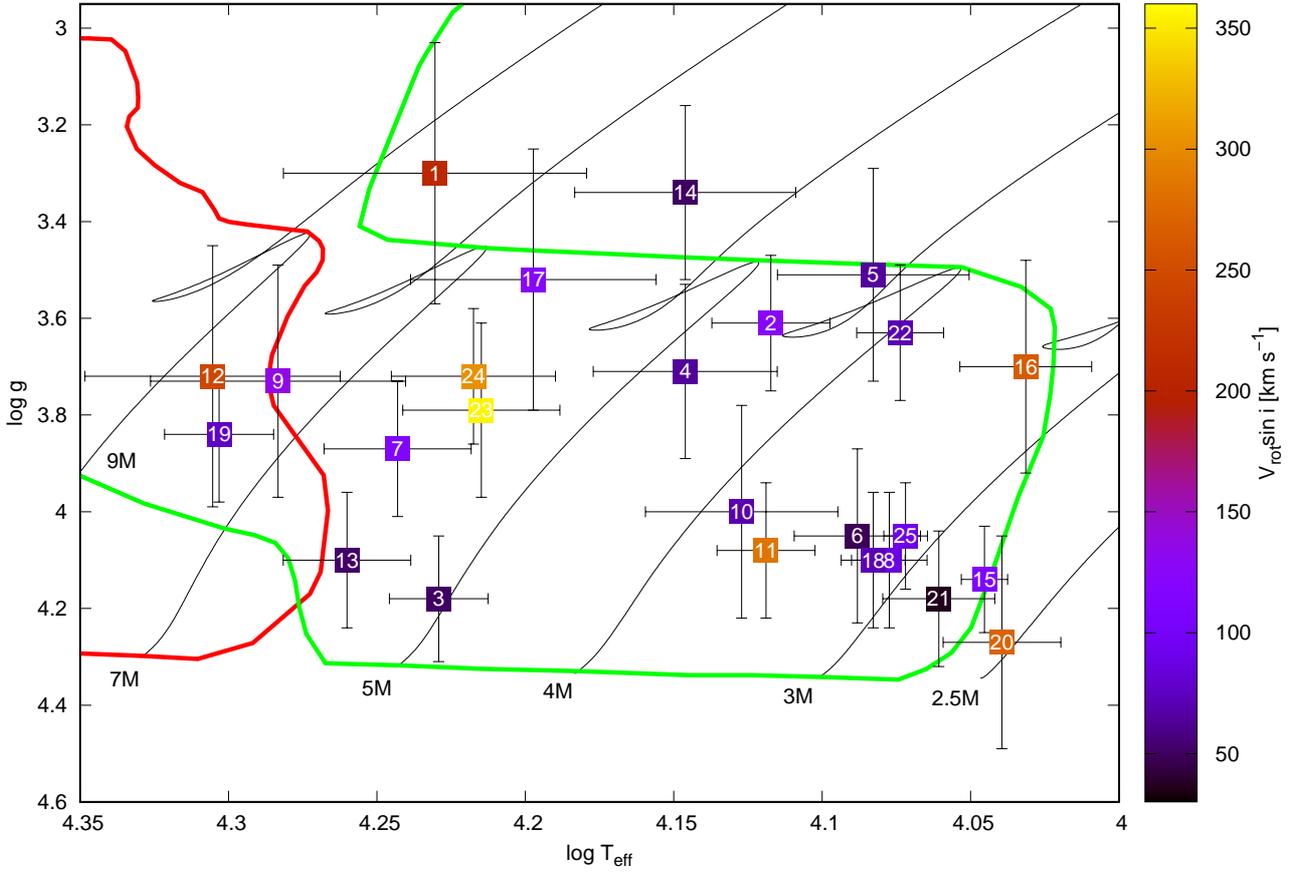}
    \caption{The position of target stars in the Kiel diagram.
    The stars are marked with numbers from the first column of Table\,\ref{tab:param}.
    The values of $V\sin i$ is colour-coded. All parameters are taken from
    H19. Green and red lines represent domains of pulsational instability 
    for the SPB and $\beta\,$Cep stars, respectively \citep{2015A&A...580L...9W}.
    Modes with the degree $\ell=0,\,1,\,2$ are considered.}
    \label{fig:kiel}
\end{figure*}

Here we analyze in details 25 stars listed by \citet[][hereafter H19]{2019AJ....157..129H}.
Two of these stars, KIC\,10526294 and KIC\,3459297
were already studied by \citet{2014A&A...570A...8P, 2017A&A...598A..74P} and regular period spacings
were found.
%However, in order  to compare our results with the results of the cited authors, we include them to our sample. 
We reanalysed these two stars and derived very similar results.   Moreover,
variability of KIC\,3839930,
%, KIC\,8167938,
KIC\,8381949, KIC\,8714886, KIC\,9964614 and KIC\,11360704
was previously investigated by \citet{2011MNRAS.413.2403B}
but only first quarters of {\it Kepler} data were used. Therefore, we
performed new analysis  these stars
using all available  data.

Stellar parameters of selected stars, as determined by H19, are given in Table\,\ref{tab:param}.
%For 17 stars from our sample they are  parameters obtained in the framework of {\it LAMOST-Kepler survey} \citep[][hereafter F16]{2016A&A...594A..39F}
%that we listed as well.
In Fig.\,\ref{fig:kiel},  we show the positions of our sample stars in the Kiel diagram. The domains of pulsational instability
for the $\beta$ Cephei and SPB stars are also depicted  \citep{2015A&A...580L...9W}.
As one can see all stars are  inside (within observational errors) SPB instability strip.
Four of them, KIC\,7630417, KIC\,8381949, KIC\,8714886, KIC\,9964614, lie also inside 
$\beta$ Cep instability strip.

\begin{figure*}
	\includegraphics[angle=270, width=1.95\columnwidth]{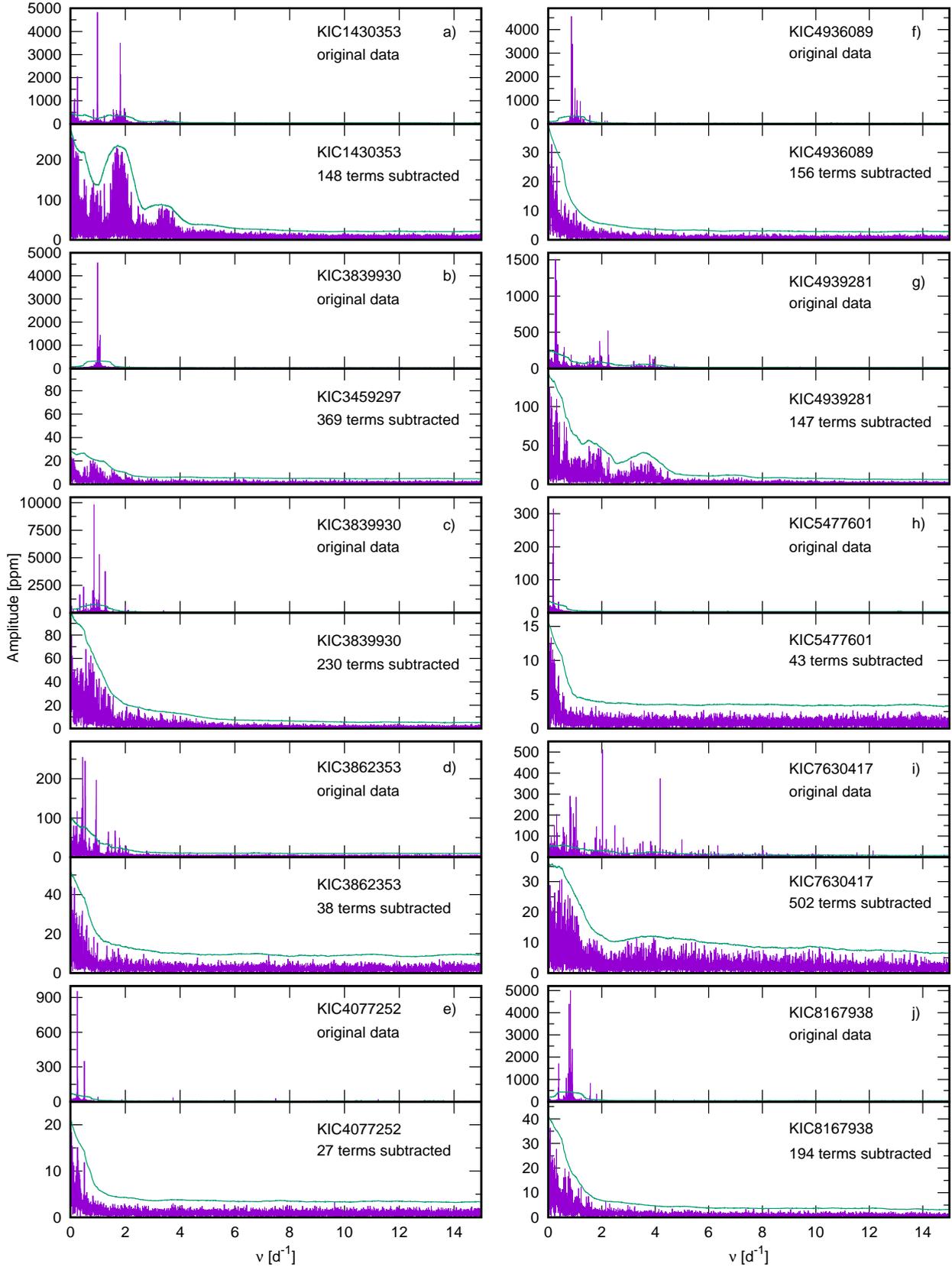}
    \caption{Fourier transforms of the original data and the data pre-whitened with  all significant frequencies (the top and bottom panel for each star, respectively).
    %The  signal-to-noise ratio is marked as a green line.
    The  $S/N=4$ level is marked as a green line.}
    \label{fig:trf1}
\end{figure*}

\begin{figure*}
	\includegraphics[angle=270, width=1.95\columnwidth]{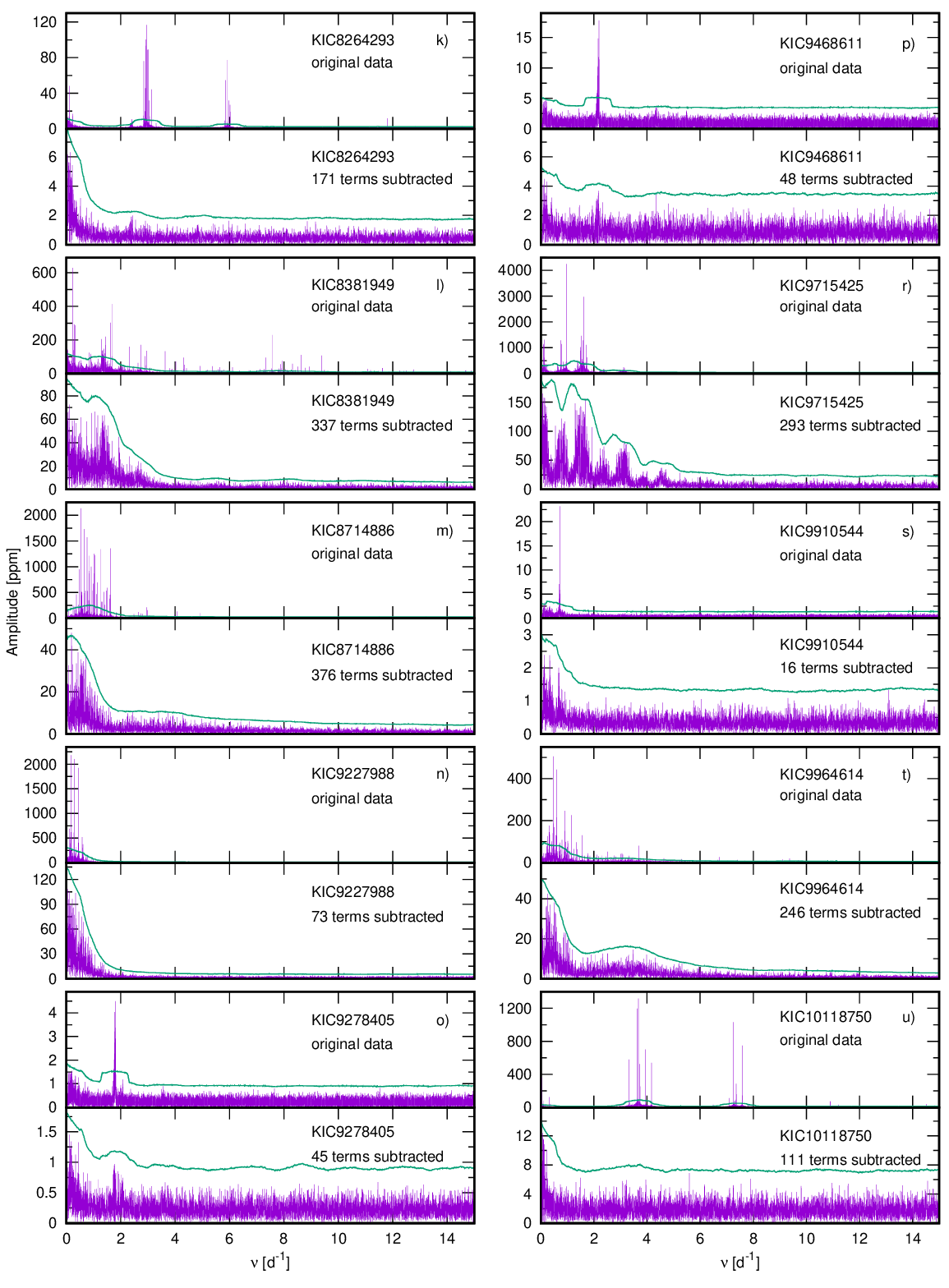}
    \caption{Continuation of Fig.\,\ref{fig:trf1}.}
    \label{fig:trf2}
\end{figure*}

\begin{figure*}
	\includegraphics[angle=270, width=1.95\columnwidth]{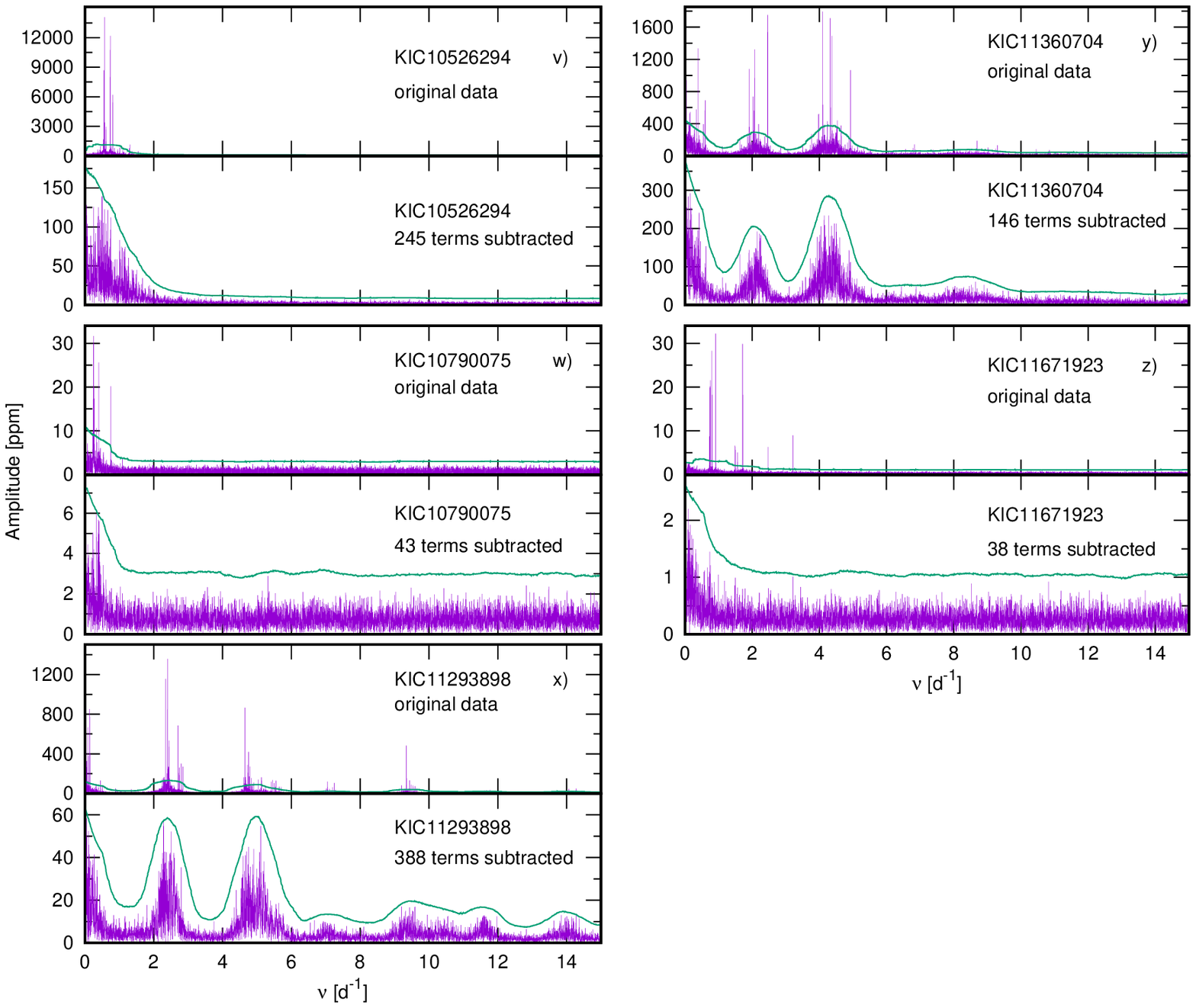}
    \caption{Continuation of Fig.\,\ref{fig:trf1}.}
    \label{fig:trf3}
\end{figure*}

\begin{figure*}
	\includegraphics[angle=270, width=2\columnwidth]{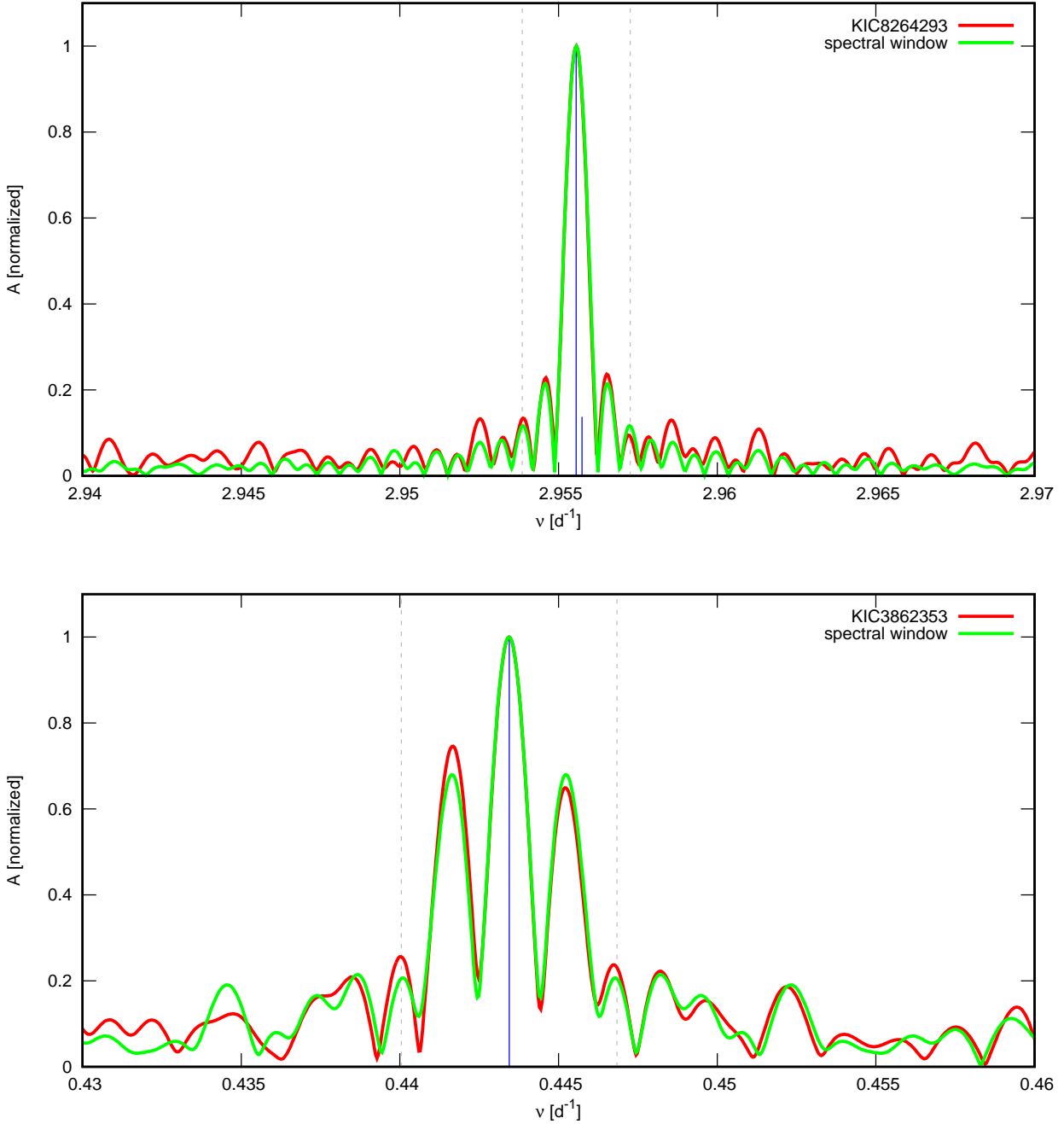}
    \caption{Normalized spectral window (green lines)  and Fourier transform of original data (red lines). Results for KIC\,8264293  and for KIC\,3862353 are displayed in the upper and bottom panel, respectively.
    Spectral window is shifted to the frequency of the dominant mode.
     There are also shown frequencies found in our periodogram analysis  that met $S/N\ge 4$ criterion (blue vertical lines).  Dashed gray lines mark $2.5/T$ resolution limit around the dominant frequency.}
    \label{fig:spec_vindow}
\end{figure*}

\begin{figure*}
	\includegraphics[angle=270, width=1.95\columnwidth]{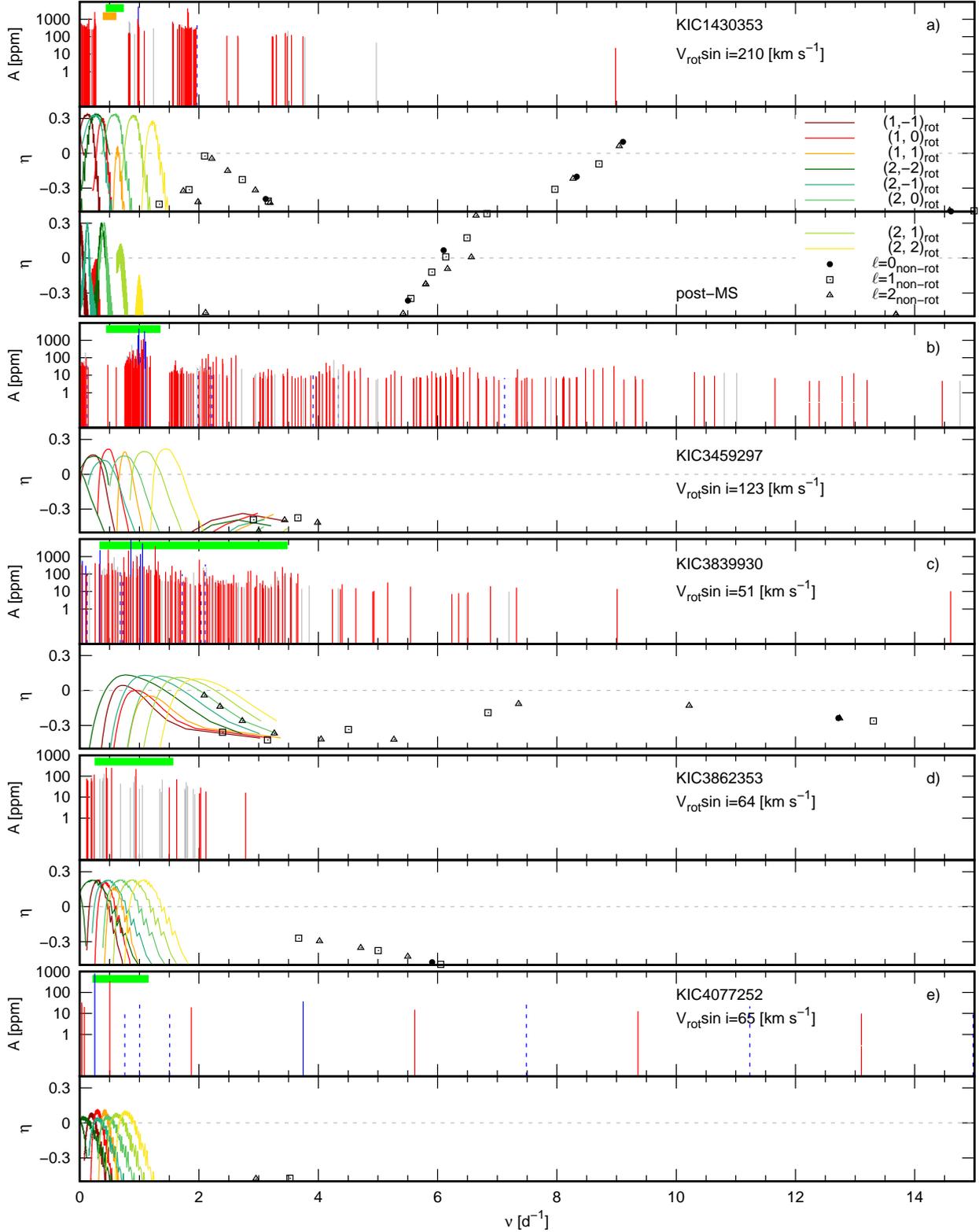}
	\vspace{-1cm}
    \caption{{\it Top panels:} All significant frequencies extracted from the observations.
    Independent frequencies are marked with vertical red lines
    and combinations with vertical gray lines. Vertical  blue continuous lines indicate independent frequencies for which harmonics are present.
    Harmonics  are marked 
    with vertical blue dashed lines. Bands (green for main sequence models and orange for post-main sequence models, if considered)
    give the range of rotation frequencies allowed by $V_\mathrm{rot}\sin i$ and critical rotation for each star.
    {\it Bottom panels:} The instability parameters, $\eta$, for radial, dipole and quadruple modes in representative models of each stars (see the text for details).
    Coloured lines depict the values of $\eta$ calculated in the framework of the traditional approximation, whereas, black symbols depict the value of  $\eta$ calculated within the zero-rotation approximation.}
    \label{fig:freq_all1}
\end{figure*}

\begin{figure*}
	\includegraphics[angle=270, width=1.95\columnwidth]{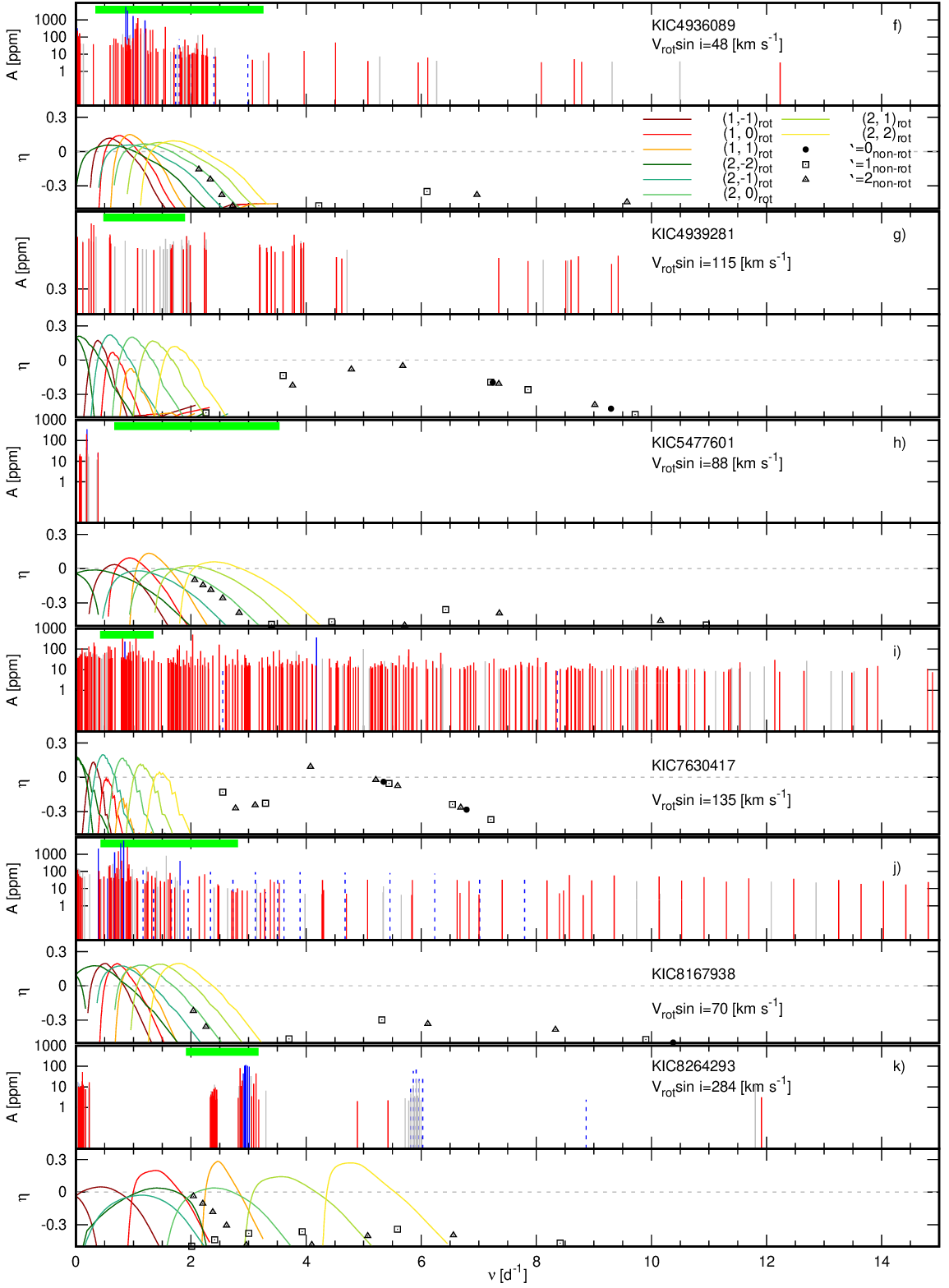}
    \caption{Continuation of Fig.\,\ref{fig:freq_all1}.}
    \label{fig:freq_all2}
\end{figure*}

\begin{figure*}
	\includegraphics[angle=270, width=1.95\columnwidth]{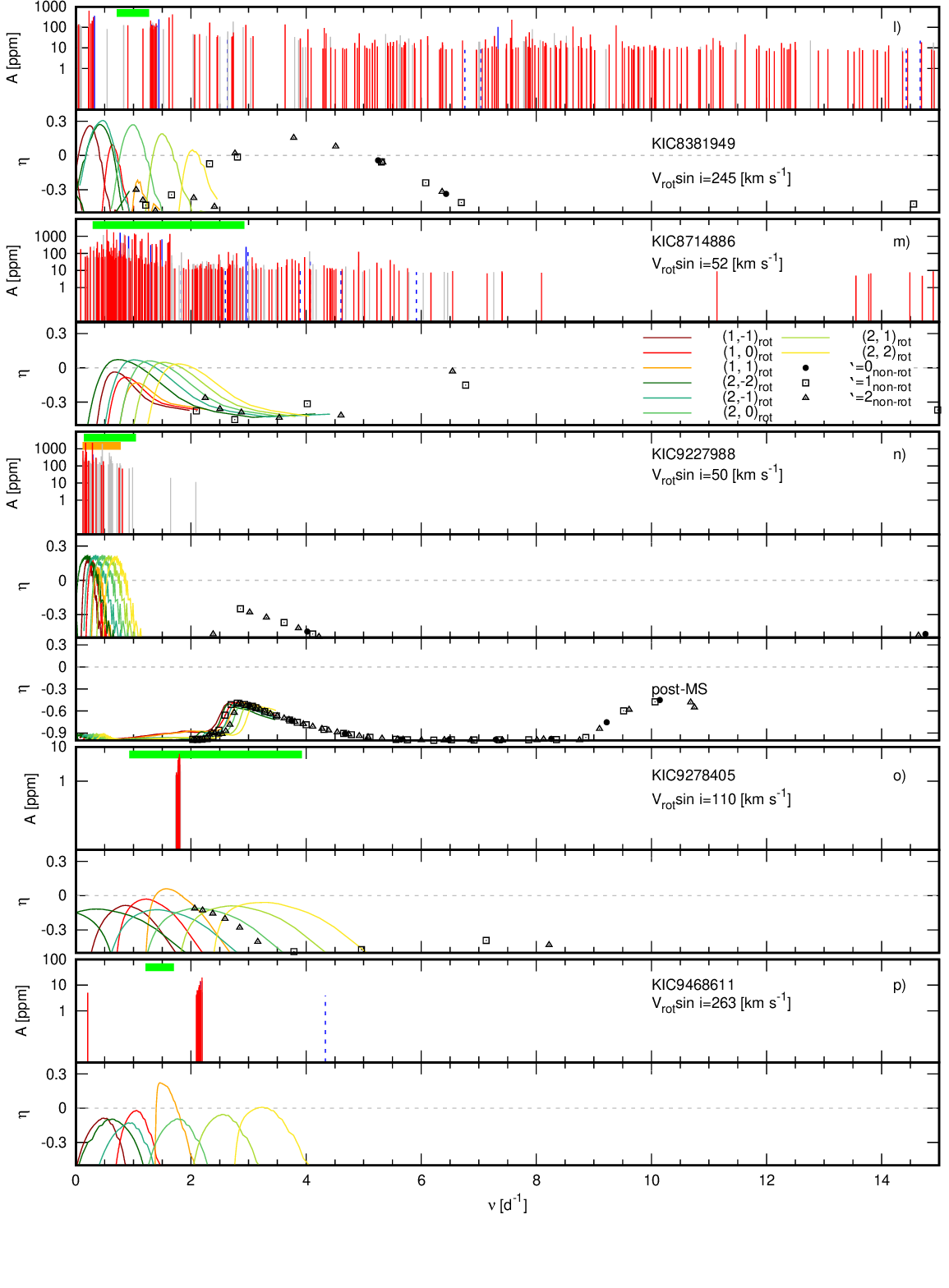}
    \caption{Continuation of Fig.\,\ref{fig:freq_all1}.}
    \label{fig:freq_all3}
\end{figure*}

\begin{figure*}
	\includegraphics[angle=270, width=1.95\columnwidth]{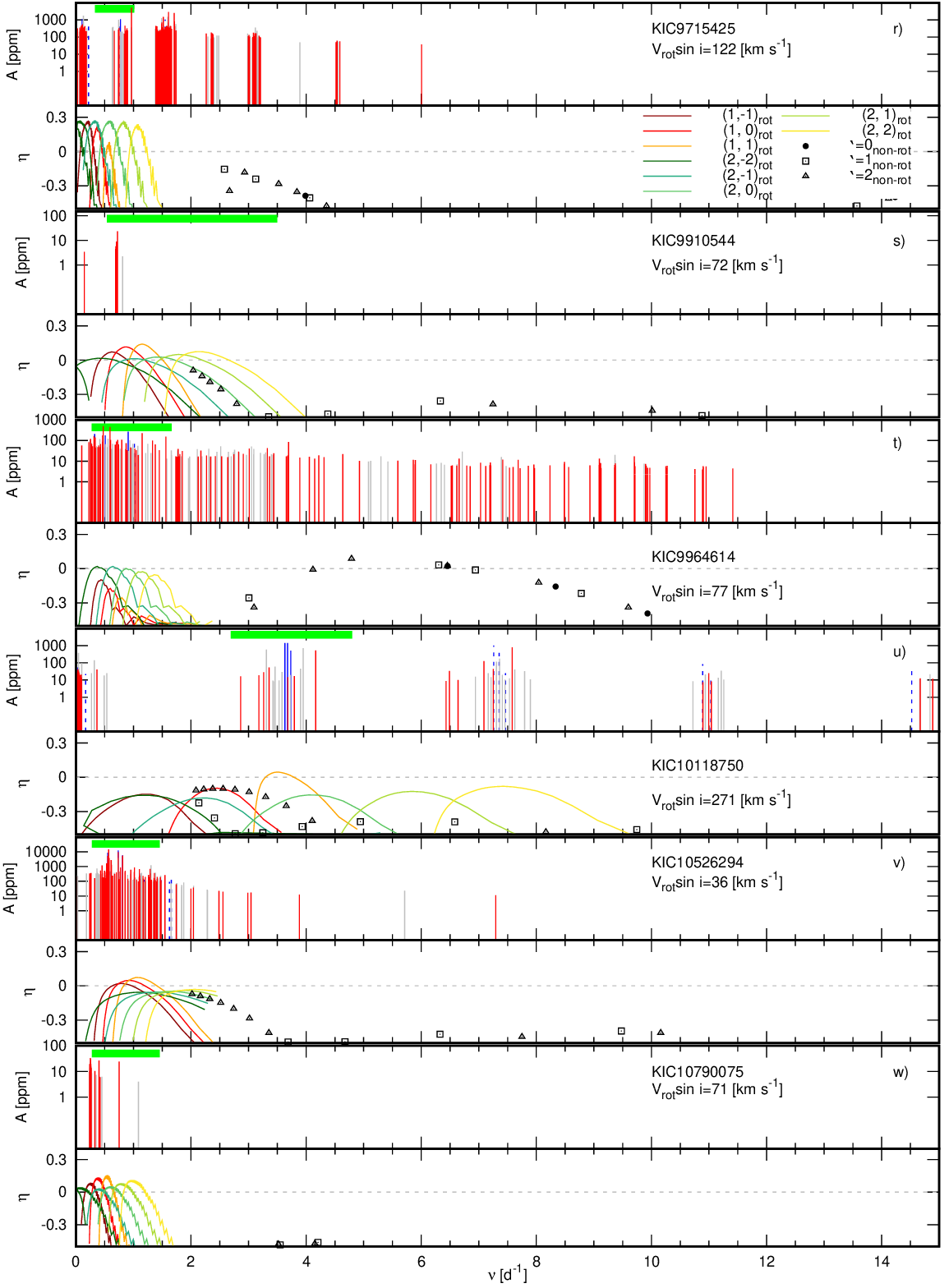}
    \caption{Continuation of Fig.\,\ref{fig:freq_all1}.}
    \label{fig:freq_all4}
\end{figure*}

\begin{figure*}
	\includegraphics[angle=270, width=1.95\columnwidth]{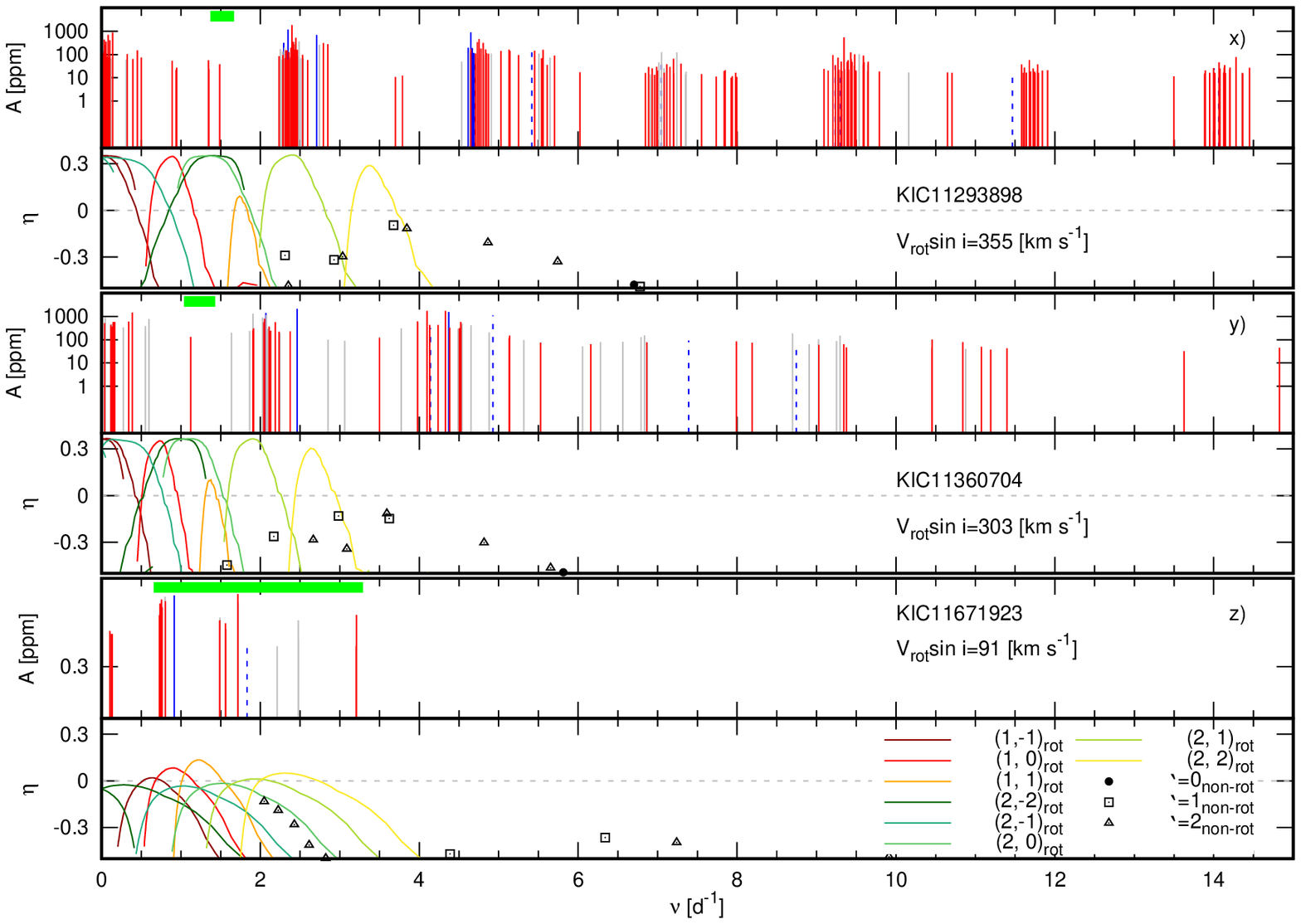}
    \caption{Continuation of Fig.\,\ref{fig:freq_all1}.}
    \label{fig:freq_all5}
\end{figure*}

\subsection{Data reduction}
\label{sec:reduction} % used for referring to this section from elsewhere

During its main mission {\it Kepler} satellite  was gathering almost continuous  photometric data  for 1470 days.
Major gaps in the observations of some stars are caused by the failures of the individual CCD chips during mission. Available data for our targets
are listed in the last column of Table\,\ref{tab:param}.
All these data are in public domain and are distributed in the form of prepared light curves and so-called target
pixel files.
Publicly available light curves were optimized for searching planets, i.e., they were extracted from the pixels that lay inside the  optimal aperture
that was selected to maximize the signal-to-noise ratio (S/N). It was shown \citep[see e.g.][]{2014A&A...570A...8P} that such definition of aperture is not
the best choice for extracting low frequencies from the light curves.
%Usually, including more pixels gives better results, especially it manifests as better long-term stability of the light curve.
Usually, including more pixels gives better results in the sense of  long-term stability of the light curve.
This is especially important in the case of long period pulsators such as the SPB stars.

Therefore, we decided to extract our own light curves from target pixel files. In general we proceeded in a similar way
as in \citet{2018MNRAS.478.2243S}.
%In the first step we defined for all target stars and all quarters customized apertures (masks) that contain all pixels with signal to noise ratio grater than 100.
In the first step, for all target stars and quarters  we defined customized apertures
(masks) that contain all pixels with signal to noise ratio grater than
100.  Next, we  performed  summing of the flux values within these masks  and  removed the obvious outliers.

An important step of the whole procedure was removing systematic trends using co-trending basis vectors \citep[e.g.\,][]{Kepler_manual}.
We started with five basis vectors and after an eye inspection we decided whether it was necessary to
add another one. Since employing too many vectors could lead to overfitting the data,  cleaning  off the real astrophysical signals
or  adding spurious signals,   this step was crucial and we paid a lot of attention to this procedure.
In the majority of cases 5 basic vectors were enough but in the most extreme ones we had to use the full set of 16 basic vectors (e.g., KIC\,11671923).
For some stars, individual quarters had to be divided into smaller subset and analyzed separately.
These steps were made with the use of PyKe 3.0 package \citep{2012ascl.soft08004S, pyke3}.

Finally, once again outliers were sought and  removed. Then, the quarters were divided by a second-order 
polynomial fit, merged and transformed into the ppm units.

\subsection{Periodogram analysis}

\begin{figure*}
	\includegraphics[angle=270, width=1.95\columnwidth]{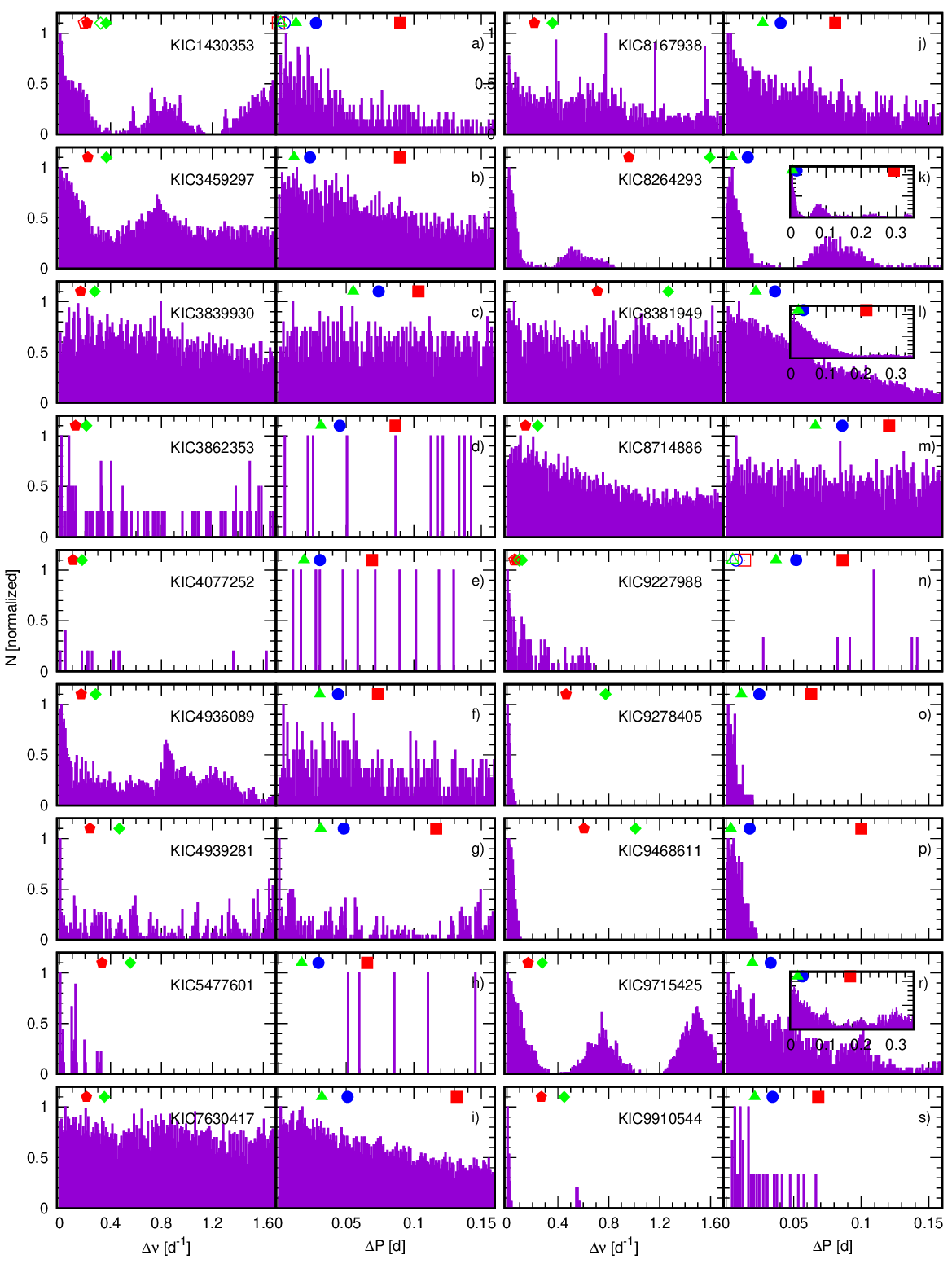}
    \caption{Normalized histograms of the differences in frequencies and periods.
    Red pentagons and
    green diamonds indicate the theoretical rotational splittings calculated for rotation rate equal to $V_\mathrm{rot}\sin i$ for $\ell=1$ and 2 modes,
    respectively.
    Open symbols are for post-main sequence models (if considered).
    Green triangles, blue circles and red squares indicate mean period
    spacing in representative
    models for dipole modes with $m=1$, 0 and $-1$, respectively.
    The insets show wider range of $\Delta \nu$ i $\Delta$ P if needed. See text for details.}
    \label{fig:delty_a}
\end{figure*}

\begin{figure*}
	\includegraphics[angle=270, width=1.95\columnwidth]{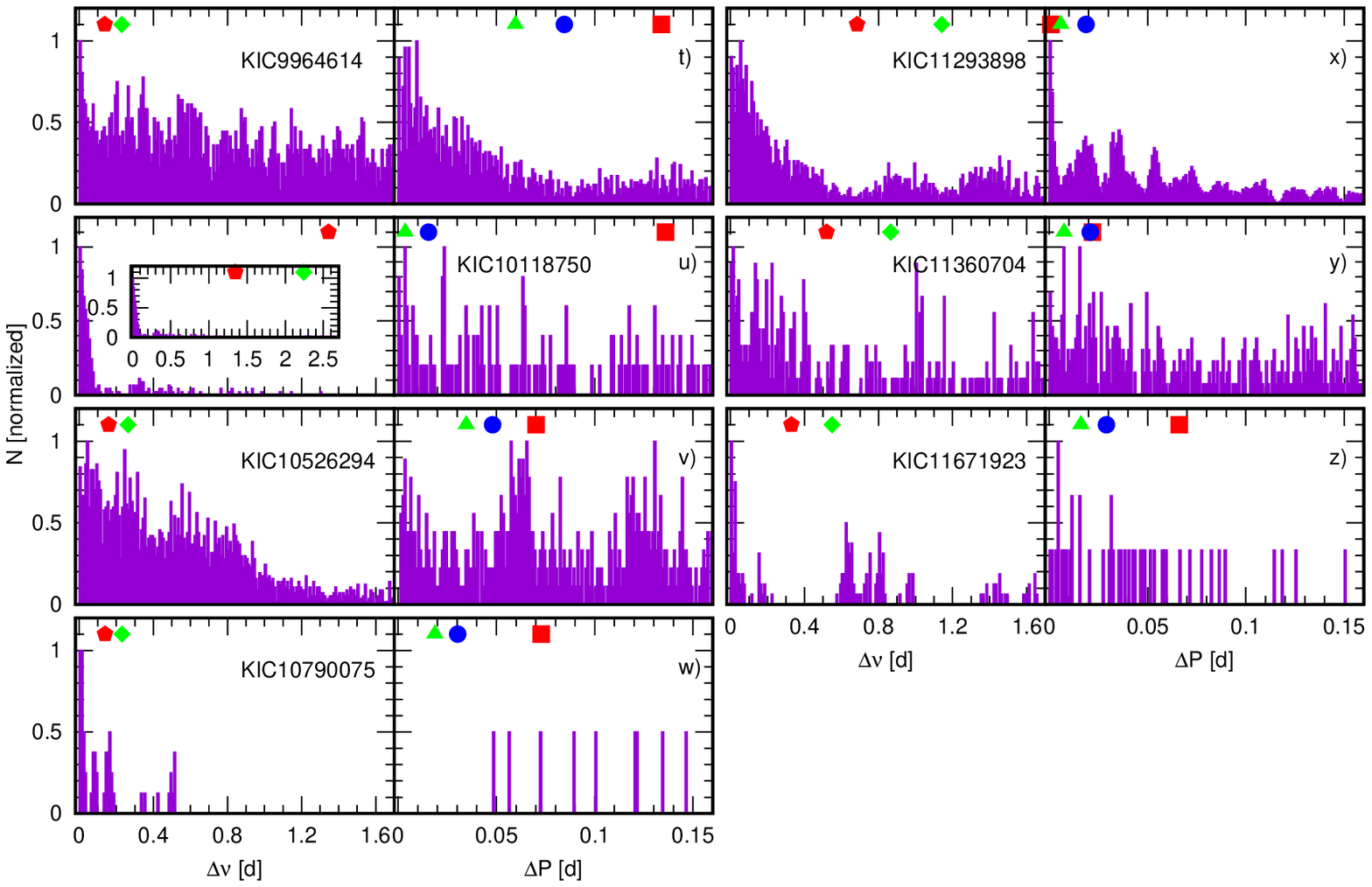}
    \caption{The continuation of Fig.\,\ref{fig:delty_a}.}
    \label{fig:delty_b}
\end{figure*}

In order to extract frequencies of the periodic light variability, we proceeded the standard prewhitening
procedure.
Amplitude periodograms were calculated up to the Nyquist frequency, i.e., to about  $24.5\,\mathrm{d}^{-1}$ with the fixed resolution of $4 \times 10^{-5}\,\mathrm{d^{-1}}$.
As a significance criterion of a given frequency peak we chose the signal-to-noise ratio $S/N\ =4$
\citep{1993spct.conf..106B,1997A&A...328..544K}. The noise was
calculated as a mean amplitude in a one day window centered at this frequency before its extraction.% \citep[for details see ][]{2018MNRAS.478.2243S}.
%However, \cite{2015MNRAS.448L..16B} argued that for {\it Kepler} data the more appropriate threshold is $S/N\approx 5$.
%Therefore, we warn that low signal-to-noise frequencies, especially those with $S/N < 5$ should be treated with caution.

Light curves of the studied stars exhibit clear and complex variability (see Fig.\,\ref{fig:lc_all} in Appendix \ref{appendix:A}).
The corresponding periodograms for all studied stars are shown in Figs.\,\ref{fig:trf1}, \ref{fig:trf2} and \ref{fig:trf3}.
%For each star we plot in two panels,  one above the other, a periodogram calculated for the original data (upper panel) and periodogram calculated for data prewhithened for all significant frequencies (bottom panel).
 For each star we plotted in two panels,  one above the other, a periodogram calculated for the original data (upper panel) and for data prewhithened for all significant frequencies (bottom panel). The number of removed terms, i.e., the number  of significant frequencies, is given as well. In addition we plotted
a spectral window for a sample star with the longest observation time span,
KI\,C8264293 (Fig.\,\ref{fig:spec_vindow}, upper panel), and for the sample 
star with the shortest observation time span, KIC\,3862353 
(Fig.\,\ref{fig:spec_vindow}, bottom panel). The spectral windows were
shifted to the frequency peaks with the highest amplitudes. Both, spectral window
and frequency amplitudes were normalized to unity.
%{\textcolor{orange}{ Spectral window
%and frequency amplitudes were normalized to unity}}

In Fig.\,\ref{fig:spec_vindow} we can see, that around dominant peaks there are high-amplitude side-lobes. 
Moreover, {\it Kepler} data have a high point-to-point precision which implies the risk of artificially introducing spurious signals in the prewhitening process. Therefore, in our analysis we decided to skip frequency with smaller amplitude from the pairs of frequencies that are separated less than  2.5 times the Rayleigh resolution ($\nu_{\rm 2.5R}=2.5/T$, where $T$ is the time span of observations). Such approach was proposed by \citet{1978Ap&SS..56..285L} as the most conservative one.  An example for KIC\,8264293 is shown in the  upper panel of Fig.\,\ref{fig:spec_vindow}. As one can see there are two close frequency peaks. The one with smaller
amplitude has been omitted in our final list of frequencies because it is separated less than our adopted criterion.

The final frequencies, amplitudes and phases were determined using the non-linear least square fit
\citep[Levenberg-Marquardt method,][]{2002nrca.book.....P} to the function in the form:
\begin{equation}
    S(t)=\sum_{i=1}^{N} A_i \sin\left(2\pi\left(\nu_i t+\phi_i\right)\right) +c,
\end{equation}
where $N$ is the number of sinusoidal components, $A_i$, $\nu_i$, $\phi_i$ are the amplitude, frequency, and phase of the $i$-th component, respectively. The offset $c$ ensures that $\int_T S(t)dt=0$, where $T$ is a time base.
We applied the correction to the formal  frequency errors as suggested by \citet[][the post mortem analysis]{1991MNRAS.253..198S} in order to account for correlated nature od {\it Kepler} data. The corrections, depending on the star, are in the range 1.4-3.9$\times$formal errors. Neglecting this effect can cause underestimation of the errors.

Next, we looked for harmonics and combinations in the form
\begin{equation}
\nu_i=m \times \nu_j + n\times \nu_k
\label{eq:combinations}
\end{equation}
with integers  $m,\,n$  from -10 to 10. For one star, KIC\,8167938, we found harmonics higher than 10. 
However, one should be aware that in  dense oscillation spectra, the criterion in Eq.\,\ref{eq:combinations}
can be easily satisfied by chance. Furthermore, it is reasonable to assume that the combinations are first formed by frequencies of higher amplitudes.
Therefore, we defined auxiliary Independence Parameter ($IP$)
 \begin{equation}
 IP=\frac{\sqrt{A_j^2+A_k^2}}{A_H},
 \end{equation}
 where $A_j$ and $A_k$ are amplitudes of supposed parents frequencies, whereas $A_H$ is the highest amplitude
 in the whole oscillation spectrum. Combinations of  frequencies with high amplitude parents have high $IP$.
 At the same time, the smaller this parameter the higher probability that condition expressed in Eq.
 \ref{eq:combinations} is satisfied by chance.
 Therefore we have considered all frequencies that not satisfy Eq.\,\ref{eq:combinations} and those that satisfy Eq.\,\ref{eq:combinations} within one times Rayleigh ($1/T$) resolution, but with $IP<1$ as independent ones.
%Therefore we introduced two terms:  firm independent frequencies, i.e., those that not satisfy Eq.\,\ref{eq:combinations}
%and probably independent frequencies, i.e., those that satisfy Eq.\,\ref{eq:combinations} within one times Rayleigh ($1/T$) resolution, but with $IP<1$.
We want to stress that $IP<1$ criterion is arbitrary. Moreover, the condition that parent frequencies should have higher amplitudes than child frequency is not strict \citep[e.g.][]{2015MNRAS.450.3015K}.  
The above procedure was applied to all stars.
Full frequencies lists are available as supplementary  online material.

In Figs.\,\ref{fig:freq_all1}--\ref{fig:freq_all5} (top panel for each star) we show all extracted frequencies. Different colours 
indicate: the independent frequencies
(solid vertical red and blue  lines), combination frequencies  (vertical gray lines) and harmonics (dashed vertical blue  lines).
Solid vertical blue lines are for independent frequencies that have harmonics.
Additionally, with green (for main-sequence evolution phase, MS) and orange bands (for post-main sequence evolution phase, post-MS)
we marked the ranges of rotational frequencies that are allowed by the value of $V\sin i$ and critical rotation rate. 
In order to calculate the rotational  frequency, we used radius for representative stellar models (for details see next section).
Notes on individual stars are given in the following sections.

In all cases, the highest amplitude signal was detected in low frequency range, typical for high-order g modes
observed in the SPB stars. In addition, for 17 stars,  we found low amplitude signals above 5 d$^{-1}$. At least in the case of four stars, KIC\,7630417, KIC\,8381949, KIC\,8714886 and KIC\,9964614, these frequencies can be interpreted as
low-order p and/or g modes typical for $\beta$ Cep stars. For a few stars (KIC\,1430353, KIC\,8264293, KIC\,9715425, KIC\,10118750,
KIC\,11293898, KIC\,11360704) the frequency grouping can be seen.

%In two panels on the right we show differences in frequency and period for all pairs of detected frequency peaks, respectively.

%\clearpage

\section{Notes about individual stars and the search for regularities in oscillation spectra}
\label{individual_stars}

For all analysed stars we search for regular patterns in their oscillation spectra, i.e.
for equally and quasi-equally spaced periods. The period difference, $\Delta P$, can be nearly constant, increase or decrease with a period $P$. The behaviour of the function $\Delta P(P)$ depends mainly on the mode degree, $\ell$, azimuthal order, $m$, and rotational velocity \citep[e.g.\,][]{2017MNRAS.469...13S}.

Here we use the standard notation. Period spacing corresponding to the mode with the $i$-th radial order 
and period $P_i$ is equal to $\Delta P_i=P_{i+1}-P_i$, where $P_{i+1}$ is period of the mode with the radial order $i+1$.
Therefore, when we write that observed sequence has an increasing  period spacing we mean that the $\Delta P_i$
is an increasing function of period and vice versa, i.e., decreasing period spacing means that $\Delta P_i$ is a decreasing function of the period.

%Of course,  the regularities found  in the observed oscillation spectrum are the result of asymptotic behaviour  only if they indeed consist of modes with consecutive values of  $n$ at a given pair ($\ell,~m$).
%In two panels on the right we show differences in frequency and period for all pairs of detected frequency peaks, respectively.

Our method of search for regular structures is a combination of an eye inspection of
the oscillation spectrum and a semi-automatic process. Firstly, we chose a few peaks from the independent frequency spectrum, that form regular structures. We generally looked for modes with the highest amplitudes, but the main stress was put on regularities. Then,  we used these manually chosen peaks as the basis for the semi-automatic search. We performed a linear regression fit to the period difference as a function of period. In the last, from  all remaining frequencies (independent and combinations) we chose a mode, that fit best to our initial relation. It was done in the sense of the smallest residual variance of the linear regression.  Such a mode was added to the initially chosen peaks and  the  procedure is repeated until there was no  a mode that follow $\Delta P$ vs. $P$ relation. The final sequence  was then manually reexamine. In some cases the frequencies that were automatically selected were manually replaced by other peaks with similar frequency but the higher amplitude. Similarly, where it was possible, combinations were replaced by independent peaks. For most stars we performed a lot of tests by choosing different initial peaks. We looked for the most regular structures, that contain high amplitude modes. Additionally, we look for regular structures also in the frequency domain. In the case of a few stars we found a regular series of the frequency difference versus frequency (the details are given in the following text).

In order to present the  expected properties of oscillations of the studied stars, we calculated linear nonadiabatic pulsations for  representative models. 
Evolutionary models were calculated with  MESA code version 11701 \citep{2011ApJS..192....3P,
2013ApJS..208....4P,2015ApJS..220...15P,2018ApJS..234...34P,2019ApJS..243...10P}. 
The MESA EOS is a blend of the OPAL \citet{Rogers2002}, SCVH
\citet{Saumon1995}, PTEH \citet{Pols1995}, HELM
\citet{Timmes2000}, and PC \citet{Potekhin2010} EOSes.
We used radiative opacities from
OPLIB opacity tables \citep{2015HEDP...14...33C,2016ApJ...817..116C},
with low-temperature data from \citet{Ferguson2005}
and the high-temperature, Compton-scattering dominated regime by
\citet{Buchler1976}.  Electron conduction opacities are from
\citet{Cassisi2007}.
Nuclear reaction rates are a combination of rates from
NACRE \citep{Angulo1999}, JINA REACLIB \citep{Cyburt2010}, plus
additional tabulated weak reaction rates \citet{Fuller1985, Oda1994,
Langanke2000}. Screening
is included via the prescription of \citet{Chugunov2007}.  Thermal
neutrino loss rates are from \citet{Itoh1996}.

We assumed the initial chemical composition
typical for B-type stars, i.e. $X=0.71$, $Z=0.014$ \citep{2012A&A...539A.143N}, 
the AGSS09 chemical mixture \citep{2009ARA&A..47..481A}
and overshooting parameter from convective core, $f_\mathrm{ov}=0.02$.
Rotational velocity was set to the value of $V_\mathrm{rot}\sin i$ as measured by H19, but, in our preliminary modelling, we did not include the rotational mixing processes.

Our representative models have parameters $\left(\log T_\mathrm{eff},\,\log g\right)$ corresponding to the central values
of the error boxes on the Kiel diagram (cf. Fig.\,\ref{fig:kiel} and Tab.\,\ref{tab:param}).
%As representative models we  selected those models whose positions on the Kiel diagram agreed with the position of the studied stars.
In two cases, KIC\,1430353 and KIC\,9227988, these parameters correspond to the post-main sequence evolutionary phase. Therefore, for these stars we decided to consider additional
models still inside their error boxes, but in the main sequence evolutionary phase.

For representative models  models we calculated the linear non-adiabatic
oscillations, both within the zero-rotation
approximation \citep[e.g.\,][]{1977AcA....27...95D} as well as taking into account the effects of rotation in the framework of
the traditional approximation for high-order g modes \citep[e.g.\,][]{1970attg.book.....C,1989nos..book.....U,1996ApJ...460..827B,1997ApJ...491..839L,2003MNRAS.340.1020T,2003MNRAS.343..125T,2007MNRAS.374..248D}.
We considered the spherical harmonic degrees, $\ell=1,\,2$ and all possible azimuthal orders, $\left|m\right|\le \ell$. 
The results are shown in Figs.\,\ref{fig:freq_all1}-\ref{fig:freq_all5} where we plotted
the instability parameter, $\eta$, as a function of the mode frequency. The value of $\eta$ is the normalized work integral
and it is positive if the mode is excited \citep{1978AJ.....83.1184S}. As one can see including the effects of rotation is indispensable   
to account, at least partly, for instability of high-order g modes in the studied stars.

In Figs.\,\ref{fig:delty_a}-\ref{fig:delty_b} we show differences between all possible pairs of independent frequencies, $\Delta \nu$, (left panels) and periods, $\Delta P$, (right panels) for all analysed stars. 
%We show differences for the set of data containing  frequencies that we call firm independent and probably independent (i.e. those with $IP<1$).
The maximum counts for certain values of $\Delta \nu$  or $\Delta P$ may indicate a large population of rotationally splitted modes or
asymptotic period spacing, respectively. Bin widths are $0.01\,\mathrm{d}^{-1}$ in the case of $\Delta \nu$ and $0.001\,\mathrm{d}$
in the case of $\Delta P$. 

In addition, we marked expected rotational splittings for dipole and quadruple modes assuming that the rotation rate is equal to $V\sin i$. These splitting were calculated according to a simple approximation
\begin{equation}
 \Delta\nu(\ell)=(1-C_{n\ell}) \nu_\mathrm{rot},
\end{equation}
where $C_\mathrm{n\ell}$ is the Ledoux constant \citep{1951ApJ...114..373L}.
For our purposes we adopted $C_\mathrm{n\ell} = 1/2$ for $\ell=1$ and
$C_\mathrm{n\ell} = 1/6$
for $\ell=2$.
 We stress that this Ledoux first-order perturbation used here to compute the rotational
splitting is only valid for slow rotators whereas stars from our sample are mostly fast rotators.
But at the moment we want to give only a general notion of the order of rotational splitting.

We also show mean $\Delta P$ for dipole retrograde, axisymmetric and prograde modes (see right panels in Figs.\,\ref{fig:delty_a}-\ref{fig:delty_b}).
The mean $\Delta P$ was calculated from all unstable high-order g modes. If there were no unstable modes, approximately 8 modes around maximum of $\eta$ corresponding to g modes were considered. Calculations were made
in the framework of the traditional approximation in order to include the effects of the Coriolis force.
The rotational velocity was set up, as previously, to the observed value of $V\sin i$. 

Brief discussions of the properties of detected oscillations are presented below for each stars separately. However, we postpone more detailed modelling and interpretation to separate papers. 
Here, the aim is only to show the first view of oscillations and their characteristic for the newly identified B-type stars
from the {\it Kepler} data.

\subsection{KIC\,1430353}
%1

\setlength{\tabcolsep}{4.8pt}
\renewcommand{\arraystretch}{0.98}

\begin{table}
	\centering
	\captionof{table}{Sequences of frequencies forming quasi-regular period spacing found in the oscillation spectrum of KIC\,1430353. In italics there are listed modes that are missing in the frequency spectrum but are required to maintain a regular structure of a series. In the last column is given frequency status (fs), 'i' stands for independent frequency and 'c' stands for combination frequency.
	%Due to lack of space, we did not enter all significant digits for the values of $P$ and $\Delta P$.
	}
	\label{tab:S1430353}
	\begin{tabular}{rrrrrrr} % four columns, alignment for each
		\hline
 ID                &    $\nu$      & $P$      &$\Delta P$& $A$     & $\frac{\mathrm S}{\mathrm N}$         &   fs\\
                   &    $(\mathrm{d}^{-1})$ & $(\mathrm{d})$  &$(\mathrm{d})$  & $(\mathrm{ppm})$    &               &  \\
\hline
\multicolumn{7}{|c|}{Sa}\\

$\nu_{46}$  & 1.768088(7)& 0.56533 & 0.00161 & 415(16) & 5.7  & i \\
$\nu_{113}$ & 1.77394(9) & 0.56372 & 0.00263 & 254(16) & 4.2  & i \\
$\nu_{65}$  & 1.78226(8) & 0.56108 & 0.00281 & 323(16) & 4.9  & i \\
--- & \textit{1.7912}     & \textit{0.55827} & \textit{0.00298} & ---     & ---  & --- \\
$\nu_{85}$  & 1.80084(8) & 0.55530 & 0.00303 & 295(16) & 4.6  & i \\
$\nu_{2}$   & 1.810730(9)& 0.55226 & 0.00351 & 4240(17)& 48 &  i\\
$\nu_{3}$   & 1.82231(2) & 0.54876 & 0.00524 & 2516(17)& 31 &  i\\
$\nu_{116}$ & 1.83986(9) & 0.54352 & 0.00542 & 254(16) & 4.2  & i \\
$\nu_{74}$  & 1.85838(8) & 0.53810 & 0.00682 & 307(16) & 4.7  &  i\\
---  & \textit{1.8822}     & \textit{0.53128} & \textit{0.00726} & ---     & ---  & --- \\
$\nu_{117}$ & 1.90830(9) & 0.52403 & 0.00734 & 250(16) & 4.2  & i \\
$\nu_{95}$  & 1.93539(8) & 0.51669 & 0.00772 & 265(16) & 4.5  & i \\
$\nu_{11}$  & 1.96475(4) & 0.50897 &    ---    & 813(16) & 12 &  i \\

\multicolumn{7}{|c|}{Sb}\\

$\nu_{89}$ & 1.63837(8) & 0.61036 & 0.00676 & 281(16) & 4.5 & i \\
$\nu_{39}$ & 1.65672(6) & 0.60360 & 0.00782 & 410(16) & 6.1 & c \\
$\nu_{28}$ & 1.67845(5) & 0.59579 & 0.00938 & 507(16) & 7.2 & i \\
$\nu_{58}$ & 1.70528(7) & 0.58641 & 0.01313 & 354(16) & 5.2 & i \\
$\nu_{80}$ & 1.74434(8) & 0.57328 & 0.01697 & 314(26) & 4.6 & c \\
$\nu_{75}$ & 1.79755(8) & 0.55631 & 0.02244 & 299(16) & 4.7 & i \\
$\nu_{63}$ & 1.87312(7) & 0.53387 & 0.02207 & 348(16) & 5.0 & i \\
$\nu_{109}$& 1.95388(9) & 0.51180 & ---     & 258(16) & 4.4 & c \\

	\end{tabular}
	\includegraphics[angle=0, width=\columnwidth]{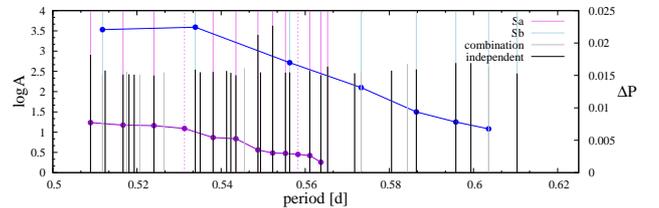}
	\vspace{-0.3cm}
    \captionof{figure}{Quasi-regular period spacings found in the oscillation spectrum of KIC\,1430353. Left y-axis indicates amplitudes of detected modes. Independent modes are marked with solid black line. Combination frequencies are marked with solid grey line. With colour lines we marked modes that form quasi-regular series (Sa -- violet, Sb -- blue). Right y-axis indicates period differences, which are marked with circles.}
    \label{fig:DP1}
	
\end{table}

KIC\,1430353  was classified as the SPB star by \citet{2012AJ....143..101M}.
\citet{2015MNRAS.451.1445B} assigned this star to the variability type SPB/ROT (where ROT marks rotational modulation)
and attributed the dominant frequency with the value of 0.9824 d$^{-1}$ to the rotational modulation.

In its {\it Kepler} light curve, we found 148 frequencies. 41 of them were rejected because of the $2.5\times$Rayleigh criterion.
%We ended up with  16 firm independent frequencies,  63 probably independent frequencies, 27 combinations and one harmonic.
We ended up with  79  independent frequencies, 27 combinations and one harmonic.

All  independent frequencies are typical for the SPB star and have values  below 4\,d$^{-1}$ (see Fig.\,\ref{fig:freq_all1}a). The one exception is small amplitude
$\nu_{148}\approx 8.98$\,d$^{-1}$.
They form four  separate groups, so we can classify the star as a frequency grouping pulsator. 
% There is also fourth group (around $3.5\,\text{d}^{-1}$) but it contains only one firm independent frequency, ie.   $\nu_{141}  =3.23775(6)$\,d$^{-1}$.
 The frequency grouping is typical for the fast rotating SPB pulsators \citep[e.g.][]{2005ApJ...635L..77W,2007CoAst.150..213D,2011MNRAS.413.2403B}. The projected rotational velocity
 of KIC\,1430353 is 210(26) km\,s$^{-1}$.
 Possible rotation frequency range resulting from $V\sin i$ and critical rotation rate  in our representative model
 is $0.43-    0.74$\,d$^{-1}$ for the  MS phase and $0.39 -  0.62$\,d$^{-1}$ for the  post-MS phase.
 Corresponding radii in our models are $9.5\,R_{\sun}$ and $10.8\,R_{\sun}$ for MS and post-MS phase, respectively.
 
 The highest amplitude frequency $\nu_1=0.982367(9)$ d$^{-1}$
 corresponds to the rotational frequency found by \citet{2015MNRAS.451.1445B}.
% The presence of the first harmonic of $\nu_1$, which is $\nu_{11}=1.964748(36)$ d$^{-1}$,  may support the rotational origin of the frequency $\nu_1$, but it can be a coincidence as well.  Moreover, if our models are correct, frequency of $\nu_1$ is too high to be of rotational origin.
 %but we do not find
 %harmonic of this frequency. The closest to the first harmonic $\nu_{11}=1.964748 \pm 0.000010 $ d$^{-1}$ differs from the first harmonic by
 %$0.000014$ what is slightly more than observatM=1,5,10M$_{\odot}$)ional error {(\bf Jagoda, mi tak wychodzi)}. Therefore $\nu_{11}$ may be of pulsational origin.
 %On the other hand, errors can be underestimated and $\nu_{11}$ can be harmonic as well.
 
 The presence of the first harmonic of $\nu_1$, which is $\nu_{11}=1.964748(36)$ d$^{-1}$, 
 supports the hypothesis of its rotational origin. However, our representative model indicates, that rotational frequency is much lower. Nevertheless, more detailed modelling is needed in order to derive the definite conclusions.

 We also calculated time dependent periodogram (TDP) by dividing the  data into 300-day subsets
 and performing the Fourier analysis in each  interval.
Each time the boundary of intervals were shifted by some amount of observational data points.
 The number of these points were chosen in order to ensure  that the intervals overlapped partly with each other.
 Such analysis shows that at least the dominant frequency is coherent, although its amplitude changes.

Then, we search for patterns in the oscillation spectrum.
We identified two sequences of consecutive frequencies with  decreasing period spacing (see Tab.\,\ref{tab:S1430353} and Fig.\,\ref{fig:DP1}).
Most of the frequencies in these sequences are from the group of independent modes. However, two frequencies in Sa sequence are missing to keep the regular structure
(we printed these supposed frequencies in italics).
The mean period spacing for Sa series  is 0.0047\,d (0.016\,d$^{-1}$ in the frequency domain). For Sb series the mean period spacing is 0.0141\,d (0.0451\,d$^{-1}$). 

From our initial analysis we concluded that the Sb sequence can be associated with dipole prograde modes. The mean period spacing is compatible for these modes in our representative model (see Fig.\,\ref{fig:delty_a}\,a). The Sa sequence seems to be made of prograde quadrupole modes.

These two sequences  explain most frequencies from the group, including high amplitude frequencies $\nu_2$ and $\nu_3$.
Asteroseismic modelling simultaneous with the mode identification is needed to confirm whether 
the sequence is of asymptotic origin or accidental, though.

The frequencies in a vicinity of 1 d$^{-1}$ do not form any regular pattern.
Similarly, for frequencies higher than about 2 d$^{-1}$  we have found only single peaks.

%\begin{figure}
%	\includegraphics[angle=0, width=\columnwidth]{DP_KIC1430353}
%    \caption{Period spacing found in the oscillation spectrum of KIC\,1430353.}
%    \label{fig:DP1}
%\end{figure}

\subsection{KIC\,3459297}
%1
KIC\,3459297 was classified as the SPB star by \citet{2012AJ....143..101M}.
Then \citet{2017A&A...598A..74P} found two period spacing series.

%In its {\it Kepler} light curve, we found as much a 369 frequencies. 97 of them were rejected due to the $2.5\times$Rayleigh criterion.  Finally, we derived 24 firm independent frequencies, 159 probably independent frequencies, 81 combinations and 8 harmonics.
In its {\it Kepler} light curve, we found as much a 369 frequencies. 97 of them were rejected due to the $2.5\times$Rayleigh criterion.  Finally, we derived 183 independent frequencies, 81 combinations and 8 harmonics.
TDP analysis indicate that at least higher amplitude peaks are coherent.

We found  independent frequencies up to 15\,d$^{-1}$ but those higher than $\sim 3$\,d$^{-1}$ have rather small amplitudes. Moreover, taking into account standard theoretical models the star is too cool to exhibit such high frequencies.
Therefore the origin of the highest frequency peaks remains unknown and requires more detailed analysis.

In the oscillation spectrum we found two regular series (see Tab.\,\ref{tab:S3459297} and Fig.\,\ref{fig:DP1.5}). The Sa sequence can be associated with dipole prograde modes, and the Sb -- with quadrupole prograde modes. The mean period spacings are 0.011\,d (0.011\,d$^{-1}$) for Sa and 0.011\,d (0.049\,d$^{-1}$) for Sb. 

The vast majority of frequencies of the Sa series overlap with the main period series found by \cite{2017A&A...598A..74P}.  Two low amplitude frequencies found by these authors differ from our determinations by more than observational errors ($\nu_{64}$ and $\nu_{117}$) and this is most probably caused by differences in data reduction.
In addition \cite{2017A&A...598A..74P} found $\nu=0.87791(4)$\,d$^{-1}$, but such frequency was not detected in our analysis. In the series Sa we included smaller number of frequencies, because, in our opinion, the frequencies lower than 0.83910 d$^{-1}$ do not form a regular structure. It can be cause by the fact, that we did not detect two other frequencies 0.82881(4)\,d$^{-1}$ and 0.79079(4)\,d$^{-1}$.

%However, we found frequencies from the list of the cited authors with the exception of their 0.82881(4)\,d$^{-1}$ and 0.79079(4)\,d$^{-1}$.

The second series of modes includes slightly higher frequencies. Although we detected all frequencies listed in a possible $l=2$ series found by \cite{2017A&A...598A..74P}, we decided to include into series Sb only a part of it. Due to this, we found more regular pattern that covers considerably larger amount of frequencies.

%The components of this sequence cover a possible $l=2$ series found by \cite{2017A&A...598A..74P}, but we found different period spacing.   Our series Sb is denser and it contains much more frequencies.

%We note that components excluded from our series are present in our frequency list.
%The differences in the values of small amplitude frequencies are most probably caused by differences in data reduction. \textcolor{red}{\bf troche pozmienialem, nie brzmi to zbyt skladnie ale sadze ze te informacje warto zamiescic}

\begin{table}
	\centering
	\captionof{table}{The same as in Tab.\,\ref{tab:S1430353} but for KIC\,3459297. 
	}
	\label{tab:S3459297}
	
	\begin{tabular}{rrrrrrr} % four columns, alignment for each
		\hline
		ID                &    $\nu$      & $P$      &$\Delta P$& $A$     & $\frac{\mathrm S}{\mathrm N}$        &   fs\\
		&    $(\mathrm{d}^{-1})$ & $(\mathrm{d})$  &$(\mathrm{d})$  & $(\mathrm{ppm})$    &               &  \\
		\hline
		\multicolumn{7}{|c|}{Sa}\\
	
	$\nu_{15}$ & 0.83910(2)  & 1.19176 & 0.00752 & 216(2) & 14 & i\\
	$\nu_{134}$& 0.84443(4)  & 1.18423 & 0.00754 & 34(2) & 4.8 & c\\
	$\nu_{44}$ & 0.84984(2)  & 1.17670 & 0.00795 & 87(2)  & 8.9 & i\\
	$\nu_{54}$ & 0.85562(3)  & 1.16874 & 0.00810 & 70(2) & 7.7 & i \\
	$\nu_{55}$ & 0.86159(3)  & 1.16065 & 0.00783 & 69(2)  & 7.4 & i\\
	$\nu_{23}$ & 0.86744(2)  & 1.15282 & 0.00746 & 102(2) & 11 & i\\
	$\nu_{7}$  & 0.873087(7) & 1.14536 & 0.00823 & 543(3) & 29 & c\\
	---        &\textit{ 0.8794  }    & \textit{1.13713} & \textit{0.00789} & --- & --- & ---\\
	$\nu_{13}$ & 0.88555(1)  & 1.12924 & 0.01004 & 258(3) & 18 & c\\
	$\nu_{12}$ & 0.89349(1)  & 1.11920 & 0.00903 & 263(2) & 17 & i\\
	$\nu_{20}$ & 0.90076(2)  & 1.11018 & 0.00933 & 142(2) & 11 & i\\
	$\nu_{11}$ & 0.90839(1)  & 1.10085 & 0.01192 & 269(2) & 17 & c\\
	$\nu_{117}$& 0.91833(5)  & 1.08893 & 0.00891 & 40(2)  & 5.0  & c\\
	$\nu_{42}$ & 0.92591(2)  & 1.08002 & 0.01234 & 89(2)  & 9.2 & i\\
	$\nu_{21}$ & 0.93661(2)  & 1.06768 & 0.00961 & 128(2) & 12& i  \\
	$\nu_{64}$ & 0.94511(3)  & 1.05808 & 0.01100 & 86(3)  & 6.5 & i \\
	$\nu_{9}$  & 0.95505(1)  & 1.04707 & 0.01536 & 317(2) & 18 & c\\
	$\nu_{8}$  & 0.96927(1)  & 1.03171 & 0.00965 & 314(2) & 19 &  i \\
	$\nu_{3}$  & 0.978418(3) & 1.02206 & 0.01200 & 1970(2)& 50 & i\\
	$\nu_{1}$  & 0.990042(3) & 1.01006 & 0.01446 & 4590(2)& 60 & i\\
	$\nu_{16}$ & 1.00443(2)  & 0.99559 & 0.01291 & 183(2) & 14 & i\\
	$\nu_{31}$ & 1.01762(2)  & 0.98269 & 0.01388 & 110(2) & 11 & i\\
	$\nu_{5}$  & 1.032195(5) & 0.96881 & 0.01672 & 1019(2)& 38 & i \\
	$\nu_{10}$ & 1.05032(1)  & 0.95209 & 0.01402 & 299(2) & 18 & i \\
	$\nu_{4}$  & 1.066017(5) & 0.93807 & 0.01690 & 1163(2)& 38 & i\\
	$\nu_{2}$  & 1.085575(3) & 0.92117 & 0.01778 & 3308(2)& 60 & i\\
	$\nu_{6}$  & 1.106939(5) & 0.90339 & 0.01713 & 859(2) & 37 & i\\
	$\nu_{27}$ & 1.12833(2)  & 0.88627 &     --- & 116(2) & 11 & i  \\
	
	\multicolumn{7}{|c|}{Sb}\\

	$\nu_{199}$ & 1.59927(7)  & 0.62529 & 0.00250 & 18(2)  & 4.9  & i \\
	$\nu_{50}$  & 1.60570(3)  & 0.62278 & 0.00414 & 70(2)  & 16 & i \\
	---  & \textit{1.61643 }    & \textit{0.61865} &\textit{ 0.00440} & --- & ---  & ---\\
	$\nu_{198}$ & 1.62800(7)  & 0.61425 & 0.00560 & 16(2)  & 5.1  & i \\
	$\nu_{218}$ & 1.64298(8)  & 0.60865 & 0.00617 & 16(2)  & 4.6  & c \\
	--- & \textit{1.65981  }   & \textit{0.60248} & \textit{0.00663} & --- & ---  & --- \\
	$\nu_{185}$ & 1.67828(6)  & 0.59585 & 0.00714 & 20(2)  & 5.7 & c \\
	$\nu_{123}$ & 1.69865(5)  & 0.58870 & 0.00733 & 25(2)  & 7.9 & i \\
	$\nu_{120}$ & 1.72006(5)  & 0.58138 & 0.00806 & 29(2)  & 8.0   & i \\
	$\nu_{231}$ & 1.74424(9)  & 0.57332 & 0.00835 & 13(2)  & 4.4 &  i\\
	--- & \textit{1.77003}     & 0\textit{.56496} & \textit{0.00862} & --- & ---  & --- \\
	$\nu_{202}$ & 1.79744(7)  & 0.55635 & 0.00992 & 18(2)  & 5.2 & i \\
	$\nu_{244}$ & 1.8301(1)   & 0.54643 & 0.01053 & 12(2)  & 4.3 &  i \\
	$\nu_{233}$ & 1.86603(9)  & 0.53590 & 0.01100 & 13(2)  & 4.4 & i \\
	$\nu_{236}$ & 1.90511(9)  & 0.52490 & 0.01128 & 13(2)  & 4.5 & i \\
	---  & \textit{1.94694}     & \textit{0.51363} & \textit{0.01355} & --- & ---  & --- \\
	$\nu_{188}$ & 1.99969(7)  & 0.50008 & 0.01342 & 18(2)  & 5.9  & i \\
	$\nu_{39}$  & 2.05484(2)  & 0.48666 & 0.01442 & 89(2)  & 20 & i \\
	$\nu_{36}$  & 2.11760(2)  & 0.47223 & 0.01613 & 93(2)  & 22   & i \\
	$\nu_{131}$ & 2.19247(7)  & 0.45611 & 0.01970 & 31(2)  & 11   & i \\
	$\nu_{29}$  & 2.29123(2)  & 0.43645 & 0.01891 & 114(2) & 24 & i \\
	$\nu_{43}$  & 2.39500(2)  & 0.41754 & 0.02390 & 86(2)  & 26 & i \\
	$\nu_{35}$  & 2.54042(2)  & 0.39364 & 0.02557 & 100(2) & 26 & i \\
	$\nu_{166}$ & 2.71688(6)  & 0.36807 &     --- & 23(2)  & 12 & c \\

	\end{tabular}
	\includegraphics[angle=0, width=\columnwidth]{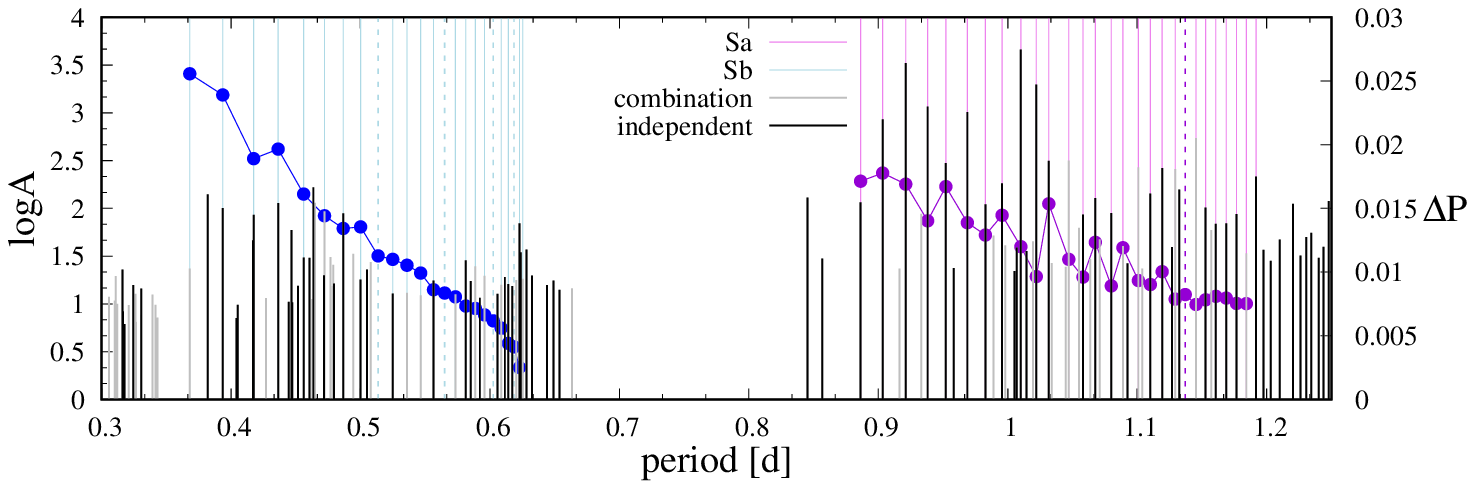}
	\vspace{-0.3cm}
	\captionof{figure}{The same as in Fig.\,\ref{fig:DP1} but for KIC\,3459297.}
	\label{fig:DP1.5}
	
\end{table}

%\begin{figure}
%	\includegraphics[angle=0, width=\columnwidth]{DP_KIC1430353}
%    \caption{Period spacing found in the oscillation spectrum of KIC\,1430353.}
%    \label{fig:DP1}
%\end{figure}

\subsection{KIC\,3839930}
%2

\begin{table}
	\centering
	\captionof{table}{The same as in Tab.\,\ref{tab:S1430353} but for KIC\,3839930.}
	\label{tab:S3839930}
	\begin{tabular}{rrrrrrr} % four columns, alignment for each
		\hline
 ID          &    $\nu$      & $P$    &$\Delta P$& $A$     & $\frac{\mathrm S}{\mathrm N}$         &   fs\\
             &    $(\mathrm{d}^{-1})$ & $(\mathrm{d})$  &$(\mathrm{d})$  & $(\mathrm{ppm})$    &               &  \\
\hline

 \multicolumn{7}{|c|}{Sa}\\            
          $\nu_{21}$ & 0.53934(3) & 1.85410 & 0.05068 & 408(12) & 11 & i \\
          $\nu_{45}$ & 0.55450(5) & 1.80340 & 0.05543 & 167(4)  & 6.5  & i \\
          $\nu_{96}$ & 0.57208(7) & 1.74800 & 0.07487 & 93(4)   & 4.1  & c \\
          $\nu_{49}$ & 0.59768(5) & 1.67310 & 0.09988 & 169(4)  & 6.4  & i \\
          $\nu_{28}$ & 0.63562(4) & 1.57330 & 0.12848 & 265(4)  & 9.7  & c \\
          $\nu_{62}$ & 0.69214(6) & 1.44480 & 0.17503 & 120(4)  & 5.5  & c \\
          $\nu_{110}$& 0.78755(9) & 1.26980 & 0.22958 & 69(4)   & 4.0  & i \\
          $\nu_{9}$  & 0.96137(2) & 1.04020 & 0.28054 & 1032(5) & 23 & i \\
          $\nu_{16}$ & 1.31641(3) & 0.75964 & 0.37232 & 530(4)  & 21 & i \\
          $\nu_{88}$ & 2.58181(7) & 0.38733 & ---     & 92(4)   & 13 & i \\
          
      \multicolumn{7}{|c|}{Sb}\\         
         $\nu_{32}$  & 0.67845(4) & 1.47400 & 0.04735 & 242(4)  & 8.5 & i \\
         $\nu_{90}$  & 0.70097(7) & 1.42660 & 0.03786 & 85(4)   & 4.5 & i \\
         $\nu_{104}$ & 0.72008(8) & 1.38870 & 0.05733 & 76(4)   & 4.1 & i \\
         $\nu_{55}$  & 0.75108(6) & 1.33140 & 0.04962 & 129(4)  & 6.3 & i \\
         $\nu_{103}$ & 0.78016(9) & 1.28180 & 0.05775 & 69(4)   & 4.0 & i \\
         $\nu_{5}$   & 0.81697(1) & 1.22400 & 0.05915 & 2244(4) & 31 & i \\
         $\nu_{1}$   & 0.85845(6) & 1.16490 & 0.04919 & 9989(5) & 54 & i \\
         $\nu_{15}$  & 0.89630(3) & 1.11570 & 0.05806 & 545(4)  & 16 & i \\
         $\nu_{8}$   & 0.94550(2) & 1.05760 & 0.06326 & 1223(7) & 26 & c \\
         $\nu_{26}$  & 1.00565(4) & 0.99439 & 0.06580 & 280(4)  & 11 & i \\
         $\nu_{34}$  & 1.07691(4) & 0.92858 & 0.06649 & 273(8)  & 9.7 & i \\
         $\nu_{24}$  & 1.15997(3) & 0.86209 & 0.07182 & 311(4)  & 13 & i \\
         $\nu_{3}$ &  1.265380(8) & 0.79027 & 0.09128 & 3928(4) & 52 & i \\
         $\nu_{137}$ & 1.4306(2)  & 0.69899 & 0.08785 & 40(4)   & 4.3 & c \\
         $\nu_{133}$ & 1.6363(1)  & 0.61114 & 0.08795 & 43(4)   & 5.4 & c \\
         $\nu_{63}$ & 1.911320(6) & 0.52320 & 0.10006 & 125(4)  & 13 & c \\
         $\nu_{140}$ & 2.3633(2)  & 0.42314 & 0.09731 & 41(4)   & 6.7 & i \\
         $\nu_{135}$ & 3.0691(1)  & 0.32583 & 0.10977 & 42(4)   & 8.2 & i \\
         $\nu_{205}$ & 4.6282(4)  & 0.21607 & 0.10508 & 16(4)   & 6.0 & i \\
         $\nu_{213}$ & 9.0102(4)  & 0.11099 & ---     & 14(4)   & 9.0 & i \\
             
         \multicolumn{7}{|c|}{Sc}\\      
        $\nu_{117}$& 0.85249(9) & 1.17300 & 0.05161 & 70(4)   & 4.0  & i \\
        $\nu_{89}$ & 0.89172(7) & 1.12140 & 0.06580 & 98(4)   & 5.1  & c \\
        $\nu_{36}$ & 0.94731(5) & 1.05560 & 0.07270 & 218(4)  & 8.4  & c \\
        $\nu_{7}$  & 1.01738(2) & 0.98292 & 0.09166 & 1396(4) & 27 & i \\
        $\nu_{41}$ & 1.12201(5) & 0.89126 & 0.11113 & 180(4)  & 9.1  & i \\
        $\nu_{40}$ & 1.28185(5) & 0.78012 & 0.13344 & 176(4)  & 10 & i \\
        $\nu_{27}$ & 1.54635(4) & 0.64669 & 0.16366 & 294(4)  & 18 & i \\
        $\nu_{31}$ & 2.07026(4) & 0.48303 & 0.20080 & 239(4)  & 20 & c \\
        $\nu_{66}$ & 3.54319(6) & 0.28223 & ---     & 118(4)  & 19 & i \\
           
	\end{tabular}
	
	\includegraphics[angle=0, width=\columnwidth]{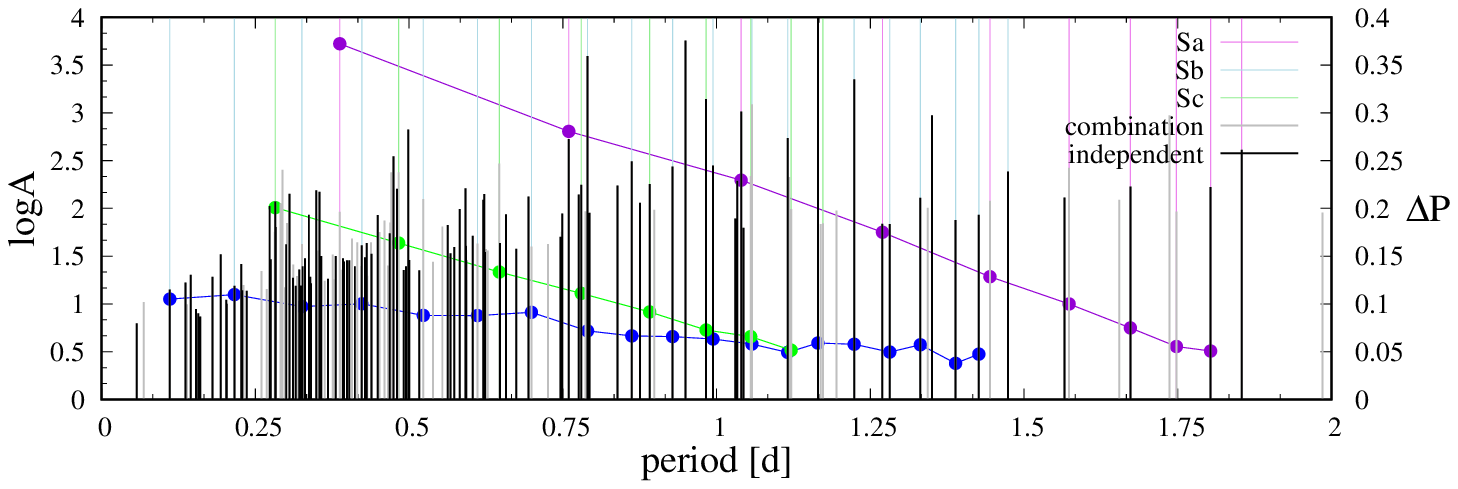}
	\vspace{-0.3cm}
    \captionof{figure}{The same as in Fig.\,\ref{fig:DP1} but for KIC\,3839930.}
    \label{fig:DP2}
	
\end{table}

The star was classified as the SPB variable by \citet{2011MNRAS.413.2403B} who gave spectroscopic
determinations of the effective temperature and surface gravity as $T_\mathrm{eff}=16500$\,K and $\log g=4.2$, respectively.
The authors mentioned also the values from the  Str\"omgren photometry $(T_\mathrm{eff}=17160\,\mathrm{K},~\log g=4.51)$
and from the fitting of spectral energy distribution (SED,  $T_\mathrm{eff}=18200\pm7000\,\mathrm{K})$. These determination of effective temperature and surface gravity are consistent with analysis of H19 (see also Table\,\ref{tab:param}).
The SPB classification was confirmed by \cite{2012AJ....143..101M}.

In the case of KIC\,3839930 we have found 230 frequencies and removed 48 from close pairs because of the criterion
for the $2.5\times$Rayleigh resolution. 
%Finally, we were left with 19 firm independent frequencies,  90 probably independent frequencies, 67 combinations and 6 harmonics.
Finally, we were left with 109 independent frequencies,  67 combinations and 6 harmonics. 

All independent frequencies, with the exception of low amplitude peak $\nu_{230}\sim17.5$\,d$^{-1}$, have the values below 9\,d$^{-1}$ and those with the highest amplitudes below 2\,d$^{-1}$.
The dominant frequency is $\nu_1=0.858446(6)$\,d$^{-1}$. %The star is SPB pulsator.
Using TDP we find that at least frequencies with the highest amplitudes are coherent.

In the oscillation spectrum of  this star we have detected three regular sequences of modes with decreasing period spacings,
Sa, Sb and Sc.
They consist of the  frequencies  listed in Tab.\,\ref{tab:S3839930} (see also Fig.\,\ref{fig:DP2}).

According to our criterion, some of the frequencies involved in the sequences may be combinations. Therefore, these sequences should be treated with caution.
The sequences have mean period spacing of the order of $\sim0.07-0.16$\,d ($\sim0.23-0.44$\,d$^{-1}$).
This is in qualitative agreement with period spacing in our representative model (see Fig.\,\ref{fig:delty_a}\,b).
These series may represent the consecutive modes of various angular numbers $(\ell,~m)$.
They cover a considerable part of the observed oscillation spectrum.

%Sa
%mean df = 0.22694
%mean dP = 0.16298

%Sb
%mean df = 0.43851
%mean dP = 0.071735

%Sc
%mean df = 0.33634
%mean dP = 0.11135

\subsection{KIC\,3862353}
%3
The star was classified as the SPB variable by \citet{2012AJ....143..101M}.
%From its Kepler light curve, we extracted 38 frequency peaks of which 8 are firm independent, 8 are probably independent and 22 are combinations.
From its Kepler light curve, we extracted 38 frequency peaks of which 16 are  independent and 22 are combinations. All  independent frequencies have the values below 2.8\,d$^{-1}$  which makes the star a typical SPB pulsator.  TDP analysis indicate that at least high amplitude peaks are coherent. 

In the oscillation spectrum of this star, we detected  a signatures of three sequences Sa, Sb, and Sc, given in Tab.\,\ref{tab:S3862353} and plotted in Fig.\,\ref{fig:DP3}.  Two of them, Sa and Sc, have slightly decreasing period spacings while $\Delta P(P)$ for Sb is nearly constant.  These series can be associated with dipole axisymmetric modes (Sa), dipole retrograde modes (Sc) and quadrupole axisymmetric modes (Sb).  The problem is that the series consist of rather small number of modes and some frequencies were identified as combination. Therefore, the presented sequences are rather uncertain.

\begin{table}
	\centering
	\captionof{table}{The same as in Table\,\ref{tab:S1430353} but for KIC\,3862353}
	\label{tab:S3862353}
	\begin{tabular}{rrrrrrr} % four columns, alignment for each
		\hline
 ID          &    $\nu$      & $P$    &$\Delta P$& $A$     & $\frac{\mathrm S}{\mathrm N}$         &   fs\\
             &    $(\mathrm{d}^{-1})$ & $(\mathrm{d})$  &$(\mathrm{d})$  & $(\mathrm{ppm})$    &               &  \\
\hline

    \multicolumn{7}{|c|}{Sa}\\        
  $\nu_{13}$ & 1.38468(5) & 0.72219 & 0.05662 & 68(3)  & 12 & c \\
 $\nu_{25}$ & 1.50248(9) & 0.66557 & 0.05066 & 29(3)  & 6.9 & i \\
 $\nu_{11}$ & 1.62626(4) & 0.61491 & 0.05491 & 72(3)  & 13 & i \\
 $\nu_{18}$ & 1.78572(6) & 0.56000 & 0.06674 & 52(3)  & 11 & c \\
 $\nu_{26}$ & 2.0273(1)  & 0.49326 & ---     & 29(3)  & 7.5 & i \\

    \multicolumn{7}{|c|}{Sb}\\  
  $\nu_{20}$ & 0.90614(6) & 1.10358 & 0.02085 & 52(3)  & 8.0 & c \\
 $\nu_{6}$ & 0.92360(3)  & 1.08272 & 0.02032 & 101(3 )& 13 & c \\
 $\nu_{3}$ & 0.94126(2)  & 1.06240 & 0.02873 & 214(3) & 24 & i \\
 --- & \textit{0.96742}    & \textit{1.03367 }&\textit{ 0.03584} &    --- & --- & --- \\
 $\nu_{30}$ & 1.0022(1)  & 0.99783 & 0.04463 & 25(3)  & 4.9 & c \\
 $\nu_{23}$ & 1.04909(7) & 0.95321 & --- & 38(3)  & 7.1 & c \\
            
     \multicolumn{7}{|c|}{Sc}\\  
     
     $\nu_{10}$ & 0.39072(4)  & 2.55936 & 0.10952 & 80(3) & 6.1 & c \\
     $\nu_{4}$  & 0.40819(3)  & 2.44983 & 0.10012 & 132(3)& 8.7 & c \\
      ---- &  \textit{0.42558}     &  \textit{2.34971} &  \textit{0.09464 }&  ---  & --- &  \\
     $\nu_{1}$  & 0.44345(2)  & 2.25507 & 0.08631 & 258(3)& 13& i \\
     $\nu_{7}$  & 0.46109(4)  & 2.16876 & 0.10157 & 82(3) & 6.5 & i \\
     --- &  \textit{0.48375}     &  \textit{2.06719} &  \textit{0.09879} & ---   & --- & --- \\
     --- &  \textit{0.50803}     &  \textit{1.96840} &  \textit{0.09233} & ---   & --- &--- \\
     $\nu_{2}$  & 0.53303(2)  & 1.87608 & --- & 249(3)& 14 & i \\

	\end{tabular}
	
	\includegraphics[angle=0, width=\columnwidth]{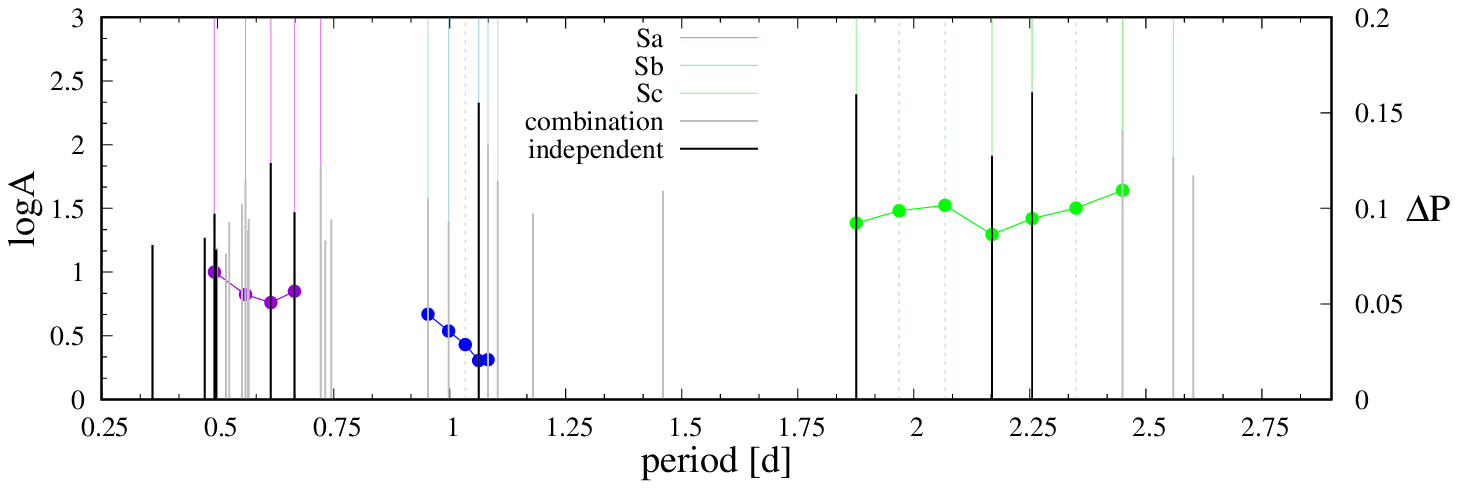}
	\vspace{-0.3cm}
    \captionof{figure}{The same as in Fig.\,\ref{fig:DP1} but for KIC\,3862353.}
    \label{fig:DP3}
	
\end{table}

%\begin{figure}
%	\includegraphics[angle=0, width=\columnwidth]{DP_KIC3862353}
%    \caption{The same as in Fig.\,\ref{fig:DP1} but for KIC\,3862353.}
%    \label{fig:DP3}
%\end{figure}

\subsection{KIC\,4077252}
%4
\citet{2012AJ....143..101M} classified the star as the SPB, whereas \citet{2015MNRAS.451.1445B} as
an ellipsoidal variable (ELL).

We found 27 frequencies of which 7  were
rejected  because of $2.5\times$Rayleigh limit. %According to our criteria, 6 are firm independent, 6 are probably independent and 8 are harmonics.
According to our criterion 12 frequencies are independent. We found also first five harmonics of $\nu_6=3.744559(17)$\,d$^{-1}$ and third, fourth and sixth harmonic of $\nu_1=0.251537(1)$\,d$^{-1}$. 
The TDP analysis showed that  at least frequencies with the highest amplitudes are coherent 
but their amplitudes decrease over time. This decrease may be instrumental as well as astrophysical in origin. In the oscillation spectrum we did not find any regular structure.

\subsection{KIC\,4936089}
%5
The classification of KIC\,4936089 as the SPB was made by \citet{2012AJ....143..101M} and this was confirmed by \citet{2015MNRAS.451.1445B}.

We found 156 frequency peaks but removed 24 of them because they are closer to adjacent peaks 
with the higher amplitudes than the $2.5\times$Rayleigh limit.
%From this subset 20 frequencies were identified as firm independent and  67  as probably independent. The 39  frequencies seems to be combinations and 6 are harmonics. Most of the frequencies
From this subset 87 frequencies were identified as  independent. The 39  frequencies seems to be combinations and 6 are harmonics. Most of the frequencies
are typical for the SPB pulsators. Harmonics of $\nu_1= 0.866262(3)$\,d$^{-1}$, $\nu_2=0.898342(3)$\,d$^{-1}$, $\nu_3=0.995864(4)$\,d$^{-1}$ and $\nu_5=1.201027(5)$\,d$^{-1}$ were  detected.
At least high amplitude frequencies are coherent. We did not find significant changes of amplitudes. Independent frequencies are present up to 12.2\,d$^{-1}$.

%Najnizsze czestotliwosci tworza krotka structure {\bf zastanowic sie nad tym, porownac z Przemkiem}.

For this star there are three series of frequencies, that may by form by modes with consecutive radial orders (see Tab.\,\ref{tab:S4936089} and Fig.\,\ref{fig:DP5}). In Series Sb and Sc one frequency seems to be missing. The mean period differences are 0.050\,d (0.110\,d$^{-1}$) for Sa, 0.071\,d (0.041\,d$^{-1}$) for Sb and 0.064\,d (0.292\,d$^{-1}$) for Sc. From our preliminary analysis we concluded that sequences can be associated with dipole axisymmetric modes (Sa), dipole prograde modes (Sb) and quadrupole prograde modes (Sc) (see also Fig.\,\ref{fig:delty_a}\,f).

\begin{table}
	\centering
	\captionof{table}{The same as in Table\,\ref{tab:S1430353} but for KIC\,4936089}
	\label{tab:S4936089}
	\begin{tabular}{rrrrrrr} % four columns, alignment for each
		\hline
 ID          &    $\nu$      & $P$    &$\Delta P$& $A$     & $\frac{\mathrm S}{\mathrm N}$         &   fs\\
             &    $(\mathrm{d}^{-1})$ & $(\mathrm{d})$  &$(\mathrm{d})$  & $(\mathrm{ppm})$    &               &  \\
\hline
   \multicolumn{7}{|c|}{Sa}\\         
   
   $\nu_{3}$  & 0.995864(4)& 1.00415 & 0.03282 & 1639(2) & 50 & i \\
   $\nu_{11}$ & 1.029510(9)& 0.97134 & 0.04134 & 297(2)  & 29 & i\\
   $\nu_{4}$  & 1.075270(4)& 0.93000 & 0.04303 & 1267(3) & 47 & i \\
   $\nu_{8}$  & 1.12743(1) & 0.88697 & 0.04434 & 318(2)  & 24 & i \\
   $\nu_{14}$ & 1.18675(2) & 0.84264 & 0.04276 & 171(2)  & 24 & c \\
   $\nu_{99}$ & 1.25018(8) & 0.79988 & 0.04981 & 16(2)   & 5.7  & i \\
   $\nu_{12}$ & 1.33320(1) & 0.75007 & 0.04756 & 264(2)  & 30 & i \\
   $\nu_{107}$& 1.42346(9) & 0.70251 & 0.05854 & 14(2)   & 6.2  & c \\
   $\nu_{7}$  & 1.55286(9) & 0.64397 & 0.05764 & 392(5)  & 34 &  i\\
   $\nu_{79}$ & 1.70550(5) & 0.58634 & 0.05841 & 24(2)   & 11 &  i\\
   $\nu_{38}$ & 1.89420(3) & 0.52793 & 0.07369 & 70(2)   & 23 & i \\
   $\nu_{17}$ & 2.20147(2) & 0.45424 & ---     & 140(2)  & 35 &  i\\

    \multicolumn{7}{|c|}{Sb}\\              
  
   $\nu_{66}$ & 0.63027(4) & 1.58663 & 0.07389 & 34(2) & 5.5 & c \\
   $\nu_{31}$ & 0.66105(2) & 1.51274 & 0.07181 & 81(2) & 11 & i \\
   $\nu_{33}$ & 0.69400(2) & 1.44093 & 0.07109 & 77(2) & 12 & i \\
   $\nu_{55}$ & 0.73001(3) & 1.36984 & 0.07142 & 45(2) & 8.0 & i \\
   --- &  \textit{0.77017}    &  \textit{1.29842} &  \textit{0.07138} & ---   & --- & --- \\
   $\nu_{16}$ & 0.81497(2) & 1.22704 & 0.06967 & 145(2)& 20 & c \\
   $\nu_{76}$ & 0.86403(5) & 1.15737 & 0.06872 & 33(2) & 6.2 & i \\
   $\nu_{47}$ & 0.91857(3) & 1.08865 & ---     & 84(3) & 12 & i \\

        \multicolumn{7}{|c|}{Sc}\\          
            
    $\nu_{89}$  & 1.52627(6) & 0.65519 & 0.07801 & 21(2) & 9.1 & i \\
    $\nu_{111}$ & 1.73256(9) & 0.57718 & 0.07049 & 13(2) & 7.2 & i \\
    $\nu_{125}$ & 1.9736(1)  & 0.50669 & 0.06342 & 9(2)  & 5.7 & i \\
    $\nu_{132}$ & 2.2560(1)  & 0.44327 & 0.05703 & 8(2)  & 5.8 & i \\
    --- &  \textit{2.5891}     &  \textit{0.38624} &  \textit{0.05150} & ---   & --- & --- \\
    $\nu_{124}$ & 2.9874(1)  & 0.33474 & ---     & 10(2) & 8.6 &  i \\

	\end{tabular}
	
		\includegraphics[angle=0, width=\columnwidth]{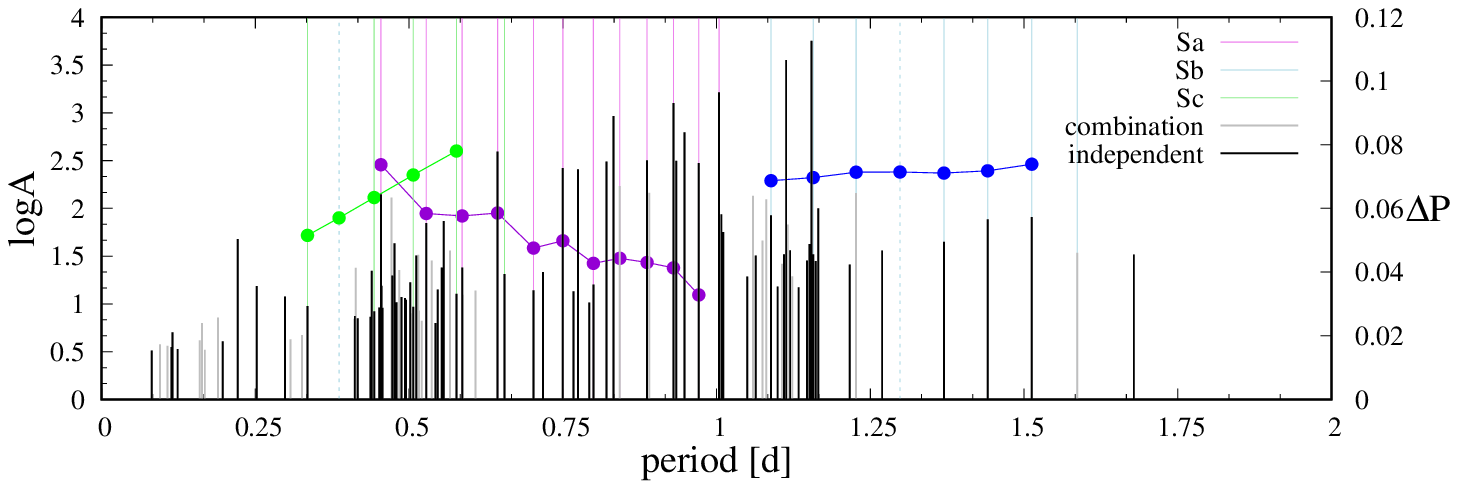}
		\vspace{-0.3cm}
    \captionof{figure}{The same as in Fig.\,\ref{fig:DP1} but for KIC\,4936089.}
    \label{fig:DP5}
	
\end{table}

%\begin{figure}
%	\includegraphics[angle=0, width=\columnwidth]{DP_KIC4936089}
%    \caption{The same as in Fig.\,\ref{fig:DP1} but for KIC\,4936089.}
%    \label{fig:DP5}
%\end{figure}

\subsection{KIC\,4939281}

\begin{table}
	\centering
	\captionof{table}{The same as in Table\,\ref{tab:S1430353} but for KIC\,4939281}
	\label{tab:S4939281}
	\begin{tabular}{rrrrrrr} % four columns, alignment for each
		\hline
		ID          &    $\nu$      & $P$    &$\Delta P$& $A$     & $\frac{\mathrm S}{\mathrm N}$         &   fs\\
		&    $(\mathrm{d}^{-1})$ & $(\mathrm{d})$  &$(\mathrm{d})$  & $(\mathrm{ppm})$    &               &  \\
		\hline
		\multicolumn{7}{|c|}{Sa}\\         
		
$\nu_{69}$ & 1.0721(1) & 0.93279 & 0.06880 & 72(6)  & 4.4 & i \\
$\nu_{80}$ & 1.1574(1) & 0.86400 & 0.04952 & 64(6)  & 4.2 & c \\
$\nu_{90}$ & 1.2278(1) & 0.81448 & 0.07533 & 60(6)  & 4.3 & c \\
$\nu_{93}$ & 1.3529(1) & 0.73915 & 0.05520 & 57(6)  & 4.2 & i \\
$\nu_{57}$ & 1.46211(9)& 0.68395 & 0.07361 & 91(6)  & 5.8 & c \\
$\nu_{34}$ & 1.63843(6)& 0.61034 & 0.06687 & 15(7)  & 8.5 & c \\
$\nu_{49}$ & 1.84004(8)& 0.54347 & 0.05136 & 220(16)& 6.7 & c \\
--- &  \textit{2.03209}   &  \textit{0.49211} &  \textit{0.04452} & ---    & --- & --- \\
$\nu_{3}$  & 2.23421(2)& 0.44759 & 0.07110 & 605(7) & 35 & i \\
--- &  \textit{2.65613}   &  \textit{0.37649} &  \textit{0.06342} & ---    & ---& --- \\
$\nu_{50}$ & 3.19424(8)& 0.31306 & 0.05790 & 115(7) & 9.8 & i \\
$\nu_{128}$& 3.9191(2) & 0.25516 & 0.04287 & 66(7)  & 4.9 & i \\
$\nu_{92}$ & 4.7106(1) & 0.21229 & ---     & 57(6)  & 16 & c \\

		\multicolumn{7}{|c|}{Sb}\\              
		
$\nu_{9}$ & 0.22514(4)  & 4.44175 & 0.72113 & 253(6)  & 6.2 & i \\
$\nu_{1}$ & 0.268772(7) & 3.72062 & 0.49214 & 1991(7) & 33 & i \\
$\nu_{2}$ & 0.309744(7) & 3.22848 & 0.37941 & 1561(6) & 32 & i \\
$\nu_{6}$ & 0.35099(3)  & 2.84906 & ---     & 336(6)  & 8.4 & c \\

	\end{tabular}

	\includegraphics[angle=0,width=\columnwidth]{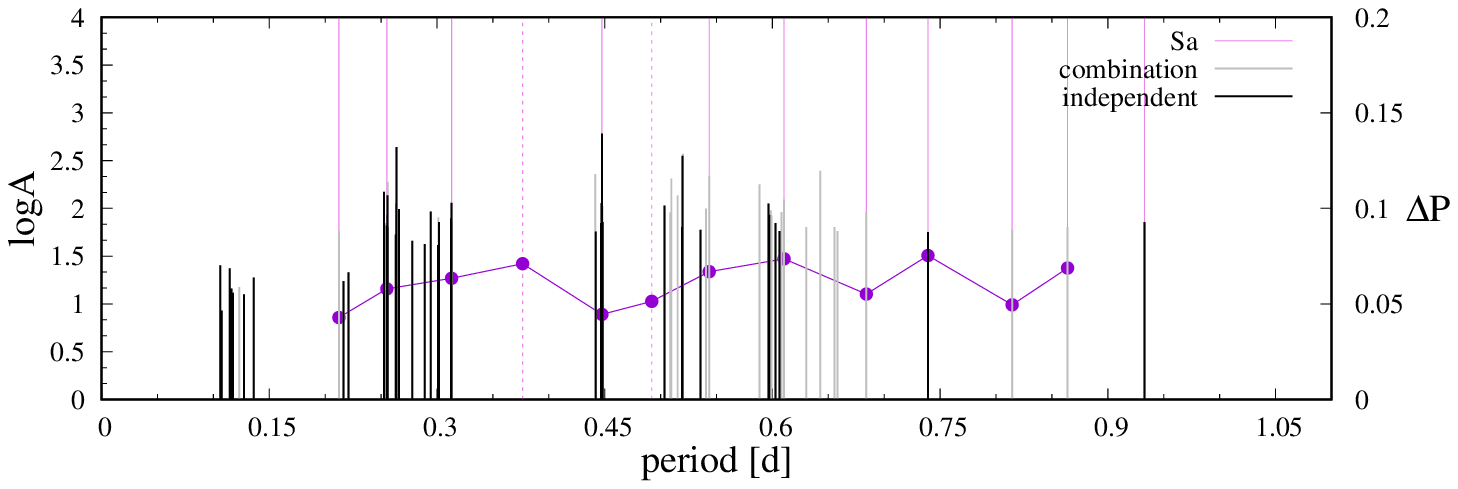}
	\includegraphics[angle=0,width=\columnwidth]{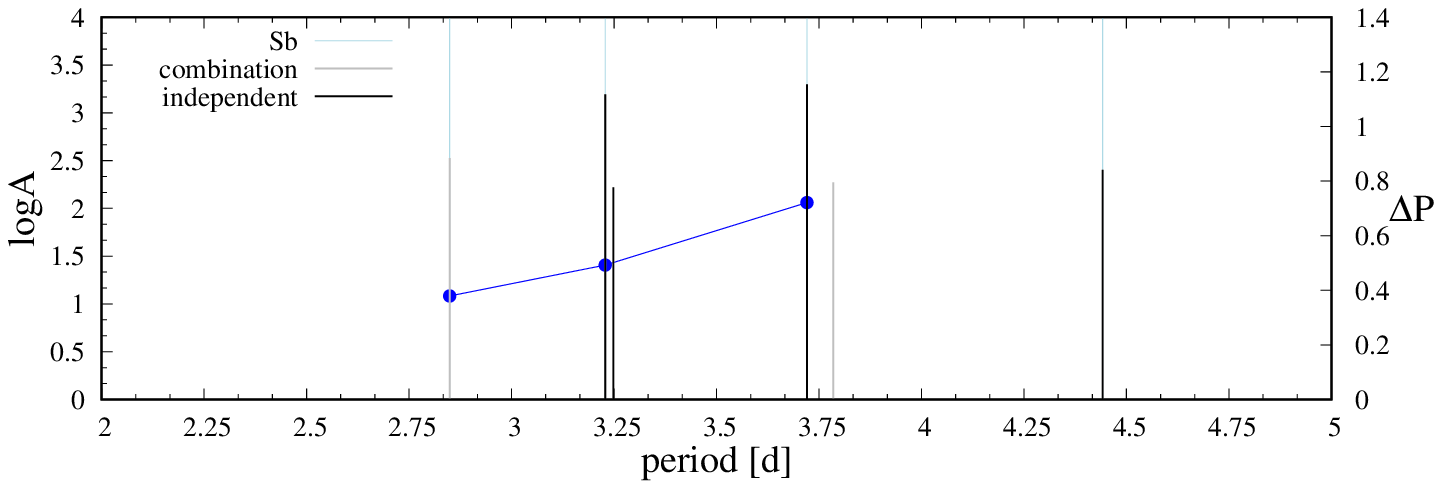}
	\vspace{-0.3cm}
	\captionof{figure}{The same as in Fig.\,\ref{fig:DP1} but for KIC\,4939281.}
	\label{fig:DP6a}
	
\end{table}

The star was recognized  to be a B spectral type by \citet{1964LS....C03....0H} and it  was classified
as the SPB variable by \citet{2012AJ....143..101M}. Then,
 \citet{2015MNRAS.451.1445B} reclassified KIC\,4939281 as the SPB/MAIA type.
The authors defined MAIA as  variable  B star with high frequencies that is cooler than the coolest
 known $\beta$ Cep variable.
Next, \citet{2016MNRAS.460.1318B} redetermined the effective temperature, 
$T_\mathrm{eff}=18900 \pm 660$\,K, surface gravity $\log g=4.16\pm0.17$ and metallicity $[\mathrm{Fe/H}]=-0.4\pm0.1$,
and suggest that high frequency variations are of the $\beta$ Cep type rather than the MAIA type.
Stellar parameters as determined by \citet{2019AJ....157..129H} place the star slightly outside $\beta$ Cep
instability strip (see Fig.\ref{fig:kiel}),
whereas parameters from \citet{2016MNRAS.460.1318B} put it just on its edge.

We found 147 frequency peaks but after removing those closer than the $2.5\times$Rayleigh limit,  we were left with 86 frequency 
peaks in total, i.e., 
%20 firm independent, 33 probably independent and 33 combinations. Those with the highest amplitudes
53  independent and 33 combinations. Those with the highest amplitudes
cover the SPB frequency range.
The group of peaks between 3\,d$^{-1}$ and 10\,d$^{-1}$ may be the manifestation of $\beta$ Cep pulsations.

Frequencies seem to be coherent but amplitudes are variable. In the middle of observational data   one can see 
a kind of outburst (see Fig.\,\ref{fig:KIC4939281lc} in Appendix\,\ref{appendix:A}).
We can not exclude an instrumental origin of this phenomenon, but it covers three quarters which means 
that it was registered on three different CCD  chips. The outburst started approximately 
at 2455530 BJD and lasted about 150 days. Interestingly, during the outburst, amplitudes of some frequencies increased, whereas of others decreased (see frequencies in the proximity of 2.225\,d$^{-1}$ and 1.925\,d$^{-1}$,
Fig.\,\ref{fig:fourier_time_depend_lc0002KIC4939281} in Appendix\,\ref{appendix:A}). This fact suggests that we are dealing rather with the astrophysical signal not the instrumental one.

In the case of this star we found two series (see Tab.\,\ref{tab:S4939281} and Fig.\,\ref{fig:DP6a}) that can be connected with consecutive radial orders of pulsational modes with the same mode degree and azimuthal number. The Sa series may correspond to the dipole axisymmetric modes (see Fig\,\ref{fig:delty_a}\,g) while Sb to the quadrupole retrograde modes. The mean period differences are 0.060\,d (0.303\,d$^{-1}$) for Sa and 0.531\,d (0.0420\,d$^{-1}$) for Sb.

%Sa:
%mean df = 0.303208
%mean dP = 0.0600419

%Sb:
%mean df = 0.0419519
%mean dP = 0.530894

Our representative model for this star (see Fig.\,\ref{fig:freq_all2}\,g) predicts unstable SPB-like modes, 
but $\beta$ Cep-like modes are stable. 
%Taking into account  observations this may be another call for opacity data revision.
% taka konluzja odnosi sie do wielu gwiazd wiec lepiej napisac to w podsumowaniu

\subsection{KIC\,5477601}

The star was classified as SPB by \citet{2012AJ....143..101M} and reclassified as ROT by
 \citet{2015MNRAS.451.1445B}.

We found 43 frequency peaks, but after rejecting those that did not meet the $2.5\times$Rayleigh criterion, 
we ended up with 23 frequencies. 
%According to our criteria, 6 are firm independent,4 are probably independent, 11 are combinations and two are harmonics.
According to our criteria, 10 are  independent, 11 are combinations and two are harmonics.
All found frequency peaks are below 0.4 d$^{-1}$. The highest amplitude frequency peak seems to be incoherent (see Fig.\,\ref{fig:fourier_time_depend_KIC5477601} in Appendix\,\ref{appendix:A}).
Successive  prewhitening leads to unveiling 20 frequencies in close proximity of $\nu_1$. They are also
incoherent. This can originate in independent modes, but most probably result from subtraction incoherent signal. 

Due to the small number of detected frequencies, which additionally seem to be incoherent, we did not look for regularities.

\subsection{KIC\,7630417}
%7

\begin{table}
	\centering
	\captionof{table}{The same as in Table\,\ref{tab:S1430353} but for KIC\,7630417}
	\label{tab:S7630417}
	\begin{tabular}{rrrrrrr} % four columns, alignment for each
		\hline
 ID          &    $\nu$      & $P$    &$\Delta P$& $A$     & $\frac{\mathrm S}{\mathrm N}$         &   fs\\
             &    $(\mathrm{d}^{-1})$ & $(\mathrm{d})$  &$(\mathrm{d})$  & $(\mathrm{ppm})$    &               &  \\
\hline
\multicolumn{7}{|c|}{Sa}\\

   $\nu_{141}$ & 1.87124(6) & 0.534406 & 0.00346 & 31(2)   & 8 & i \\
   ---  &  \textit{1.88344}    &  \textit{0.530944} &  \textit{0.00372} &     --- & --- & --- \\
   $\nu_{264}$ & 1.8967(1)  & 0.527225 & 0.00415 & 17(2)   & 4.8 & i \\
   --  &  \textit{1.91177}    &  \textit{0.523075} &  \textit{0.00388} &     --- & --- & --- \\
   $\nu_{32}$  & 1.92606(3) & 0.519194 & 0.00622 & 84(2)   & 17 & i \\
   --  &  \textit{1.94941}    &  \textit{0.512976} &  \textit{0.00498} &     --- & --- & --- \\
   $\nu_{340}$ & 1.9685(1)  & 0.507996 & 0.00733 & 13(2)   & 4.2 & i \\
   $\nu_{84}$  & 1.99735(5) & 0.500663 & 0.00735 & 43(2)   & 10 & i \\
   $\nu_{1}$   & 2.027100(7)& 0.493315 & 0.01064 & 515(3)  & 66 & i \\
   $\nu_{267}$ & 2.0718(1)  & 0.482672 & 0.00997 & 16(2)   & 5 & i \\
   $\nu_{120}$ & 2.11550(6) & 0.472700 & 0.01197 & 34(2)   & 9.2 & i \\
   $\nu_{34}$  & 2.17046(3) & 0.460731 & 0.01383 & 81(2)   & 18 & i \\
   $\nu_{97}$  & 2.23761(5) & 0.446904 & 0.01221 & 38(2)   & 11 & i \\
   $\nu_{191}$ & 2.30049(8) & 0.434690 & 0.01447 & 22(2)   & 7.5 & i \\
   $\nu_{76}$  & 2.37971(4) & 0.420219 & 0.01834 & 48(2)   & 14 & i \\
   $\nu_{9}$   & 2.48832(2) & 0.401877 & 0.01717 & 165(2)  & 34 & i \\
   $\nu_{73}$  & 2.59936(4) & 0.384710 & 0.01796 & 49(2)   & 14 & i \\
   $\nu_{135}$ & 2.72662(6) & 0.366754 & 0.02025 & 33(2)   & 11 & i \\
   $\nu_{235}$ & 2.8860(1)  & 0.346502 & ---     & 17(2)   & 5.8 & i \\
        
           \multicolumn{7}{|c|}{Sb}\\  
      $\nu_{162}$ & 0.79443(6) & 1.25877 & 0.02847 & 29(2)  & 4.0 & i \\
      $\nu_{4}$   & 0.81281(1) & 1.23030 & 0.03358 & 327(2) & 27 & i \\
      $\nu_{109}$ & 0.83561(5) & 1.19673 & 0.02166 & 39(2)  & 4.8 & i \\
      $\nu_{5}$   & 0.85101(1) & 1.17507 & 0.03033 & 230(2) & 21 & i \\
      $\nu_{147}$ & 0.87356(6) & 1.14474 & 0.02301 & 30(2)  & 4.2 & i \\
      $\nu_{78}$  & 0.89148(5) & 1.12173 & 0.03015 & 47(2)  & 5.9 & i \\
      $\nu_{53}$  & 0.91611(4) & 1.09158 & 0.02553 & 67(2)  & 7.8 & i \\
      $\nu_{7}$   & 0.93804(1) & 1.06605 & 0.03032 & 231(2) & 20 & i \\
      $\nu_{168}$ & 0.96550(7) & 1.03573 & 0.03510 & 27(2)  & 4.1 & i \\
      $\nu_{31}$  & 0.99937(3) & 1.00063 & 0.04113 & 78(2)  & 10 & i \\
      $\nu_{3}$   & 1.04220(1) & 0.95950 & 0.04431 & 325(2) & 30 & i \\
      $\nu_{33}$  & 1.09266(3) & 0.91520 & 0.05356 & 85(2)  & 11 & i \\
      $\nu_{99}$  & 1.16058(5) & 0.86164 & 0.06843 & 38(2)  & 6.0 & i \\
      $\nu_{123}$ & 1.26070(6) & 0.79321 & 0.05509 & 35(2)  & 5.8 & i \\
      $\nu_{35}$  & 1.35480(3) & 0.73812 & 0.06414 & 81(2)  & 12 & i \\
      $\nu_{209}$ & 1.48373(8) & 0.67398 & 0.06727 & 21(2)  & 4.9 & i \\
      $\nu_{146}$ & 1.64825(6) & 0.60670 & 0.08751 & 30(2)  & 7.1 & i \\
      $\nu_{32}$  & 1.92606(3) & 0.51919 & 0.09897 & 84(2)  & 17 & i \\
      $\nu_{76}$  & 2.37971(4) & 0.42022 & ---     & 48(2)  & 14 & i \\
   
 	\end{tabular}

%	\includegraphics[angle=0, width=\columnwidth]{KIC7630417_P_spec_Sab.eps}
%    \includegraphics[angle=0, width=\columnwidth]{KIC7630417_P_spec_Sc.eps}
%    \captionof{figure}{The same as in Fig.\,\ref{fig:DP1} but for KIC\,7630417.}
%    \label{fig:DP7}

\end{table}
 
\setcounter{table}{7}
  
 \begin{table}
 	\centering
 	\captionof{table}{Continued.}
 	\label{tab:S7630417b}
 	\begin{tabular}{rrrrrrr} % four columns, alignment for each
 		\hline
 		ID          &    $\nu$      & $P$    &$\Delta P$& $A$     & $\frac{\mathrm S}{\mathrm N}$         &   fs\\
 		&    $(\mathrm{d}^{-1})$ & $(\mathrm{d})$  &$(\mathrm{d})$  & $(\mathrm{ppm})$    &               &  \\
 		\hline
             
        \multicolumn{7}{|c|}{Sc}\\         
    $\nu_{50}$ & 0.15394(3) & 6.49590 & 0.24213 & 69(2)   & 6.2 & i \\
    $\nu_{87}$ & 0.15990(5) & 6.25377 & 0.22717 & 43(2)   & 4.4 & c \\
    --- & 0.16593    & 6.02660 & 0.23471 &     --- & --- & --- \\
    $\nu_{77}$ & 0.17266(4) & 5.79189 & 0.25152 & 45(2)   & 4.9 & i \\
    $\nu_{45}$ & 0.18049(3) & 5.54037 & 0.32718 & 74(2)   & 6.5 & c \\
    $\nu_{100}$& 0.19182(5) & 5.21319 & 0.26774 & 38(2)   & 4.2 & i \\
    $\nu_{80}$ & 0.20221(5) & 4.94546 & 0.26412 & 45(2)   & 4.7 & i \\
    --- & 0.21361    & 4.68133 & 0.23660 &   ---   & --- & --- \\
    $\nu_{13}$ & 0.22499(2) & 4.44474 & 0.23911 & 134(3)  & 12 & i \\
    --- & 0.23778    & 4.20563 & 0.23720 &    ---  & --- & --- \\
    $\nu_{41}$ & 0.25199(3) & 3.96842 & 0.16805 & 72(2)   & 6.9 & i \\
    --- & 0.26313    & 3.80037 & 0.14973 &  ---    & --- & --- \\
    $\nu_{88}$ & 0.27392(5) & 3.65064 & 0.15974 & 41(2)   & 4.5 & i \\
    $\nu_{68}$ & 0.28646(4) & 3.49091 & 0.14496 & 54(2)   & 5.5 & c \\
    $\nu_{58}$ & 0.29887(4) & 3.34594 & 0.15479 & 61(2)   & 5.8 & c \\
    $\nu_{6}$  & 0.31337(1) & 3.19115 & 0.16862 & 208(2)  & 17 & i \\
    $\nu_{17}$ & 0.33085(2) & 3.02253 & 0.13392 & 135(2)  & 11 & i \\
    --- & 0.34619    & 2.88861 & 0.14308 &   ---   & --- & --- \\
    $\nu_{61}$ & 0.36423(4) & 2.74553 & 0.12082 & 56(2)   & 5.7 & c \\
    --- & 0.38099    & 2.62472 & 0.09992 &   ---   & --- & --- \\
    $\nu_{81}$ & 0.39607(5) & 2.52479 & 0.14271 & 44(2)   & 4.7 & i \\
    --- & 0.41980    & 2.38208 & 0.14812 &   ---   & --- & --- \\
    $\nu_{159}$& 0.44764(5) & 2.23396 & 0.14306 & 34(2)   & 4.0 & c \\
    $\nu_{67}$ & 0.47827(4) & 2.09089 & 0.11231 & 55(2)   & 5.5 & c \\
    $\nu_{52}$ & 0.50541(3) & 1.97858 & 0.10569 & 65(2)   & 6.2 & i \\
    $\nu_{44}$ & 0.53393(3) & 1.87289 & 0.09566 & 70(2)   & 6.5 & c \\
    $\nu_{12}$ & 0.56267(2) & 1.77723 & 0.08887 & 136(2)  & 12 & i \\
    $\nu_{15}$ & 0.59229(2) & 1.68836 & 0.10309 & 131(2)  & 12 & i \\
    $\nu_{112}$& 0.63081(5) & 1.58527 & 0.11528 & 37(2)   & 4.4 & i \\
    $\nu_{18}$ & 0.68028(2) & 1.46999 & ---     & 121(2)  & 11 & i \\
                          
	\end{tabular}
	
	\includegraphics[angle=0, width=\columnwidth]{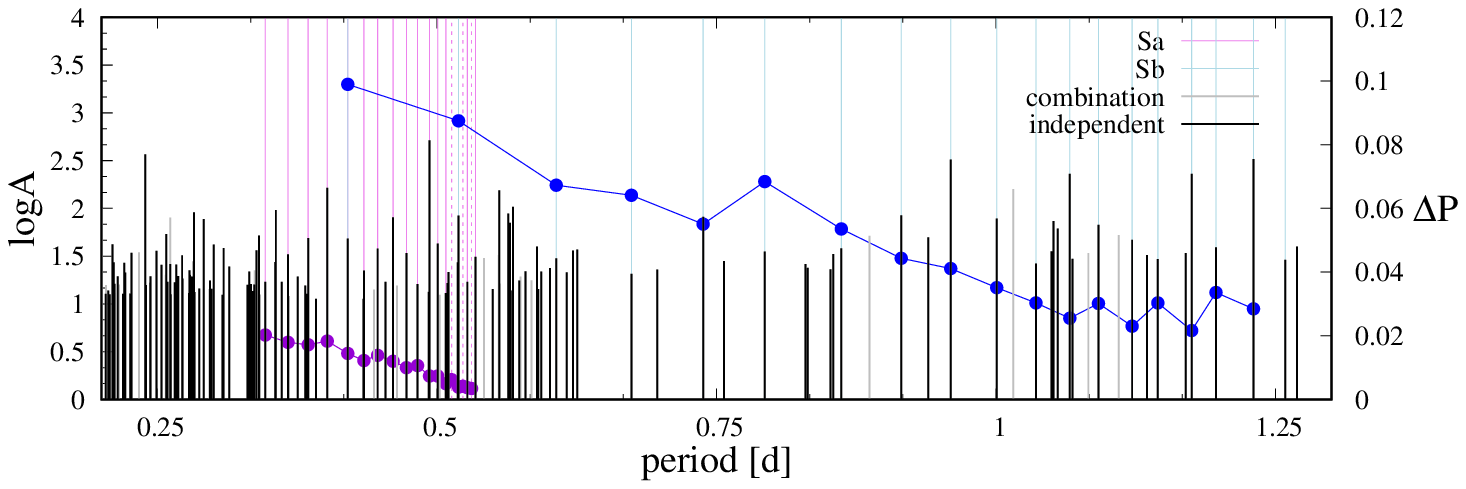}
    \includegraphics[angle=0, width=\columnwidth]{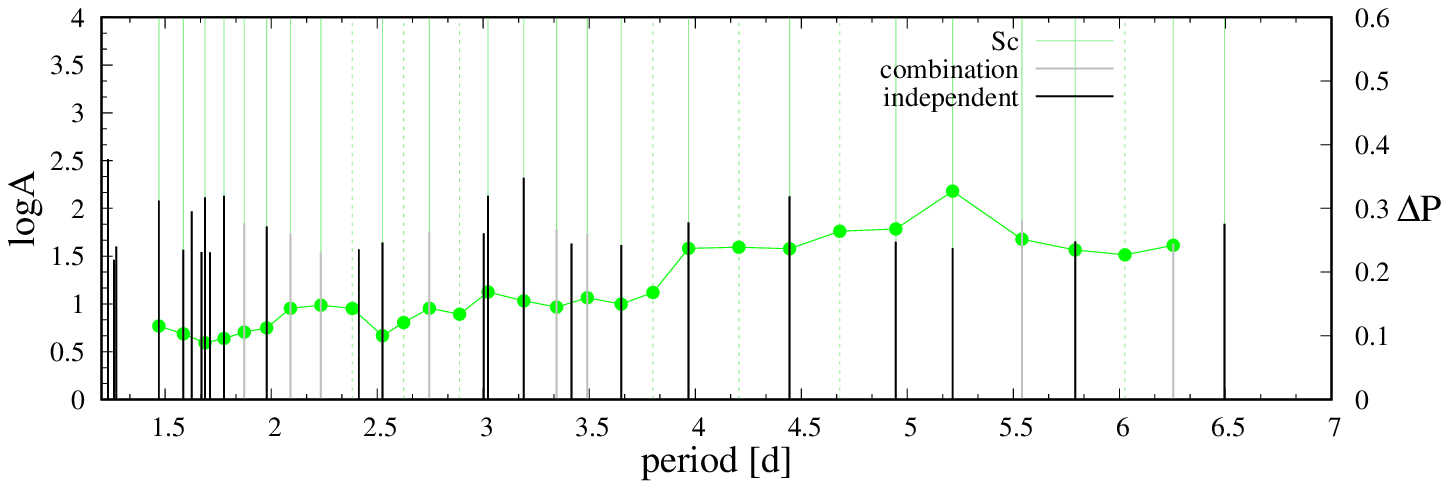}
    \vspace{-0.3cm}
    \captionof{figure}{The same as in Fig.\,\ref{fig:DP1} but for KIC\,7630417.}
    \label{fig:DP7}
	
\end{table}

KIC\,7630417 was classified as the SPB variable  by \citet{2012AJ....143..101M}.
%In its {\it Kepler} light curve we found 502 frequency peaks; 33 firm independent, 233 probably  independent,
In its {\it Kepler} light curve we found 502 frequency peaks; 266 independent,
75 combinations and 3 harmonics.
We rejected 158 peaks because of the $2.5\times$Rayleigh limit.
Independent frequency peaks are seen up to 16.3\,d$^{-1}$ with the most prominent peak
at $\nu_1=2.027102(7)$ d$^{-1}$. Taking into account the range of observed frequencies and the fact
that the star lies on the border of $\beta$ Cep instability strip
we classified it as the SPB/$\beta$ Cep hybrid. 
The extracted frequencies seems to be coherent, although the amplitudes are variable.

For KIC\,7630417 we have found three sequences of frequencies with increasing and decreasing  period spacings.
The frequencies involved in the series are listed in Tab.\,\ref{tab:S7630417} and \ref{tab:S7630417b} (see also Fig.\,\ref{fig:DP7}).

The Sa series seems to be a sequence of prograde modes, but the mode degree identification required more detailed modelling. Its mean period difference is 0.0104\,d (0.0564\,d$^{-1}$). The Sb series, with the mean period difference 0.0466\,d (0.0881\,d$^{-1}$),  fits to dipole axisymmetric modes. The Sc series may be made of dipole retrograde modes. Its mean period difference is 0.1733\,d (0.01815\,d$^{-1}$).

The values of  mean period differences are similar to the mean period spacings  for our representative model presented in Fig.\,\ref{fig:delty_a}\,i. Due to the presence of three sequences containing a lot of frequencies the star seems to be a very promising target for a detailed asteroseismic modelling.

%\begin{figure}
%\includegraphics[angle=0, width=\columnwidth]{KIC7630417_P_spec_Sab.eps}
%\includegraphics[angle=0, width=\columnwidth]{KIC7630417_P_spec_Sc.eps}
%    \caption{The same as in Fig.\,\ref{fig:DP1} but for KIC\,7630417.}
%    \label{fig:DP7}
%\end{figure}

\subsection{KIC\,8167938}
%8

\begin{table}
	\centering
	\captionof{table}{The same as in Table\,\ref{tab:S1430353} but for  KIC\,8167938}
	\label{tab:S8167938}
	\begin{tabular}{rrrrrrr} % four columns, alignment for each
		\hline
 ID          &    $\nu$      & $P$    &$\Delta P$& $A$     & $\frac{\mathrm S}{\mathrm N}$         &   fs\\
             &    $(\mathrm{d}^{-1})$ & $(\mathrm{d})$  &$(\mathrm{d})$  & $(\mathrm{ppm})$    &               &  \\
\hline
       \multicolumn{7}{|c|}{Sa}\\      
      $\nu_{54} $ & 1.31827(4) & 0.75857 & 0.01698 & 55(2) & 12 & i \\
      $\nu_{150}$ & 1.3485(1)  & 0.74159 & 0.02694 & 14(2) & 4.7 & i \\
      $\nu_{27} $ & 1.39930(3) & 0.71465 & 0.03354 & 107(2)& 21 & c \\
      $\nu_{110}$ & 1.46820(6) & 0.68111 & 0.04436 & 27(2) & 10 & i \\
      $\nu_{7}$   & 1.570480(6)& 0.63675 & 0.05505 & 819(2)& 64 & c \\
      $\nu_{156}$ & 1.7191(1)  & 0.58169 & 0.06855 & 12(2) & 5.6 & i \\
      $\nu_{106}$ & 1.94877(6) & 0.51315 & 0.08553 & 31(2) & 15 & i \\
      $\nu_{31}$  & 2.33852(3) & 0.42762 & 0.10924 & 94(2) & 32 & i \\
      $\nu_{138}$ & 3.14093(8) & 0.31838 & 0.14149 & 20(2) & 13 & c \\
      $\nu_{191}$ & 5.6534(4)  & 0.17689 & ---     & 4(2)  & 4.3 & c \\
             
       \multicolumn{7}{|c|}{Sb}\\      
              
      $\nu_{72}$  & 1.37913(4) & 0.72509 & 0.02007 & 43(2) & 12 & i \\
      $\nu_{152}$ & 1.4184(1)  & 0.70502 & 0.02508 & 13(2) & 4.8 & c \\
      $\nu_{158}$ & 1.4707(1)  & 0.67994 & 0.03067 & 12(2) & 4.2 & c \\
      $\nu_{128}$ & 1.54018(7) & 0.64927 & 0.03838 & 25(2) & 9.5 & i \\
      $\nu_{35}$  & 1.63694(3) & 0.61090 & 0.05784 & 81(2) & 23 & i \\
      $\nu_{9}$   & 1.80814(1) & 0.55306 & 0.08635 & 405(2) & 62 & i \\
      $\nu_{57}$  & 2.14269(4) & 0.46670 & 0.10017 & 52(2) & 22 & c \\
      $\nu_{94}$  & 2.72830(6) & 0.36653 & ---     & 33(2) & 18 & i \\
             
         \multicolumn{7}{|c|}{Sc}\\
         
    $\nu_{65}$ & 0.70638(4)  & 1.41568 & 0.01589 & 52(2)  & 6.7 & i \\
    $\nu_{17}$ & 0.71440(2)  & 1.39978 & 0.02052 & 154(2) & 15 & i \\
    $\nu_{43}$ & 0.72502(3)  & 1.37926 & 0.01772 & 68(2)  & 8.9 & c \\
    $\nu_{-1}$ & 0.73446     & 1.36154 & 0.02352 &  ---    & --- & --- \\
    $\nu_{96}$ & 0.74737(6)  & 1.33802 & 0.02785 & 35(2)  & 5.3 & c \\
    $\nu_{21}$ & 0.76326(2)  & 1.31018 & 0.02731 & 134(2)  & 15 & c \\
    $\nu_{2}$  & 0.77951(4)  & 1.28286 & 0.02380 & 4363(2) & 57 & i \\
    $\nu_{8}$  & 0.79424(1)  & 1.25906 & 0.02673 & 423(2) & 27 & i \\
    $\nu_{81}$ & 0.81147(5)  & 1.23233 & 0.02716 & 41(2)  & 6.5 & c \\
    $\nu_{1}$  & 0.829755(4) & 1.20517 & 0.02966 & 6191(2) & 58 & i \\
    $\nu_{-1}$ & 0.85069     & 1.17551 & 0.02917 &   --- & --- & --- \\
    $\nu_{124}$& 0.87234(6)  & 1.14635 & 0.03055 & 26(2)( & 5.2 & i \\
    $\nu_{3}$  & 0.896222(5) & 1.11580 & 0.03184 & 2784(2)& 58 & i \\
    $\nu_{10}$ & 0.92254(2)  & 1.08396 & 0.03373 & 254(2)  & 18 & i \\
    $\nu_{56}$ & 0.95218(4)  & 1.05023 & 0.03513 & 50(2)  & 8.3 & i \\
    $\nu_{127}$& 0.98513(7)  & 1.01509 & 0.04292 & 27(2) & 5.5 & c \\
    $\nu_{92}$ & 1.02862(5)  & 0.97218 & 0.04084 & 34(2) & 7.1 & c \\
    $\nu_{119}$& 1.07373(6)  & 0.93134 & ---     & 27(2) & 6.1 & i \\

	\end{tabular}
	
	\includegraphics[angle=0, width=\columnwidth]{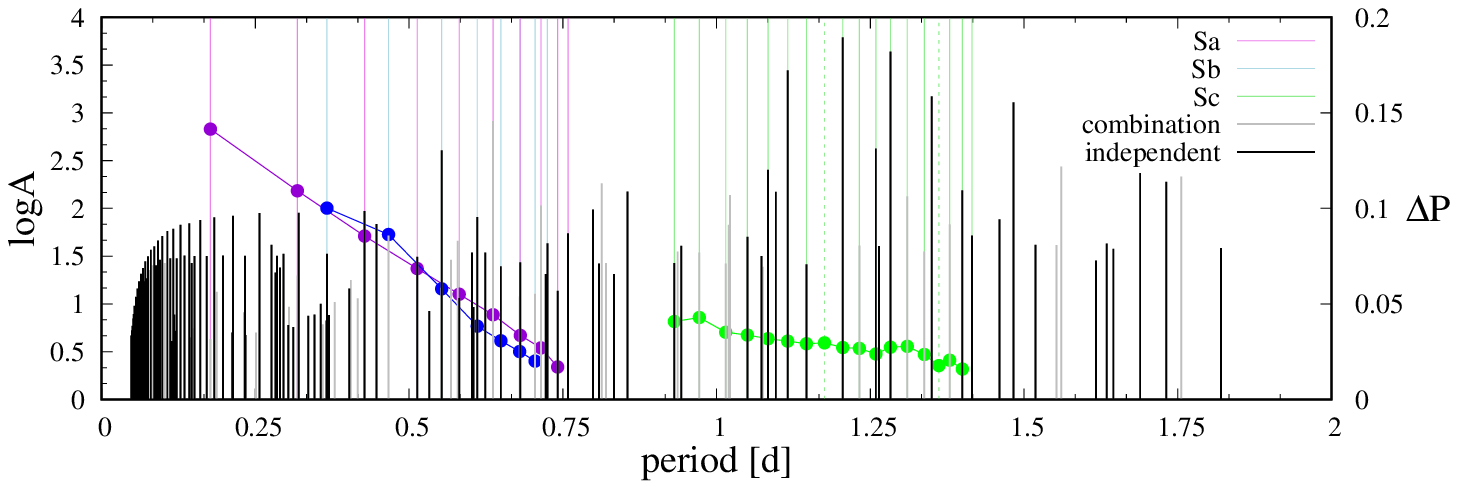}
	\vspace{-0.3cm}
    \captionof{figure}{The same as in Fig.\,\ref{fig:DP1} but for KIC\,8167938.}
    \label{fig:DP8}
	
\end{table}

\citet{2011AJ....142..160S} list the star as an eclipsing binary.  KIC\,8167938 was classified as SPB by \citet{2012AJ....143..101M}. This classification was confirmed by \citet{2015MNRAS.451.1445B}.
Radial velocities, as determined by \citet{2016A&A...594A..39F}
seem to be variable.
For the three spectra the authors obtained $16.0\pm17.8$\,km\,s$^{-1}$,
$-30.6\pm24.4$\,km\,s$^{-1}$ and $-23.1\pm31.5$\,km\,s$^{-1}$, respectively.
However, large uncertainties do not allow for certain conclusions.

We found 194 frequency peaks of which 18 were rejected because of the $2.5\times$Rayleigh criterion.% 20 frequencies were found to be firm independent, 88  probably independent, 52 combinations and 16 harmonics.
 108 frequencies were found to be  independent, 52 combinations and 16 harmonics.
The dominant frequency peak is 
$\nu_1=0.829755(4)$ d$^{-1}$ . For the second one, 
$\nu_2=0.779507(4)$ d$^{-1}$, we found a set of eight harmonics and one subharmonic 
($\nu_4=0.389756(5)$\,d$^{-1}$).

The peak at $\nu_4$ is the  orbital frequency as given by \citet{2011AJ....142..160S}.  We also see many combinations with the orbital frequency. There are  independent frequencies up to 20.7\,d$^{-1}$. Moreover, we see the first harmonic of $\nu_1$,
ie. $\nu_{95}=1.65950(6)$\,d$^{-1}$, second  harmonic of  $\nu_6=0.674252(5)$\,d$^{-1}$,
second harmonic of $\nu_9=1.80814(1)$\,d$^{-1}$ and sixth harmonic of $\nu_{86}=0.54936(5)$\,d$^{-1}$.
Based on TDP we conclude that frequencies seem to be coherent.

After removing variations associated with pulsations and phasing data with $\nu_4$ there are seen two eclipses. 
Therefore the star is of particular interest because it is the eclipsing binary with the SPB component.

Furthermore, three series of regular patterns in the oscillation spectrum were identified
(see Tab.\,\ref{tab:S8167938} and Fig.\,\ref{fig:DP8}). The mean period differences are 0.065\,d (0.482\,d$^{-1}$) for Sa, 0.051\,d (0.193\,d$^{-1}$) for Sb and 0.025\,d (0.0216\,d$^{-1}$) for Sc.
The Sc sequence fits well to the dipole axisymmetric modes. The others two series can be prograde modes, but the more precise identification requires enhanced modelling. We are planning to study the star thoroughly in a separate paper.

%Sa:
%mean df = 0.481678
%mean dP = 0.0646318

%Sb:
%mean df = 0.192738
%mean dP = 0.0512236

%Sc:
%mean df = 0.0216088
%mean dP = 0.0284906

%\begin{figure}
%	\includegraphics[angle=0, width=\columnwidth]{DP_KIC8167938}
%    \caption{The same as in Fig.\,\ref{fig:DP1} but for KIC\,8167938.}
%    \label{fig:DP8}
%\end{figure}

\subsection{KIC\,8264293}

\begin{table}
	\centering
	\captionof{table}{The same as in Table\,\ref{tab:S1430353} but for  KIC\,8264293}
	\label{tab:S8264293}
	\begin{tabular}{rrrrrrr} % four columns, alignment for each
		\hline
 ID          &    $\nu$      & $P$    &$\Delta P$& $A$     & $\frac{\mathrm S}{\mathrm N}$         &   $fs$\\
             &    $(\mathrm{d}^{-1})$ & $(\mathrm{d})$  &$(\mathrm{d})$  & $(\mathrm{ppm})$    &               &  \\
\hline
\multicolumn{7}{|c|}{Sa}\\

$\nu_{133}$ & 2.82140(9)  & 0.35444 & 0.00183 & 3.0(4)   & 5.1 & i \\
$\nu_{40}$  & 2.83603(3)  & 0.35261 & 0.00197 & 11.1(4)  & 13 & c \\
$\nu_{6}$   & 2.851930(6) & 0.35064 & 0.00211 & 82(2)    & 52 & i \\
$\nu_{41}$  & 2.86917(3)  & 0.34853 & 0.00227 & 11.2(4)  & 14 & i \\
$\nu_{19}$  & 2.88794(2)  & 0.34627 & 0.00244 & 19.7(4)  & 23 & i \\
$\nu_{10}$  & 2.90846(1)  & 0.34382 & 0.00263 & 44(1)    & 43 & i \\
$\nu_{2}$   & 2.930918(6) & 0.34119 & 0.00284 & 107.4(9) & 49 & i \\
$\nu_{1}$   & 2.95556(6)  & 0.33835 & 0.00308 & 115.0(9) & 48 & i \\
$\nu_{3}$   & 2.982687(6) & 0.33527 & 0.00333 & 106.3(9) & 57 & i \\
$\nu_{4}$   & 3.012605(6) & 0.33194 & 0.00332 & 94.8(4)  & 60 & i \\
$\nu_{11}$  & 3.04307(1)  & 0.32862 & 0.00469 & 33.4(4)  & 43 & c \\
$\nu_{27}$  & 3.08709(2)  & 0.32393 & 0.00439 & 14.3(4)  & 22 & i \\
$\nu_{9}$   & 3.12948(1)  & 0.31954 & 0.00489 & 44.7(4)  & 44 & i \\
$\nu_{156}$ & 3.1781(1)   & 0.31465 & 0.00563 & 2.4(4)   & 4.9 & i \\
---  & \textit{3.23601}     & \textit{0.30902} & \textit{0.00607} &   ---    & --- & --- \\
$\nu_{86}$  & 3.30087(4)  & 0.30295 & ---     & 6.6(4)   & 12 & c \\

	\end{tabular}
	
	\includegraphics[angle=0, width=\columnwidth]{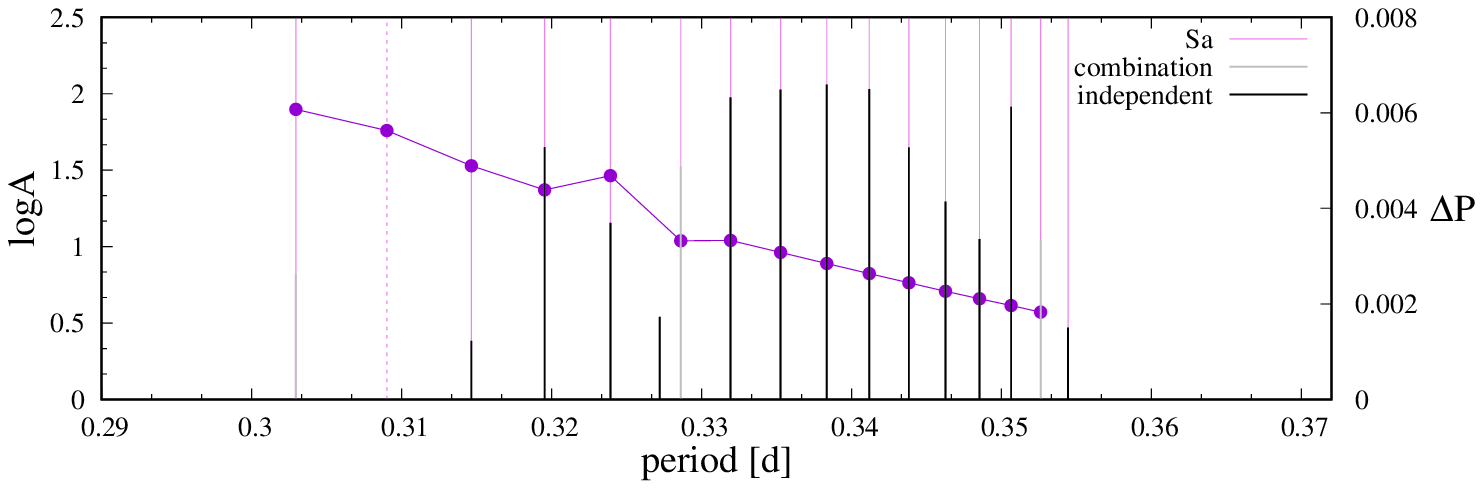}
	\vspace{-0.3cm}
    \captionof{figure}{The same as in Fig.\,\ref{fig:DP1} but for KIC\,8264293.}
    \label{fig:DP9}
	
\end{table}

KIC\,8264293 lies in the field of the open cluster NGC\,6866. The membership probability 
was estimated by several author and it is $79\%$  according to \citet{2008AJ....136..118F},
$90\%$ according to \citet{2009AcA....59..193M} and $43\%$ as given by \citet{2015MNRAS.453.1095B}.

\citet{2012AJ....143..101M} classified the star as a hybrid $\beta$\,Cep/SPB pulsator, whereas \citet{2015MNRAS.451.1445B}
as a variable caused by the rotational modulation (ROT).

We found 171 frequency peaks and rejected 69 due to the $2.5\times$Rayleigh limit. 
%22 frequencies we classified as firm independent, 40 as probably  independent, 33 as combinations and 2 as harmonics.
62 frequencies we classified as independent, 33 as combinations and 7 as harmonics.
We detected coherent as well as incoherent frequencies. The first ones
are associated with pulsational variability. The second ones (in the range $2.33-2.45$\,d$^{-1}$) may come from the variability in the disk around the star. 
This is confirmed by the weak emission  in H$\alpha$ (D. Mo{\'z}dzierski, private communication).
Thus, KIC\,8264293 is the SPB pulsator with the Be phenomenon.

The oscillation spectrum of the star shows the frequency grouping typical for the fast rotating SPB stars. There are four distinct groups of frequencies (see Fig.\ref{fig:freq_all2}\,k).

The star has very regular structures in the frequency spectrum.
All frequencies in the third group form nearly straight, decreasing  line in
the period spacing $vs.$ period diagram.
Only two modes are slightly shifted from the linear dependence, see Tab.\,\ref{tab:S8264293} and Fig.\ref{fig:DP9}.
In the fourth group one can also identify a sequence of frequencies with decreasing period spacing, but it is a bit less regular.
In other frequency group we could not find any significant regularities. 
The mean period difference is 0.0034\,d (0.03197\,d$^{-1}$).

The star is indeed very interesting and requires more detailed and advanced studies. The research is ongoing 
and will be published in a separate paper.

\subsection{KIC\,8381949}

%\FloatBarrier
\begin{table}
%\begin{table}
	\centering
	\captionof{table}{The same as in Table\,\ref{tab:S1430353} but for  KIC\,8381949}
	\label{tab:S8381949}
	\begin{tabular}{rrrrrrr} % four columns, alignment for each
		\hline
 ID          &    $\nu$      & $P$    &$\Delta P$& $A$     & $\frac{\mathrm S}{\mathrm N}$         &   fs\\
             &    $(\mathrm{d}^{-1})$ & $(\mathrm{d})$  &$(\mathrm{d})$  & $(\mathrm{ppm})$    &               &  \\
\hline
  \multicolumn{7}{|c|}{Sa}\\          
  $\nu_{1}$ & 0.22945(1) & 4.35816 & 0.48459 & 636(5) & 24 & i\\
  $\nu_{13}$& 0.25816(5) & 3.87357 & 0.35453 & 154(5) & 6.7 & i \\
  $\nu_{8}$ & 0.28417(4) & 3.51904 & 0.23628 & 207(5) & 8.8 & i \\
  $\nu_{4}$ & 0.30462(3) & 3.28275 & 0.18910 & 312(5) & 13 & i \\
  $\nu_{3}$ & 0.32324(2) & 3.09365 & ---     & 363(5) & 15 & i \\
    \multicolumn{7}{|c|}{Sb}\\ 
    $\nu_{9}$  & 1.29847(4) & 0.77014 & 0.00105 & 212(5) & 8.7 & i \\
    $\nu_{33}$ & 1.30024(6) & 0.76909 & 0.00194 & 109(5) & 5.3 & c \\
    $\nu_{10}$ & 1.30352(4) & 0.76715 & 0.00279 & 178(5) & 8.0 & c \\
    $\nu_{40}$ & 1.30828(6) & 0.76436 & 0.00443 & 118(5) & 5.0 & i \\
    $\nu_{24}$ & 1.31591(6) & 0.75993 & 0.00735 & 138(5) & 5.7 & i \\
    $\nu_{31}$ & 1.32876(6) & 0.75258 & 0.00830 & 121(5) & 5.5 & i \\
    $\nu_{174}$& 1.34358(8) & 0.74428 & 0.01133 & 81(5)  & 4 & i \\
    $\nu_{23}$ & 1.36435(6) & 0.73295 & 0.01536 & 136(5) & 5.9 & c \\
    $\nu_{14}$ & 1.39355(5) & 0.71759 & 0.02444 & 155(5) & 6.8 & i \\
    $\nu_{6}$  & 1.44268(3) & 0.69316 & 0.03504 & 245(5) & 11 & i \\
    $\nu_{64}$ & 1.51949(8) & 0.65812 & 0.06202 & 81(5)  & 4.3 & c \\
    $\nu_{2}$  & 1.67758(2) & 0.59610 & 0.13056 & 425(5) & 20 & i \\
    $\nu_{76}$ & 2.1480(1)  & 0.46554 & 0.19052 & 46(5)  & 4.7 & i \\
    $\nu_{19}$ & 3.63612(5) & 0.27502 & ---     & 138(5) & 39 & i \\
    \multicolumn{7}{|c|}{Sc}\\           
   $\nu_{35}$ & 0.02911(6) & 34.3479 & 8.76010 & 118(5) & 4.8 & c \\
   $\nu_{30}$ & 0.03908(6) & 25.5878 & 4.99624 & 120(5) & 5.1 & c \\
   $\nu_{34}$ & 0.04857(6) & 20.5915 & 4.05143 & 115(5) & 4.8 & i \\
   --- &\textit{ 0.06046 }   & \textit{16.5401} & \textit{3.24861} &   ---  & --- & --- \\
   $\nu_{28}$ & 0.07524(6) & 13.2915 & 2.62799 & 131(5) & 5.1 & c \\
   $\nu_{39}$ & 0.09378(6) & 10.6635 &  ---    & 110(5) & 4.7 & c \\
	\end{tabular}
	\includegraphics[angle=0, width=\columnwidth]{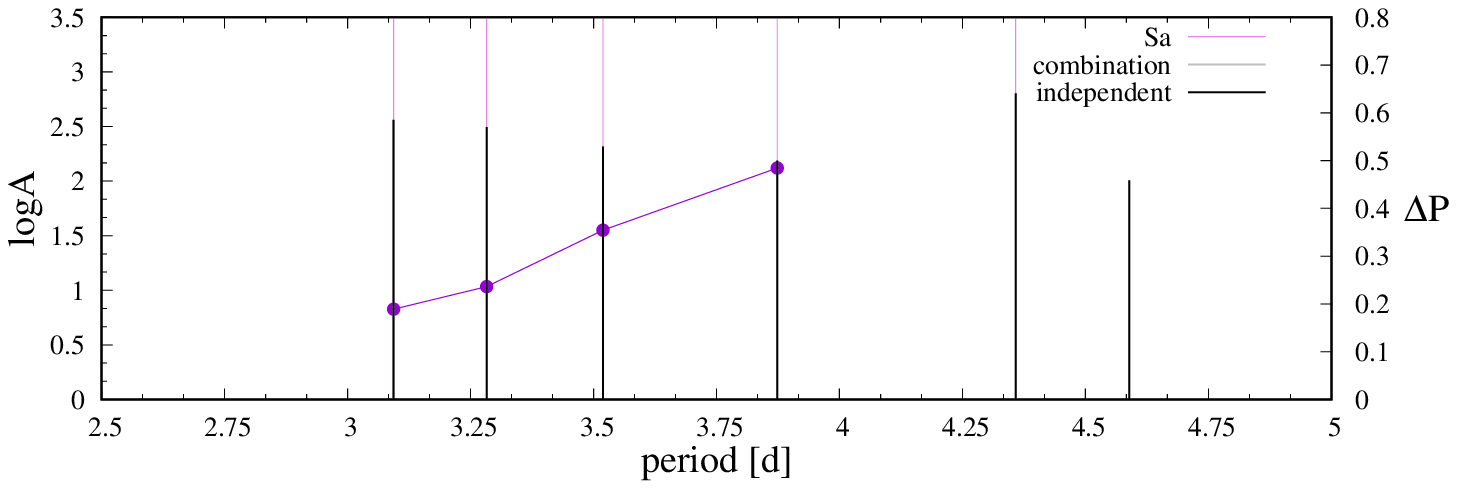}	
	\includegraphics[angle=0, width=\columnwidth]{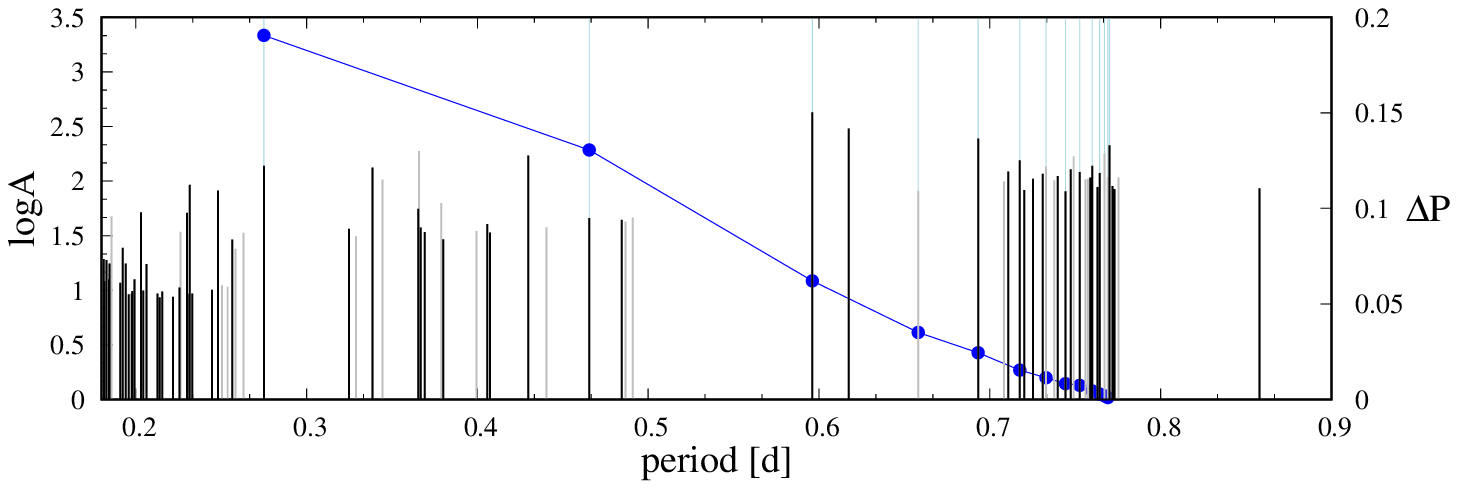}
	\includegraphics[angle=0, width=\columnwidth]{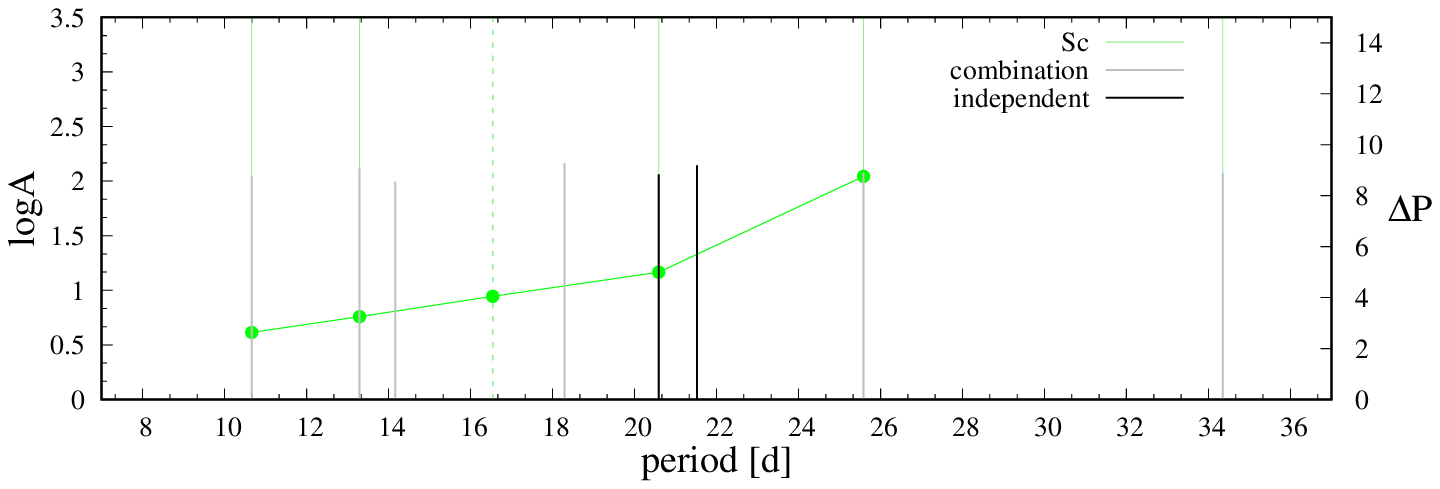}
	\vspace{-0.3cm}
    \captionof{figure}{The same as in Fig.\,\ref{fig:DP1} but for KIC\,8381949.}
    \label{fig:DP10}
\end{table}
%\FloatBarrier

%\FloatBarrier
\citet{1998ApJS..115..271R}  assigned to the star  the spectral type OB.
\citet{2011MNRAS.413.2403B} and  \citet{2015MNRAS.451.1445B} included it to the hybrid SPB/$\beta$\,Cep pulsators.
\citet{2011MNRAS.413.2403B} derived stellar parameters to be
$T_\mathrm{eff}=24500\,\mathrm{K}$, $\log g=4.3$ (from spectroscopy), $T_\mathrm{eff}=21000\,\mathrm{K}$,, $\log g=3.82$ (from
Str\"omgren photometry) and $T_\mathrm{eff}=22 400\pm7400\,\mathrm{K}$ (from SED fitting).

We found 337 frequency peaks and rejected 67 because of the $2.5\times$Rayleigh resolution. %Then, 40 frequencies were identified as firm independent,  156  as probably independent, 68  as combinations and 6  as harmonics.
Then, 196 frequencies were identified as  independent, 68  as combinations and 6  as harmonics.
One can see the rich oscillation spectrum for this star, both in the high and low frequency range (Fig.\,\ref{fig:delty_a}\,l).
Given that the star lies at an overlap of the SPB and $\beta$ Cep instability strips, we concluded that KIC\,8381949 is the hybrid SPB/$\beta$ Cep pulsator.

We can see coherent and incoherent frequencies as well as variations of amplitudes. 
The incoherence mainly concerns the frequencies in the vicinity of 1.3\,d$^{-1}$--1.4\,d$^{-1}$.

For this fast rotating star we detected three regular series of frequencies (see Tab.\,\ref{tab:S8381949} and Fig.\,\ref{fig:DP10}). The Sb series can be associated with quadrupole prograde modes. Its mean period difference is 0.038\,d (0.180\,d$^{-1}$). The Sa sequence can be explained by dipole retrograde modes. The mean period difference is 4.737\,d (0.013\,d$^{-1}$). The Sa series is very interesting. Its increasing period spacings fits to the retrograde modes whose frequencies, in a co-rotating frame of reference, are lower that the rotational frequency. This, however, requires more detailed analysis. The mean period difference for Sa series is  0.316\,d (0.024\,d$^{-1}$).

\subsection{KIC\,8714886}
%11

\begin{table}
	\centering
	\captionof{table}{The same as in Table\,\ref{tab:S1430353} but for  KIC\,8714886}
	\label{tab:S8714886}
	\begin{tabular}{rrrrrrr} % four columns, alignment for each
		\hline
 ID          &    $\nu$      & $P$    &$\Delta P$& $A$     & $\frac{\mathrm S}{\mathrm N}$         &   fs\\
             &    $(\mathrm{d}^{-1})$ & $(\mathrm{d})$  &$(\mathrm{d})$  & $(\mathrm{ppm})$    &               &  \\
\hline

 \multicolumn{7}{|c|}{Sa}\\
 $\nu_{96}$ & 0.63144(5) & 1.58368 & 0.02379 & 68(2)   & 5.2 & i \\
 $\nu_{48}$ & 0.64107(3) & 1.55988 & 0.03548 & 147(2)  & 8.6 & i \\
 $\nu_{2}$  & 0.65599(1) & 1.52441 & 0.05570 & 1721(2) & 32 & i \\
 $\nu_{38}$ & 0.68087(3) & 1.46871 & 0.05365 & 208(2)  & 11 & i \\
 $\nu_{15}$ & 0.70668(2) & 1.41507 & 0.06340 & 624(2)  & 22 & i \\
 $\nu_{110}$& 0.73983(6) & 1.35166 & 0.04941 & 66(2)   & 5.0 & i \\
 $\nu_{3}$  & 0.76790(1) & 1.30225 & 0.07353 & 1621(2) & 31 & i \\
 $\nu_{183}$& 0.81386(8) & 1.22872 & 0.06359 & 38(2)   & 4.0 & i \\
 $\nu_{29}$ & 0.85827(2) & 1.16513 & 0.07225 & 326(2)  & 17 & i \\
 $\nu_{11}$ & 0.91501(1) & 1.09288 & 0.07191 & 1008(2) & 30 & i \\
 $\nu_{14}$ & 0.97946(2) & 1.02097 & 0.07303 & 660(3)  & 23 & i \\
 $\nu_{81}$ & 1.05491(5) & 0.94794 & 0.07807 & 84(2)   & 7.7 & i \\
 $\nu_{12}$ & 1.14959(2) & 0.86988 & 0.07447 & 710(2)  & 26 & i \\
 $\nu_{4}$  & 1.25722(1) & 0.79541 & 0.07423 & 1621(2) & 35 & i \\
 $\nu_{202}$& 1.3866(1)  & 0.72117 & 0.07911 & 28(2) & 5.7 & c \\
 $\nu_{228}$& 1.5575(1)  & 0.64207 & 0.06135 & 22(2) & 5.8 & i \\
 $\nu_{292}$& 1.7220(2)  & 0.58072 & 0.05805 & 13(2) & 4.0 & i \\
 $\nu_{309}$& 1.9133(2)  & 0.52267 & 0.05771 & 12(2) & 4.1 & i \\
 $\nu_{236}$& 2.1507(1)  & 0.46497 & 0.05328 & 20(2) & 6.2 & i \\
 $\nu_{285}$& 2.4291(2)  & 0.41168 & 0.04830 & 13(2) & 4.5 & i \\
 $\nu_{223}$& 2.7519(1)  & 0.36339 & 0.04361 & 22(2) & 6.8 & c \\
 $\nu_{205}$& 3.1272(1)  & 0.31978 & 0.03993 & 26(2) & 8.1 & i \\
 $\nu_{259}$& 3.5734(2)  & 0.27984 & 0.03454 & 16(2) & 5.5 & i \\
 $\nu_{324}$& 4.0766(2)  & 0.24530 & 0.03195 & 11(2) & 4.4 & i \\
 $\nu_{248}$& 4.6870(2)  & 0.21336 & ---     & 17(2) & 7.2 & c \\

	\end{tabular}
	
	\includegraphics[angle=0, width=\columnwidth]{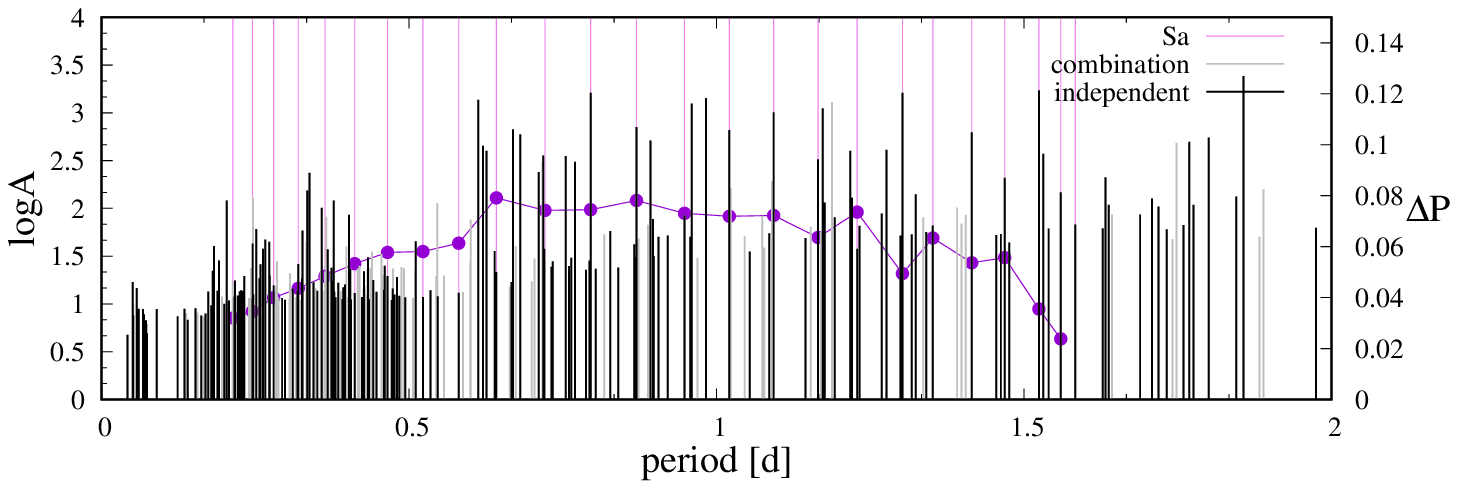}
	\includegraphics[angle=0, width=\columnwidth]{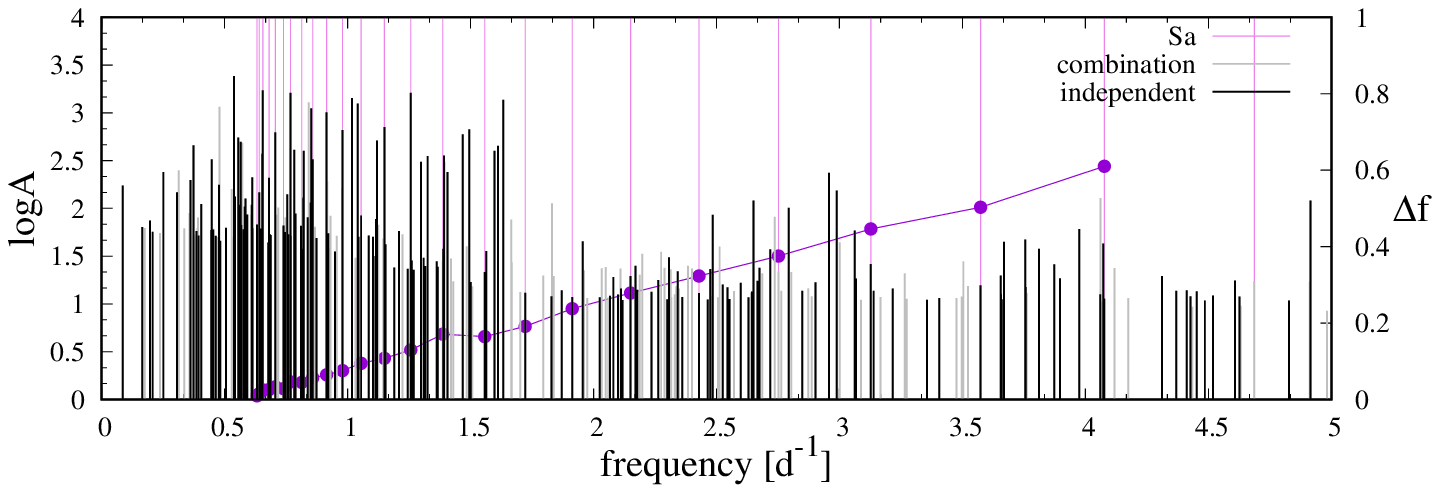}
	\vspace{-0.3cm}
    \captionof{figure}{Upper panel: the same as in Fig.\,\ref{fig:DP1} but for KIC\,8714886. Bottom panel: Quasi-regular frequency spacings found in the oscillation spectrum of KIC\,8714886.
     Frequency spacing is constituted from  the same modes as in the upper panel.}
    \label{fig:DP11}

\end{table}

The star was first classified as F spectral type by \citet{1993yCat.3135....0C}. Based on the analysis of the Kepler light curve,
\citet{2011MNRAS.413.2403B} classified the star among  hybrid $\beta$\,Cep/SPB pulsators with frequency 
grouping. Later \citet{2015MNRAS.451.1445B} classified it as the SPB/MAIA star.
Atmospheric parameters as determined by \citet{2011MNRAS.413.2403B}
are
$T_\mathrm{eff}=19000\,\mathrm{K}$, $\log g=4.3$ (from spectroscopy), $T_\mathrm{eff}=18505\,\mathrm{K}$, $\log g=4.49$
(from Str\"omgren photometry) and $T_\mathrm{eff}=18000\pm3400\,\mathrm{K}$ (from SED fitting).

In the light curve of KIC\,8714886, we found 376 frequency peaks and rejected 73 of them because of the $2.5\times$Rayleigh limit.
%26 peaks were assigned  as firm independent, 166 as probably independent, 103 as combinations and 8 as harmonics.
192 peaks were assigned  as  independent, 103 as combinations and 8 as harmonics.
The highest amplitude frequencies have the values typical for SPB pulsators but there are also small amplitude peaks up
to 23.7\,d$^{-1}$.

The star is located inside the SPB instability strip and close to the $\beta$ Cep instability strip (see Fig.\,\ref{fig:kiel}). Therefore, we classified it as the SPB/$\beta$ Cep hybrid pulsator.
The TDP analysis indicated that  at least highest amplitude frequencies are coherent. 

In the oscillation spectrum, we identified one sequence of frequencies, that may origin from mode of the same degree and azimuthal number. The period differences resemble a parabola, but the frequency differences are nearly straight line (see Tab.\,\ref{tab:S8714886} and Fig.\,\ref{fig:DP11}).  The mean period spacing is 0.057\,d (0.169\,d$^{-1}$).
The series can be reproduce by retrograde dipole modes or axisymmetric dipole modes. The detailed seismic modelling is needed to clarify the identification.  The biggest problem is that the series do not explain most of the others frequencies. And we do not see another regularities in the frequency spectrum. Therefore, the series may be accidental and has to be considered as uncertain.

\subsection{KIC\,9227988}
%12
The star was classified as the SPB-type variable by \citet{2012AJ....143..101M}. This type was
confirmed by \citet{2015MNRAS.451.1445B}. 
We found 73 frequency peaks and rejected 10 because of the $2.5\times$Rayleigh limit. %From the rest we identified 8 firm independent frequencies, 10 probably independent, 44 combinations and one harmonic.
From the rest we identified 18  independent frequencies,  44 combinations and one harmonic.
All  independent frequencies are below 0.8\,d$^{-1}$. High amplitude modes appear to be coherent, whereas 
those with small amplitude may be incoherent.

For this star we  notice  two frequency patterns  in the oscillation spectrum  (Tab.\,\ref{tab:S9227988}).
The Sa sequence has decreasing period spaces with the mean period difference 0.0224\,d (0.0083\,d$^{-1}$).
The second series has more-less constant period spacing with the mean period difference  0.2415\,d
(0.0169 d$^{-1}$). Both series are shown in Fig.\,\ref{fig:DP12}.

\begin{table}
	\centering
	\captionof{table}{The same as in Table\,\ref{tab:S1430353} but for  KIC\,9227988}
	\label{tab:S9227988}
	\begin{tabular}{rrrrrrr} % four columns, alignment for each
		\hline
 ID          &    $\nu$      & $P$    &$\Delta P$& $A$     & $\frac{\mathrm S}{\mathrm N}$         &   fs\\
             &    $(\mathrm{d}^{-1})$ & $(\mathrm{d})$  &$(\mathrm{d})$  & $(\mathrm{ppm})$    &               &  \\
\hline
\multicolumn{7}{|c|}{Sa}\\
$\nu_{6}$  & 0.58071(2) & 1.72204 & 0.01495 & 587(5) & 17 & c \\
$\nu_{9}$  & 0.58579(3) & 1.70709 & 0.01204 & 377(5) & 12 & c \\
$\nu_{51}$ & 0.58995(6) & 1.69505 & 0.01835 & 141(5) & 5.5 & i \\
$\nu_{18}$ & 0.59641(4) & 1.67670 & 0.02092 & 238(6) & 8.2 & c \\
$\nu_{65}$ & 0.60395(8) & 1.65578 & 0.02436 & 88(5)  & 4.0 & c \\
$\nu_{12}$ & 0.61296(3) & 1.63142 & 0.03110 & 356(5) & 12 & c \\
--- & \textit{0.62487}    & \textit{1.60033} & \textit{0.03527} & ---    & --- & --- \\
$\nu_{48}$ & 0.63895(6) & 1.56506 & ---     & 139(5) & 6.1 & c \\

 \multicolumn{7}{|c|}{Sb}\\

$\nu_{32}$ & 0.15303(5) & 6.53472 & 0.29889 & 174(5) & 4.8 & i \\
$\nu_{16}$ & 0.16036(3) & 6.23582 & 0.31984 & 300(5) & 7.3 & i \\
$\nu_{1}$  & 0.16903(1) & 5.91599 & 0.22561 & 2174(5)& 31 & i \\
$\nu_{49}$ & 0.17574(6) & 5.69038 & 0.24683 & 135(5) & 4.1 & i \\
$\nu_{5}$  & 0.18370(2) & 5.44355 & 0.26467 & 666(5) & 14 & i \\
--- & \textit{0.19309}    & \textit{5.17888} & \textit{0.24242} & ---    & --- & --- \\
--- & \textit{0.20257}    & \textit{4.93646} & \textit{0.22195} & ---    & --- & --- \\
$\nu_{24}$ & 0.21211(5) & 4.71451 & 0.26900 & 196(5) & 5.2 & c \\
$\nu_{23}$ & 0.22495(5) & 4.44551 & 0.25180 & 187(5) & 5.3 & c \\
$\nu_{26}$ & 0.23845(5) & 4.19371 & 0.29987 & 191(5) & 5.3 & c \\
$\nu_{41}$ & 0.25682(6) & 3.89384 & 0.19570 & 141(5) & 4.4 & c \\
--- & \textit{0.27041}    & \textit{3.69814} & \textit{0.22514} &  ---   & --- & --- \\
$\nu_{2}$  & 0.28794(8) & 3.47301 & 0.23860 & 2164(5)& 37 & i \\
$\nu_{30}$ & 0.30918(5) & 3.23440 & 0.23403 & 175(5) & 5.4 & c \\
--- & \textit{0.33329}    & \textit{3.00037} & \textit{0.22577} & --     & --- & --- \\
$\nu_{17}$ & 0.36041(4) & 2.77460 & 0.24403 & 268(5) & 7.4 & c \\
--- & \textit{0.39517}    & \textit{2.53057} & \textit{0.18508} &   ---  & --- & --- \\
$\nu_{40}$ & 0.42635(5) & 2.34549 & 0.15715 & 141(5) & 5.1 & c \\
$\nu_{3}$  & 0.456968(7)& 2.18834 & ---     & 1948(5)& 43 & c \\

	\end{tabular}	
    \includegraphics[angle=0, width=\columnwidth]{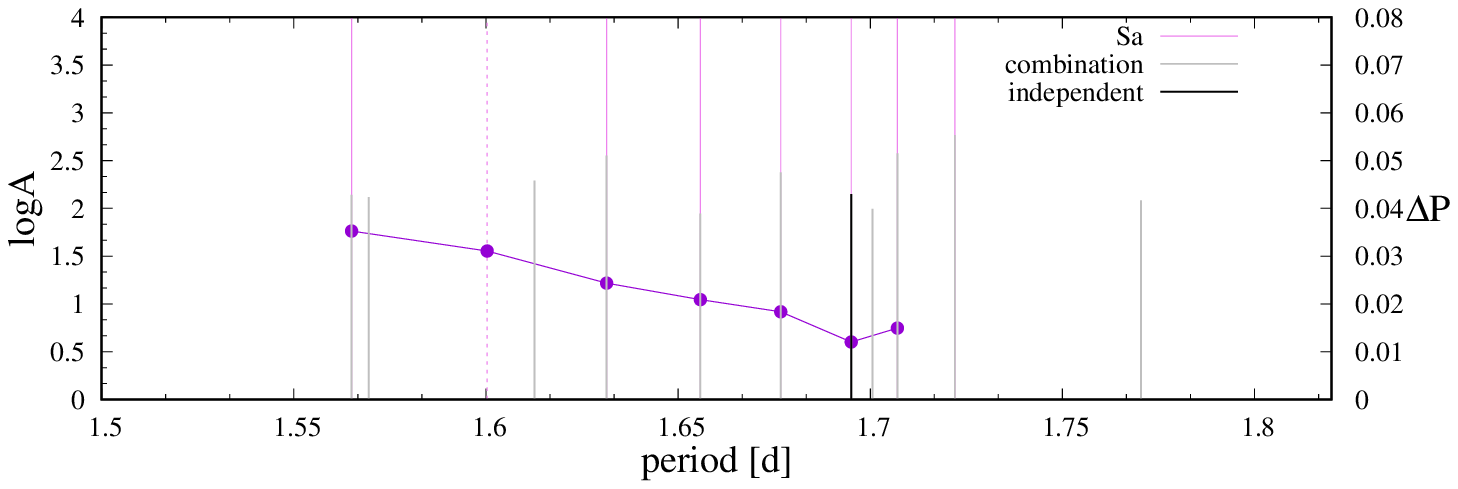}
    \includegraphics[angle=0, width=\columnwidth]{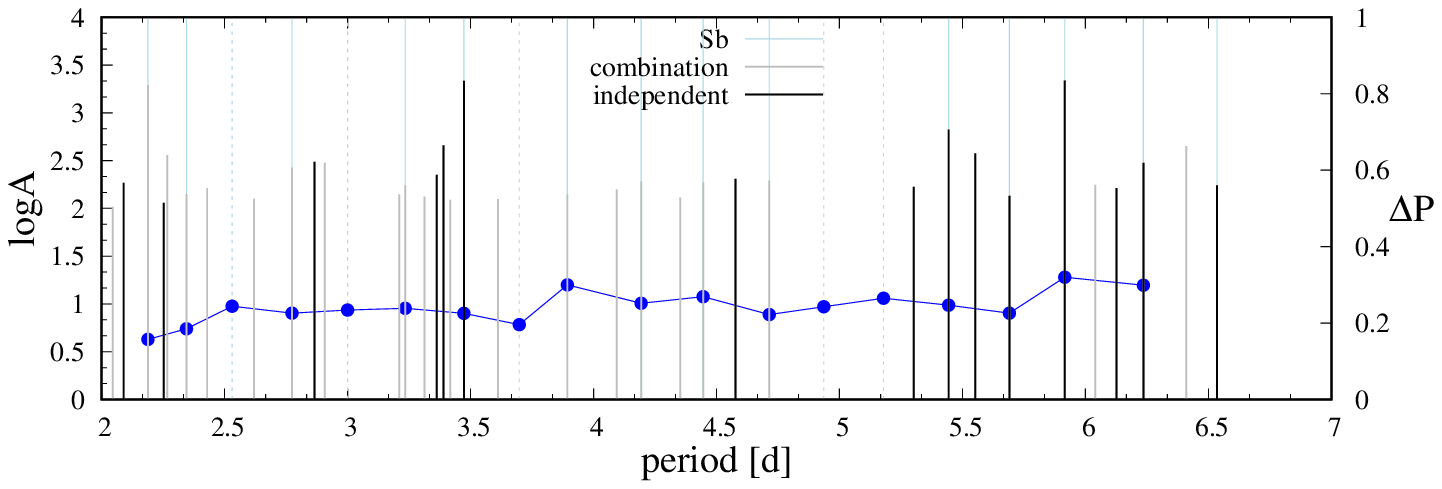}
    \vspace{-0.3cm}
    \captionof{figure}{The same as in Fig.\,\ref{fig:DP1} but for KIC\,9227988.}
    \label{fig:DP12}
	
\end{table}

%\begin{figure}
%	\includegraphics[angle=0, width=\columnwidth]{DP_KIC9227988}
%    \caption{The same as in Fig.\,\ref{fig:DP1} but for KIC\,9227988.}
%    \label{fig:DP12}
%\end{figure}

\subsection{KIC\,9278405}
%13
The star was classified as a low amplitude SPB pulsator by \citet{2012AJ....143..101M}.
\citet{2015MNRAS.451.1445B}  attributed the star's variability to the rotational origin. Recently, \citet{2019MNRAS.485.3457B} reclassified it as the SPB/ROT variable.

In the {\it Kepler} light curve, we found 45 closely spaced frequency peaks around 1.8 $\mathrm{d}^{-1}$. 
21 of them were rejected because of the $2.5\times$Rayleigh limit and, according to our criterion,
%7 frequency peaks were categorized as firm independent,  6 as probably independent and 11 as combinations.
13 frequency peaks were categorized as  independent  and 11 as combinations.
These frequencies fit well into the range of unstable prograde dipole modes predicted by our representative model (see Fig.\,\ref{fig:freq_all3}\,o).
The TDP analysis showed that signal is strongly incoherent (see Fig.\,\ref{fig:fourier_time_depend_KIC9278405} in
Appendix\,\ref{appendix:A}).
This phenomenon can be caused by differential rotation or originates in a variable disc around the star. Incoherent signal prevented us from searching regularities in the frequency spectrum.

\subsection{KIC\,9468611}
%14

The star was classified as the SPB variable with a single frequency grouping by  \citet{2012AJ....143..101M}
and reclassified  by \citet{2015MNRAS.451.1445B, 2019MNRAS.485.3457B}
as a rotational variable (ROT).

The oscillation spectrum of the star is similar to the case of KIC\,9278405.
We found 48 frequency peaks and rejected 18 because of the $2.5\times$Rayleigh limit.
%Then, 9 peaks were identified as firm independent, 13 as probably independent, 7 as combinations and 1 as harmonic. Frequencies
Then, 22 peaks were identified as  independent, 7 as combinations and 1 as a harmonic. Frequencies 
are slightly higher than unstable dipole prograde modes predicted by our representative 
model (see Fig.\,\ref{fig:freq_all3}\,p). The TDP analysis revealed that  variability is incoherent. Therefore, as it was in the case of KIC\,9278405, we did not search for regularities in the frequency spectrum.

\subsection{KIC\,9715425}
The star was classified as the SPB variable with a frequency grouping and low
frequency noise by \citet{2012AJ....143..101M}. This classification was confirmed by \citet{2015MNRAS.451.1445B}.

We found 293 frequency peaks and rejected 124 because of the $2.5\times$Rayleigh limit.
%We identified 20 firm independent frequencies, 92 probably independent, 54 combinations and 3 harmonics. Firm independent frequencies are observed
We identified 112 independent frequencies,  54 combinations and 3 harmonics. Independent frequencies are observed
up to $6\,\mathrm{d^{-1}}$. There is also seen a clear frequency grouping. In the Kepler light curve,  we
could see a kind of outbursts and the amplitude variability. The TDP analysis
indicated that most of frequencies are coherent but amplitudes are strongly variable. 
Most of the amplitudes increase
significantly in the second part of observations. 
The standout exception is $\nu_{16} = 1.70832(5)$\,d$^{-1}$. In this case amplitude strongly 
decreases in the second part of observations
(see Fig.\,\ref{fig:fourier_time_depend_KI9715425} in Appendix\,\ref{appendix:A}).

\begin{figure}
    \centering
	\includegraphics[angle=0, width=\columnwidth]{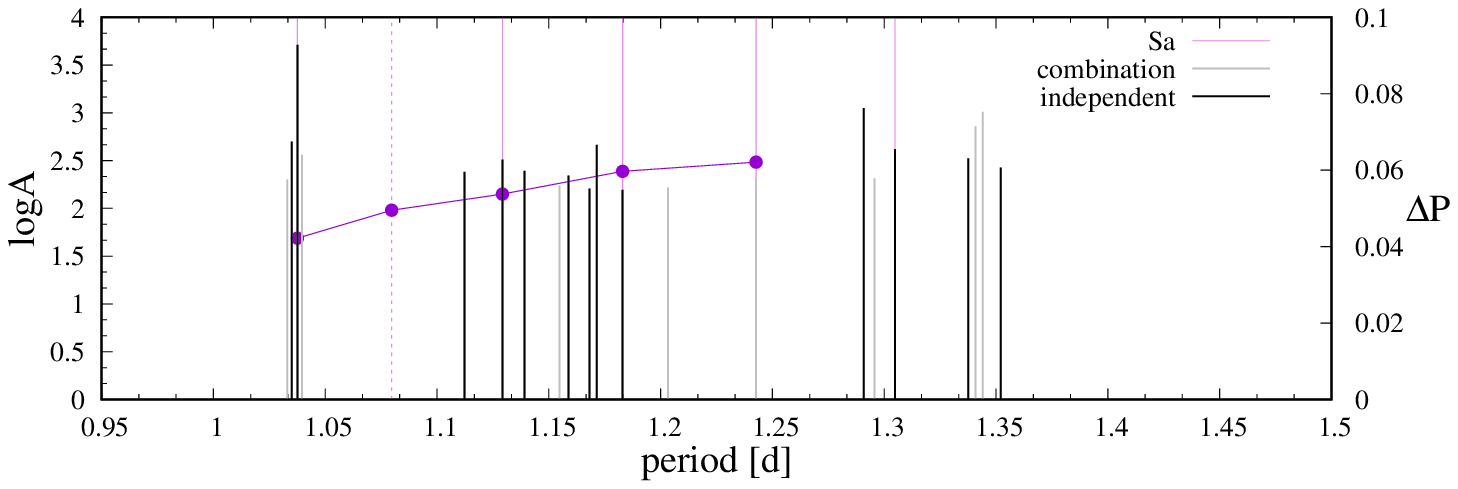}
	\includegraphics[angle=0, width=\columnwidth]{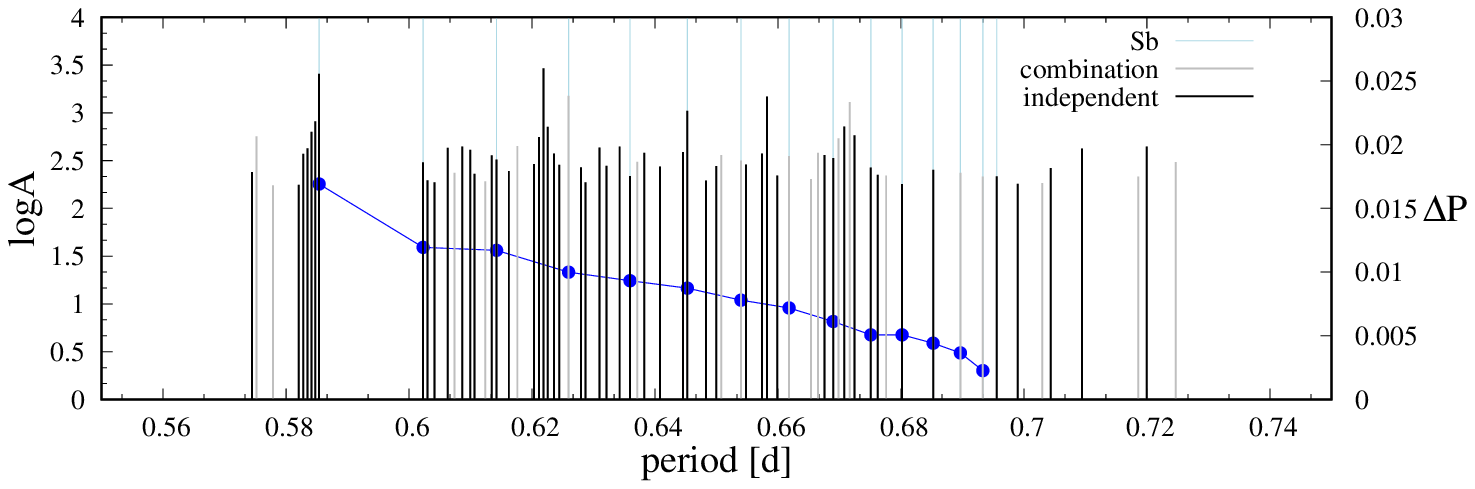}
	\includegraphics[angle=0, width=\columnwidth]{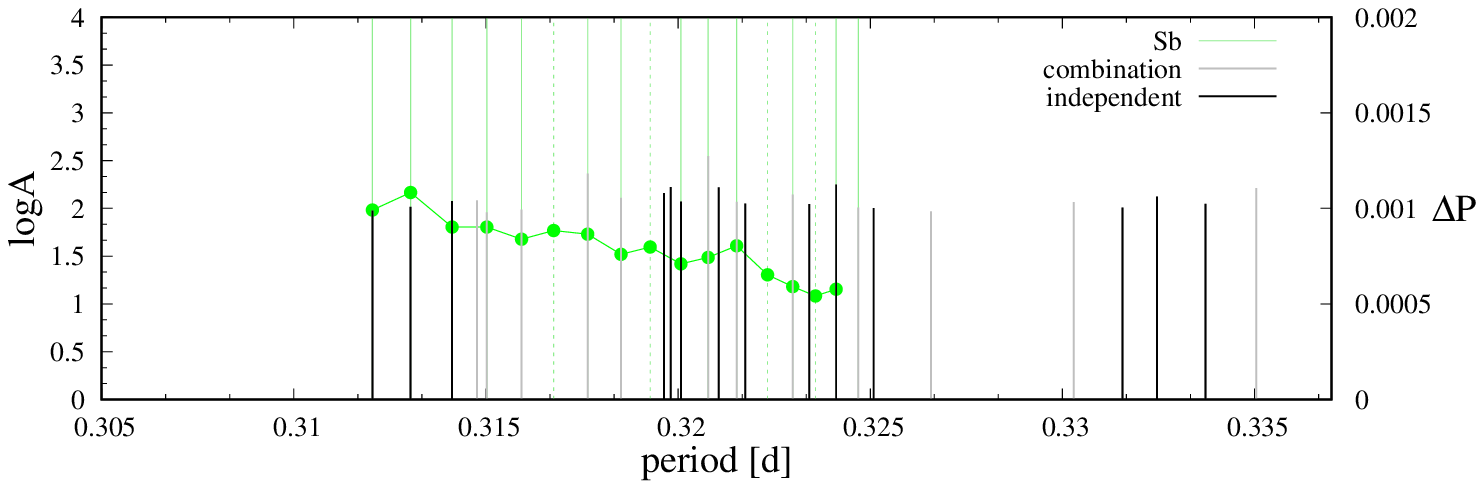}
	\vspace{-0.3cm}
	\captionof{figure}{The same as in Fig.\,\ref{fig:DP1} but for KIC\,9715425.}
	\label{fig:DP13}
\end{figure}

In the frequency spectrum there are six group of frequencies. We found regular series in the second, third and fifth group (see Tab.\ref{tab:S9715425} and Fig.\,\ref{fig:DP13}). The mean period differences are 0.0535\,d (0.0295\,d$^{-1}$) for Sa, 0.0079\,d (0.0193\,d$^{-1}$) for Sb and 0.0008\,d (0.0078\,d$^{-1}$) for Sc. The Sa and Sb sequences can be quadrupole axisymmetric and prograde modes, respectively. The Sc series seems to be associated with higher mode degree than $\ell=2$.

\begin{table}
	\centering
	\captionof{table}{The same as in Table\,\ref{tab:S1430353} but for  KIC\,9715425}
	\label{tab:S9715425}
	\begin{tabular}{rrrrrrr} % four columns, alignment for each
		\hline
		ID          &    $\nu$      & $P$    &$\Delta P$& $A$     & $\frac{\mathrm S}{\mathrm N}$         &   fs\\
		&    $(\mathrm{d}^{-1})$ & $(\mathrm{d})$  &$(\mathrm{d})$  & $(\mathrm{ppm})$    &               &  \\
		\hline
		\multicolumn{7}{|c|}{Sa}\\
$\nu_{49}$  & 0.76640(5) & 1.30480 & 0.06210 & 418(13) & 8.7 & i \\
$\nu_{190}$ & 0.80469(8) & 1.24271 & 0.05969 & 204(13) & 5.0 & c\\
$\nu_{241}$ & 0.8453(1)  & 1.18302 & 0.05375 & 156(13) & 4.1 & i\\
$\nu_{106}$ & 0.88553(7) & 1.12927 & 0.04951 & 323(13) & 6.1 & i\\
---  &\textit{ 0.92614}    & \textit{1.07975} & \textit{0.04220} &     --- & --- &  \\
$\nu_{1}$   & 0.963802(8)& 1.03756 & ---     & 5156(23)& 55& i \\
			
		\multicolumn{7}{|c|}{Sb}\\
$\nu_{182}$ & 1.43765(8) & 0.69558 & 0.00226 & 217(13) & 4.4 & i \\
$\nu_{179}$ & 1.44233(8) & 0.69332 & 0.00365 & 216(13) & 4.3 & c \\
$\nu_{212}$ & 1.4500(1)  & 0.68967 & 0.00442 & 235(14) & 4.1 & c \\
$\nu_{184}$ & 1.45933(8) & 0.68525 & 0.00508 & 254(13) & 4.4 & i \\
$\nu_{223}$ & 1.47021(9) & 0.68017 & 0.00507 & 179(13) & 4.1 & i \\
$\nu_{131}$ & 1.48125(7) & 0.67511 & 0.00613 & 268(13) & 4.8 & i \\
$\nu_{143}$ & 1.49481(7) & 0.66898 & 0.00719 & 335*14) & 4.8 & i \\
$\nu_{88}$  & 1.51104(6) & 0.66179 & 0.00780 & 353(13) & 5.4 & c \\
$\nu_{76}$  & 1.52907(6) & 0.65399 & 0.00874 & 317(14) & 5.5 & c \\
$\nu_{23}$  & 1.54977(4) & 0.64526 & 0.00932 & 1050(25)& 9.4 & i \\
$\nu_{194}$ & 1.57248(8) & 0.63594 & 0.01000 & 218(13) & 4.4 & i \\
$\nu_{9}$   & 1.59760(3) & 0.62594 & 0.01171 & 1510(14)& 14 & c \\
$\nu_{145}$ & 1.62805(7) & 0.61423 & 0.01195 & 324(13) & 4.9 & i \\
$\nu_{99}$  & 1.66033(7) & 0.60229 & 0.01692 & 304(16) & 5.2 & i \\
$\nu_{16}$  & 1.70832(5) & 0.58537 & ---     & 2558(27)& 12 & i \\

			\multicolumn{7}{|c|}{Sc}\\
			
$\nu_{272}$ & 3.0799(2) & 0.32469 & 0.0005770 & 102(13) & 4.2 & c \\
$\nu_{240}$ & 3.0853(1) & 0.32411 & 0.0005412 & 178(13) & 6.0 & i \\
$\nu_{-1}$  & 3.0905    & 0.32357 & 0.0005905 &    ---  & --- &  \\
$\nu_{252}$ & 3.0961(1) & 0.32298 & 0.0006528 & 140(13) & 5.2 & c \\
$\nu_{-1}$  & 3.1024    & 0.32233 & 0.0008042 & ---     & --- &  \\
$\nu_{257}$ & 3.1102(1) & 0.32153 & 0.0007433 & 117(13) & 5.1 & c \\
$\nu_{123}$ & 3.11738(7)& 0.32078 & 0.0007097 & 354(16) & 9.8 & c \\
$\nu_{269}$ & 3.1243(1) & 0.32007 & 0.0007983 & 117(13) & 4.5 & i \\
$\nu_{-1}$  & 3.1321    & 0.31927 & 0.0007598 & ---     & --- &  \\
$\nu_{260}$ & 3.1396(1) & 0.31851 & 0.0008648 & 129(13) & 4.9 & c \\
$\nu_{142}$ & 3.14812(7)& 0.31765 & 0.0008844 & 232(13) & 9.0 & c \\
$\nu_{-1}$  & 3.1569    & 0.31677 & 0.0008387 & ---     & --- &  \\
$\nu_{275}$ & 3.1653(2) & 0.31593 & 0.0009025 & 97(13)  & 4.3 & c \\
$\nu_{284}$ & 3.1744(2) & 0.31502 & 0.0009021 & 91(13)  & 4.0 & c \\
$\nu_{271}$ & 3.1835(2) & 0.31412 & 0.0010833 & 120(13) & 4.2 & i \\
$\nu_{263}$ & 3.1945(1) & 0.31304 & 0.0009916 & 104(13) & 4.7 & i \\
$\nu_{283}$ & 3.2046(2) & 0.31205 & ---       & 95(13)  & 4.1 & i \\

	\end{tabular}	
%	\includegraphics[angle=0, width=\columnwidth]{KIC9715425_2P_spec_Sa.eps}
%	\includegraphics[angle=0, width=\columnwidth]{KIC9715425_2P_spec_Sb.eps}
%	\includegraphics[angle=0, width=\columnwidth]{KIC9715425_2P_spec_Sc.eps}
	%\includegraphics[angle=0, width=\columnwidth]{KIC9227988_2frec_spec_Sa.eps}
	%\includegraphics[angle=0, width=\columnwidth]{KIC9227988_2frec_spec_Sb.eps}
%	\vspace{-0.3cm}
%	\captionof{figure}{The same as in Fig.\,\ref{fig:DP1} but for KIC\,9715425.}
%	\label{fig:DP13}
	
\end{table}

\subsection{KIC\,9910544}
As KIC\,9468611, this star was classified as the SPB variable with frequency grouping and low
frequency noise by \citet{2012AJ....143..101M}.

%We found 16 frequency peaks, of which 6 are firm independent  and 3 are probably independent and 2 are combinations.
We found 16 frequency peaks, of which 9 are  independent and 2 are combinations.
All frequencies have the values below $0.9$\,d$^{-1}$ and most of them (except 2) concentrate around $0.7$\,d$^{-1}$.
Based on the TDP analysis we  concluded that at least the dominant frequency
 $\nu_1=0.723517(7)$\,d$^{-1}$ is coherent.
 The other may be incoherent but we are no able to make any reliable conclusion. Due to the small number of frequencies we could not find any regularities in the frequency spectrum for this star.

\subsection{KIC\,9964614}
%15
The star was classified as hybrid $\beta$\,Cep/SPB pulsator
by \citet{2011MNRAS.413.2403B, 2015MNRAS.451.1445B}.
The authors derived  the following atmospheric parameters: 
$T_\mathrm{eff} = 20300\,\mathrm{K}$, $\log g=3.9$ (from spectroscopy),
$T_\mathrm{eff} = 19471\,\mathrm{K}$, $\log g=3.75$ (from Str\"omgren photometry)
and $T_\mathrm{eff} = 23400\pm7400\,\mathrm{K}$ (from SED fitting).

Our Fourier analysis of the Kepler light curve revealed 246 frequency peaks of which we rejected 41 because of the $2.5\times$Rayleigh limit.
%From the rest we identified 28 firm independent frequencies, 97 probably independent, 76 combinations and 4 harmonics.
From the rest we identified 125 independent frequencies,  76 combinations and 4 harmonics. 
The highest amplitude signal is very low, $\nu_1 = 0.47356(1)$\,d$^{-1}$,
but independent frequencies are seen up to $\sim 11.4$\,d$^{-1}$ (see Fig.\,\ref{fig:freq_all4}\,t).
Moreover, the star is located where  the SPB and $\beta$\,Cep instability strips overlap. 
Therefore, it is undoubtedly a hybrid pulsator of the SPB/$\beta$\,Cep type. The TDP analysis implies that frequencies are
coherent, but amplitudes are variable.

For this star there exists a very interesting sequence of frequencies with increasing period spacing.
The whole series covers the range from 0.3 d$^{-1}$ to 1.7 d$^{-1}$
(see Tab.\,\ref{tab:S9964614} and Fig.\,\ref{fig:DP15}).
However, the two frequencies seem to be missing.
The mean period spacing is equal to 0.123\,d   which corresponds to the  frequency difference of 0.064\,d$^{-1}$.
The series can be reproduce with dipole retrograde modes. This star seems to be  very promising target for asteroseismic analysis.

\begin{table}
	\centering
	\captionof{table}{The same as in Table\,\ref{tab:S1430353} but for  KIC\,9964614}
	\label{tab:S9964614}
	\begin{tabular}{rrrrrrr} % four columns, alignment for each
		\hline
 ID          &    $\nu$      & $P$    &$\Delta P$& $A$     & $\frac{\mathrm S}{\mathrm N}$         &   fs\\
             &    $(\mathrm{d}^{-1})$ & $(\mathrm{d})$  &$(\mathrm{d})$  & $(\mathrm{ppm})$    &               &  \\
\hline
\multicolumn{7}{|c|}{Sa}\\
         &        &         &         &\\

  $\nu_{9}$  & 0.31329(3) & 3.19198 & 0.12658 & 167(2) & 9.9 & i \\
  $\nu_{6}$  & 0.32622(2) & 3.06540 & 0.11755 & 208(2) & 12 & i \\
  $\nu_{75}$ & 0.33923(6) & 2.94785 & 0.14161 & 45(2)  & 4.1 & i \\
  $\nu_{52}$ & 0.35635(6) & 2.80624 & 0.13046 & 53(2)  & 4.5 & c \\
 --- & \textit{0.37372}    & \textit{2.67578} & \textit{0.15229} & ---    & --- & --- \\
  $\nu_{48}$ & 0.39628(5) & 2.52349 & 0.12747 & 57(2)  & 4.6 & i \\
  $\nu_{74}$ & 0.41736(6) & 2.39602 & 0.13634 & 48(2)  & 4.1 & i \\
  --- & \textit{0.44254}    & \textit{2.25969} & \textit{0.14803} & ---    & --- & --- \\
  $\nu_{1}$  & 0.47356(1) & 2.11166 & 0.14630 & 503(3) & 25 & i \\
  $\nu_{8}$  & 0.50881(2) & 1.96536 & 0.13919 & 179(2) & 11 & i \\
  $\nu_{59}$ & 0.54760(6) & 1.82617 & 0.13451 & 48(2)  & 4.5 & c \\
  $\nu_{2}$  & 0.59114(1) & 1.69166 & 0.12249 & 459(2) & 25 & i \\
  $\nu_{36}$ & 0.63728(5) & 1.56916 & 0.14670 & 69(2)  & 5.9 & c \\
  $\nu_{25}$ & 0.70301(4) & 1.42246 & 0.12600 & 89(2)  & 7.3 & i \\
  $\nu_{69}$ & 0.77133(7) & 1.29646 & 0.14007 & 40(2)  & 4.7 & c \\
  $\nu_{71}$ & 0.86476(7) & 1.15638 & 0.11627 & 38(2)  & 5.1 & i \\
  $\nu_{61}$ & 0.96144(6) & 1.04011 & 0.08846 & 46(2)  & 6.6 & c \\
  $\nu_{81}$ & 1.05081(8) & 0.95165 & 0.08440 & 33(2)  & 6.2 & i \\
  $\nu_{5}$  & 1.15307(2) & 0.86725 & 0.10169 & 231(2) & 25 & i \\
  $\nu_{99}$ & 1.3062(1)  & 0.76556 & 0.07605 & 24(2)  & 5.8 & c \\
  $\nu_{84}$ & 1.45030(8) & 0.68951 & 0.08660 & 34(2)  & 8.0 & i \\
  $\nu_{86}$ & 1.65861(8) & 0.60292 & ---     & 33(2)  & 8.2 & c \\

 	\end{tabular}
 	
	\includegraphics[angle=0, width=\columnwidth]{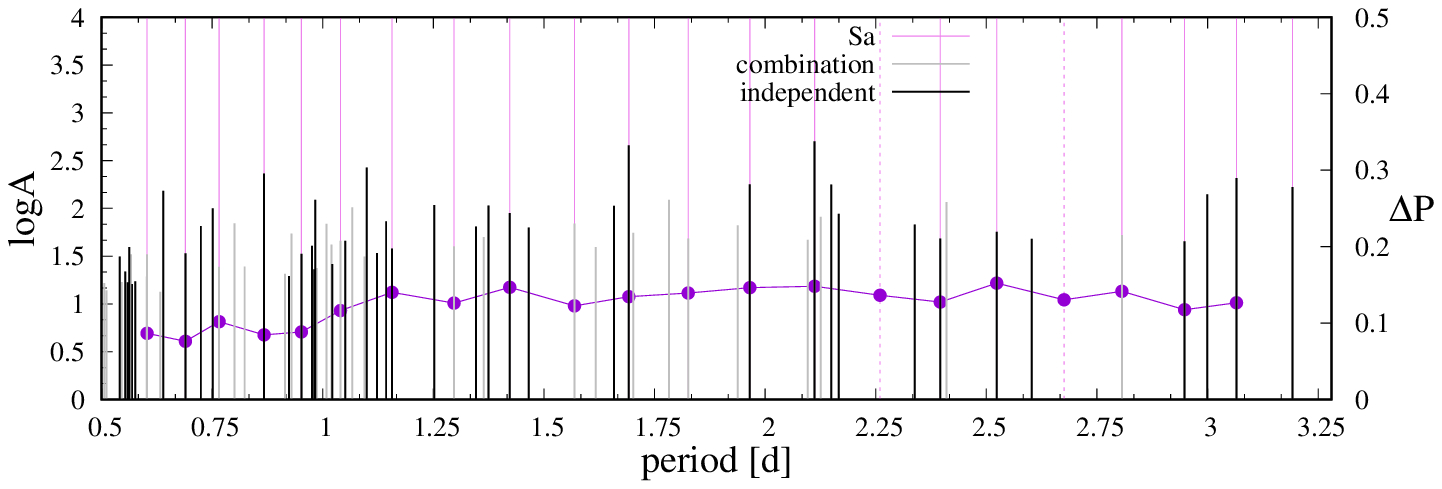}
	\includegraphics[angle=0, width=\columnwidth]{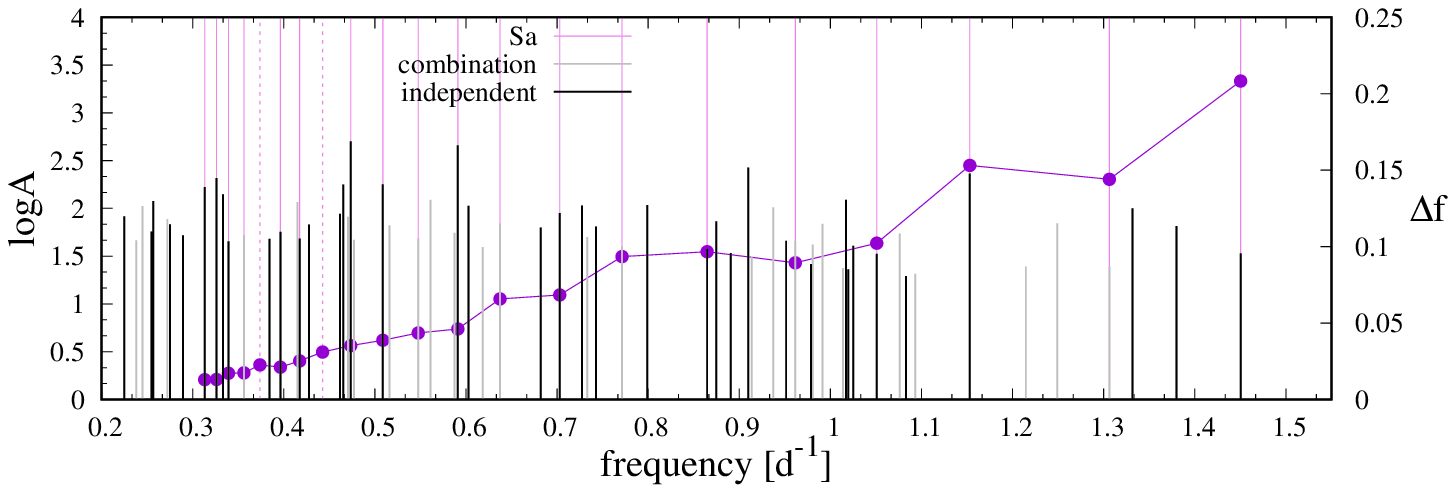}
	\vspace{-0.3cm}
    \captionof{figure}{The same as in Fig.\,\ref{fig:DP11} but for KIC\,9964614.}
    \label{fig:DP15}
 	
\end{table}

%\begin{figure}
%	\includegraphics[angle=0, width=\columnwidth]{DP_KIC9964614}
%    \caption{The same as in Fig.\,\ref{fig:DP1} but for KIC\,9964614.}
%    \label{fig:DP15}
%\end{figure}

\subsection{KIC\,10118750}
The star was classified as the $\beta$\,Cep variable  with frequency grouping by \citet{2012AJ....143..101M}.
\citet{2015MNRAS.450.3015K}  classified it as the SPB star, whereas \citet{2015MNRAS.451.1445B}
classified it as the SPB/ROT-type variable.

In its light curve we found 111 frequency peaks and  rejected 25 because of the $2.5\times$Rayleigh limit.
%From the rest, we identified 17 firm independent frequencies,  20 probably independent, 41 combinations and 8 harmonics.
From the rest, we identified 37 independent frequencies, 41 combinations and 8 harmonics.
The highest amplitude peak was found in the group around $\sim 3.5$\,d$^{-1}$. Taking into account the
position of the star in the Kiel diagram (Fig.\,\ref{fig:kiel}) and the high value of  $V \sin i$, we
concluded that the star is the SPB pulsator. The TDP analysis indicates that  at least high amplitude  frequencies are coherent.

In the frequency spectrum of KIC\,10118750 a clear grouping is visible.
The lowest frequency group can be of instrumental origin, but there is a very regular pattern among these peaks that can be associated with dipole retrograde modes. The others hypothetical series are
found in the second and third frequency groups (see Tab.\,\ref{tab:S10118750} and Fig.\,\ref{fig:DP100}). The mean period differences are 0.01367\,d (0.1630\,d$^{-1}$) for Sa, 0.0030\,d (0.1564\,d$^{-1}$) for Sb and 4.4128\,d (0.0062\,d$^{-1}$) for Sc.

\begin{table}
	\centering
	\captionof{table}{The same as in Table\,\ref{tab:S1430353} but for  KIC\,10118750}
	\label{tab:S10118750}
	\begin{tabular}{rrrrrrr} % four columns, alignment for each
		\hline
 ID          &    $\nu$      & $P$    &$\Delta P$& $A$     & $\frac{\mathrm S}{\mathrm N}$         &   fs\\
             &    $(\mathrm{d}^{-1})$ & $(\mathrm{d})$  &$(\mathrm{d})$  & $(\mathrm{ppm})$    &               &  \\
\hline
\multicolumn{7}{|c|}{Sa}\\ 
$\nu_{80}$ & 2.86288(6)  & 0.34930 & 0.01893 & 16(2)  & 8.0 & i \\
--- & \textit{3.02694}     & \textit{0.33037} & \textit{0.01595} & ---    & --- & --- \\
$\nu_{64}$ & 3.18052(5)  & 0.31441 & 0.01235 & 19(2)  & 9.3 & i \\
$\nu_{6}$  & 3.310570(4) & 0.30206 & 0.01309 & 590(2) & 83 & c \\
$\nu_{17}$ & 3.46048(2)  & 0.28898 & 0.01351 & 59(2)  & 24 & c \\
$\nu_{1}$  & 3.630231(4) & 0.27547 & 0.01187 & 1427(4)& 67 & i \\
$\nu_{72}$ & 3.79363(5)  & 0.26360 & 0.01043 & 17(2)  & 8.3 & i \\
$\nu_{5}$  & 3.949911(4) & 0.25317 & 0.01319 & 715(2) & 76 & c \\
$\nu_{7}$  & 4.167051(4) & 0.23998 & ---     & 540(2) & 86 & i \\

\multicolumn{7}{|c|}{Sb}\\ 
$\nu_{33}$  & 6.49157(3)  & 0.15405 & 0.00347 & 33(2) & 16 & i \\
$\nu_{101}$ & 6.64094(9)  & 0.15058 & 0.00307 & 10(2) & 5.3 & i \\
---  & \textit{6.77935}     & \textit{0.14751} & \textit{0.00343} &   --- & ---& --- \\
$\nu_{78}$  & 6.94077(5)  & 0.14408 & 0.00305 & 16(2) & 8.1 & c \\
$\nu_{13}$  & 7.090722(8) & 0.14103 & 0.00330 & 126(2)& 47 & i \\
$\nu_{3}$   & 7.260475(3) & 0.13773 & 0.00277 & 1033(2)& 85 & i \\
$\nu_{62}$  & 7.40957(4)  & 0.13496 & 0.00304 & 20(2) & 10 & c \\
$\nu_{4}$   & 7.580143(4) & 0.13192 & 0.00249 & 802(2)& 98 & i \\
---  & \textit{7.72589}     & \textit{0.12944} & \textit{0.00284} & ---   & --- & --- \\
$\nu_{97}$  & 7.89907(8)  & 0.12660 & ---     & 11(2) & 5.9 & c \\

\multicolumn{7}{|c|}{Sc}\\ 
$\nu_{58}$ & 0.01473(4) & 67.8996 & 14.4196 & 22(2) & 5.2 & i \\
$\nu_{39}$ & 0.01870(3) & 53.4800 & 11.2878 & 27(2) & 6.5 & i \\
$\nu_{51}$ & 0.02370(4) & 42.1922 & 6.68357 & 23(2) & 5.7 & i \\
$\nu_{20}$ & 0.02816(2) & 35.5086 & 5.11122 & 53(2) & 9.6 & i \\
$\nu_{40}$ & 0.03290(3) & 30.3974 & 3.81029 & 26(2) & 6.6 & i \\
$\nu_{36}$ & 0.03761(3) & 26.5871 & 2.97682 & 35(2) & 6.5 & i \\
--- & \textit{0.04235}    & \textit{23.6103} & \textit{2.50001} & ---   & --- & --- \\
$\nu_{9}$  & 0.047370(4)& 21.1103 & 2.72900   & 385(2)& 51 & c \\
$\nu_{18}$ & 0.05440(2) & 18.3813 & 1.92143 & 53(2) & 11 & c \\
--- & \textit{0.06075}    & \textit{16.4599} & \textit{1.77676} & ---   & --- & --- \\
$\nu_{79}$ & 0.06811(5) & 14.6831 & 1.38362 & 17(2) & 4.4 & i \\
$\nu_{65}$ & 0.07519(5) & 13.2995 & 1.48104 & 19(2) & 4.9 & i \\
$\nu_{81}$ & 0.08461(6) & 11.8185 & 1.28483 & 16(2) & 4.4 & i \\
$\nu_{84}$ & 0.09493(5) & 10.5336 & ---     & 17(2) & 4.6 & c \\

 	\end{tabular}
 	
\includegraphics[angle=0, width=\columnwidth]{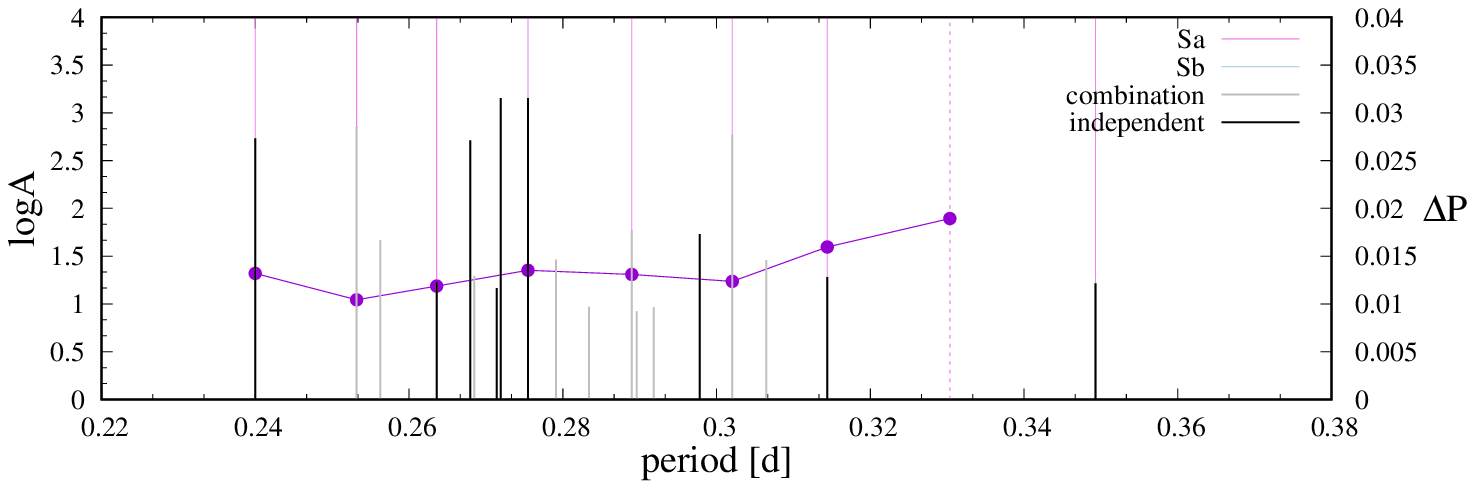}
\includegraphics[angle=0, width=\columnwidth]{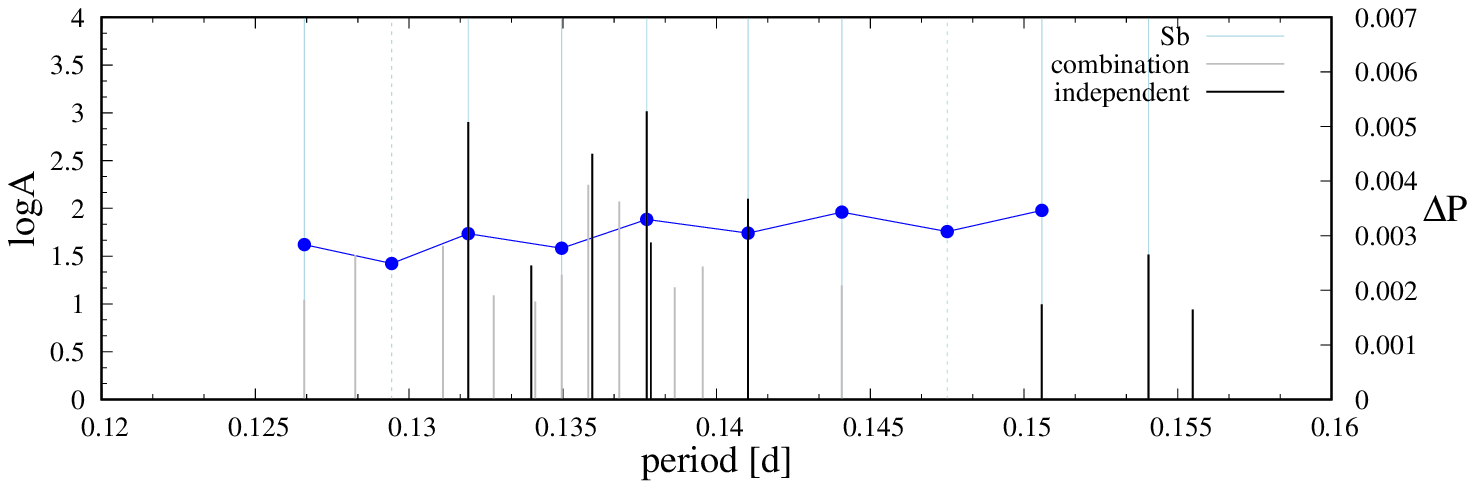}
\includegraphics[angle=0, width=\columnwidth]{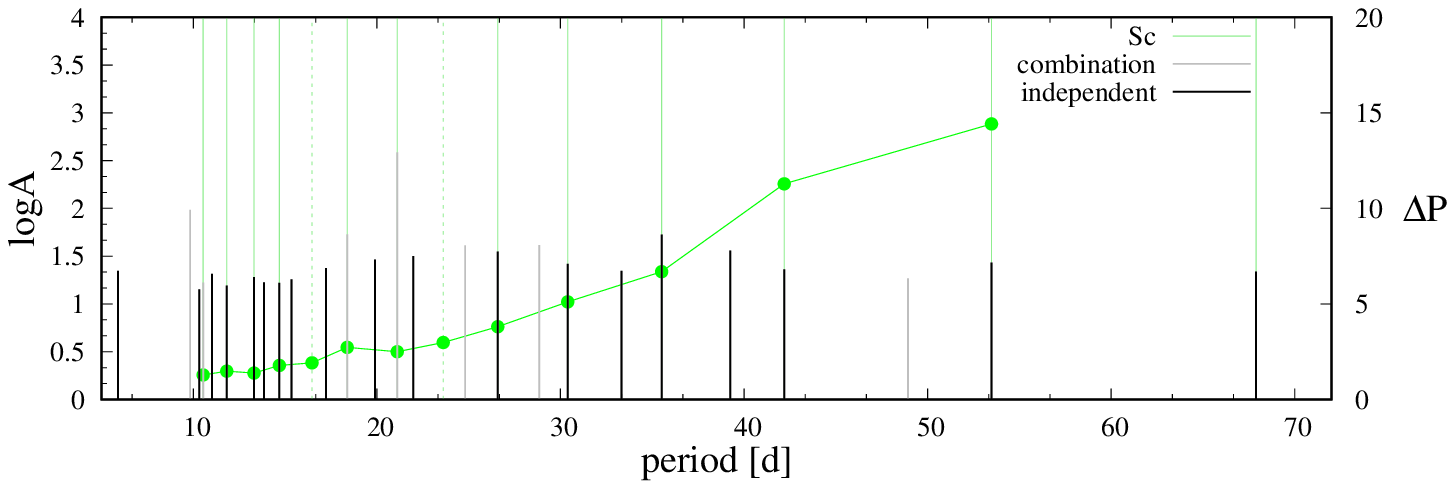}
\vspace{-0.3cm}
    \captionof{figure}{The same as in Fig.\,\ref{fig:DP1} but for KIC\,10118750.}
    \label{fig:DP100}
 	
\end{table}

%\begin{figure}
%	\includegraphics[angle=0, width=\columnwidth]{DP_KIC10118750}
%    \caption{The same as in Fig.\,\ref{fig:DP1} but for KIC\,10118750.}
%    \label{fig:DP100}
%\end{figure} 

\subsection{KIC\,10526294}
The star was the first time classified as SPB by
\citet{2011A&A...529A..89D}. This was confirmed by \citet{2012AJ....143..101M}.
Then \citet{2014A&A...570A...8P} found regular period spacing.

In the light curve of KIC\,10526294 we found 245 frequency peaks. We  rejected 88 of them because of the $2.5\times$Rayleigh limit. 
%We ended up with 16 firm independent frequencies, 62 probably independent, 74 combinations and 5 harmonics.
We ended up with 78 independent frequencies, 74 combinations and 5 harmonics.

In the frequency spectrum we found one long series of frequencies, Sa (Tab.\,\ref{tab:S10526294}) with nearly constant period differences and slightly increasing frequency differences (see Fig.\,\ref{fig:DP17_5}). The mean period difference is 0.0625\,d (0.0335\,d$^{-1}$). This sequence can be reproduce by dipole axisymmetric modes.
In general, our series overlaps with those found by \citet{2014A&A...570A...8P}.
However, our series contains more frequencies. But we do not find 
$\nu=0.472220$\,d$^{-1}$ reported by the cited authors. Other small differences are for frequencies mentioned by \citet{2014A&A...570A...8P} as do not fulfilling their significance criterion.

\begin{table}
	\centering
	\captionof{table}{The same as in Table\,\ref{tab:S1430353} but for  KIC\,10526294}
	\label{tab:S10526294}
	\begin{tabular}{rrrrrrr} % four columns, alignment for each
		\hline
		ID          &    $\nu$      & $P$    &$\Delta P$& $A$     & $\frac{\mathrm S}{\mathrm N}$         &   fs\\
		&    $(\mathrm{d}^{-1})$ & $(\mathrm{d})$  &$(\mathrm{d})$  & $(\mathrm{ppm})$    &               &  \\
		\hline
		\multicolumn{7}{|c|}{Sa}\\
	
$\nu_{174}$ & 0.39317(7) & 2.54344 & 0.07887 & 160(8) & 4.0 & c \\
$\nu_{81}$  & 0.40575(5) & 2.46458 & 0.06673 & 298(8) & 5.8 & c \\
---  & \textit{0.41704}    & \textit{2.39784} & \textit{0.06077} &   ---  & --- & --- \\
$\nu_{37}$  & 0.42789(4) & 2.33708 & 0.07186 & 533(8) & 7.9 & i \\
---  & \textit{0.44146}    & \textit{2.26522} & \textit{0.06867} & ---    & ---& --- \\
$\nu_{22}$  & 0.45526(3) & 2.19655 & 0.06698 & 1160(8)& 13 & i \\
$\nu_{144}$ & 0.46958(7) & 2.12957 & 0.07278 & 184(8) & 4.3 & i \\
$\nu_{151}$ & 0.48619(7) & 2.05680 & 0.06045 & 163(8) & 4.1 & c \\
$\nu_{51}$  & 0.50092(5) & 1.99635 & 0.06330 & 464(8) & 7.0 & i \\
$\nu_{142}$ & 0.51732(7) & 1.93305 & 0.06565 & 166(8) & 4.5 & i \\
$\nu_{16}$  & 0.53343(2) & 1.87470 & 0.0584  & 1593(1)& 17  & i \\
$\nu_{4}$   & 0.552610(9)& 1.80960 & 0.06125 & 8742(8)& 49 & i \\
$\nu_{11}$  & 0.57197(1) & 1.74835 & 0.06375 & 3384(10)& 28 & i \\
$\nu_{59}$  & 0.59361(5) & 1.68460 & 0.05983 & 442(8)  & 7.2 & i \\
$\nu_{18}$  & 0.61547(2) & 1.62477 & 0.07036 & 1309(10)& 16 & i \\
---  & \textit{0.64333}    & \textit{1.55441} & \textit{0.06331} & ---     & ---& --- \\
$\nu_{163}$ & 0.67065(8) & 1.49110 & 0.07000 & 165(8) & 4.1 & i \\
$\nu_{63}$  & 0.70368(5) & 1.42111 & 0.06002 & 389(8) & 6.8 & i \\
$\nu_{3}$   & 0.73471(1) & 1.36109 & 0.06644 & 9461(8)& 46 & i \\
$\nu_{97}$  & 0.77241(6) & 1.29465 & 0.06455 & 256(8) & 5.8 & i \\
$\nu_{5}$   & 0.81294(1) & 1.23010 & 0.06236 & 5740(9)& 36 & i \\
$\nu_{38}$  & 0.85636(4) & 1.16774 & 0.05656 & 514(8) & 8.4 & i \\
$\nu_{96}$  & 0.89995(6) & 1.11118 & 0.06308 & 267(8) & 6.0 & c \\
$\nu_{35}$  & 0.95412(4) & 1.04809 & 0.06134 & 599(8) & 9.2 & i \\
$\nu_{65}$  & 1.01343(5) & 0.98675 & 0.05896 & 416(8) & 7.5 & i \\
---  & \textit{1.07783}   & \textit{0.92779} & \textit{0.05301} & ---    & --- & --- \\
$\nu_{68}$  & 1.14314(5) & 0.87478 & 0.05188 & 369(8) & 8.8 & c \\
$\nu_{198}$ & 1.2152(1)  & 0.82291 & 0.04768 & 105(8) & 4.5 & c \\
$\nu_{61}$  & 1.28995(5) & 0.77522 & 0.04289 & 360(8) & 11 & c \\
$\nu_{62}$  & 1.36550(5) & 0.73234 & ---     & 363(8) & 11 & i \\	
	
	\end{tabular}
	
	\includegraphics[angle=0, width=\columnwidth]{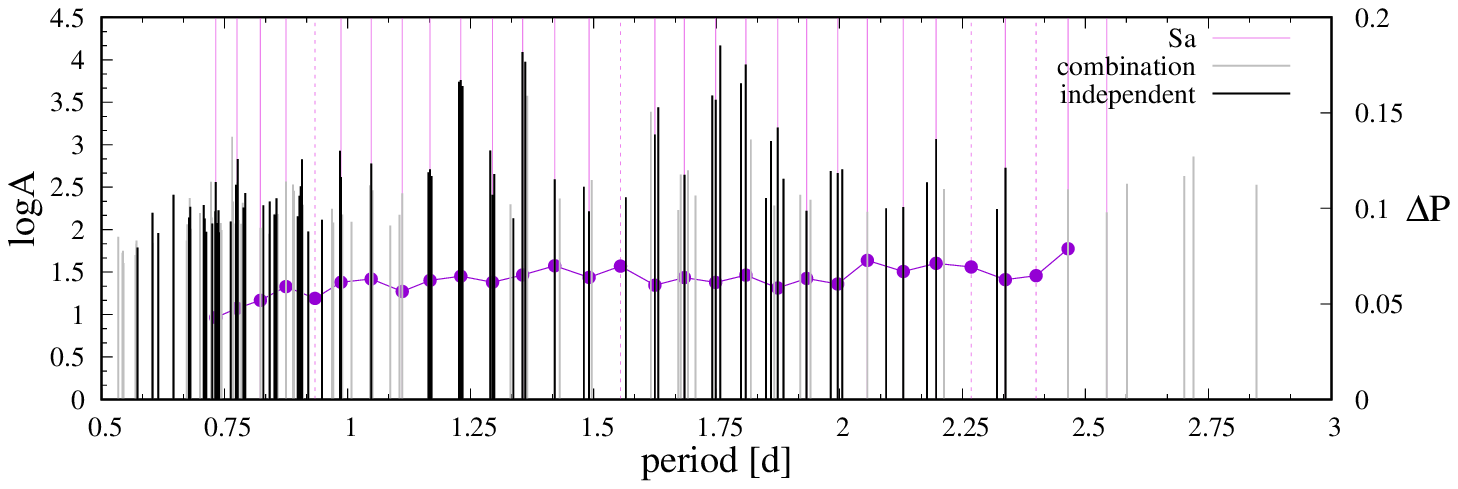}
		\includegraphics[angle=0, width=\columnwidth]{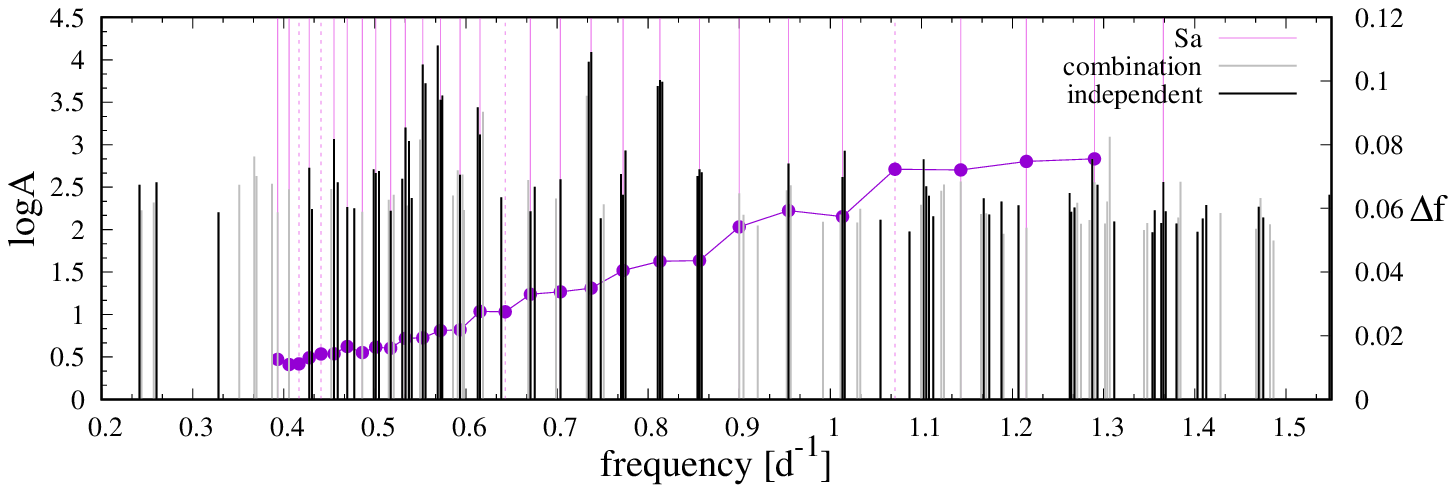}
		\vspace{-0.3cm}
	\captionof{figure}{The same as in Fig.\,\ref{fig:DP11} but for KIC\,10526294.}
	\label{fig:DP17_5}
	
\end{table}

\subsection{KIC\,10790075}
%17
The star was classified as the SPB  variable with a low frequency noise by \citet{2012AJ....143..101M},
whereas, \citet{2015MNRAS.451.1445B}  classified it  as a rotational variable with the question mark. 

We found in its light curve 43 frequency peaks and rejected 21 because of the $2.5\times$Rayleigh limit. 
%Ultimately, we were left with 8  firm independent frequencies, 3 probably independent and 11 combinations.
Ultimately, we were left with 11 independent frequencies and 11 combinations.
All found frequencies are below $1.1\,\mathrm{d^{-1}}$. The TDP analysis showed that the lowest frequency signal 
is strongly incoherent (see Fig.\,\ref{fig:fourier_time_depend_KIC10790075} in Appendix\,\ref{appendix:A}).
It is hard to state whether we do see incoherent behaviour or an interaction between two close frequencies. 
%Near 0.4\,$^{-1}$ and 0.75\,$^{-1}$ we found two pairs of close frequencies. 

The star lies inside the SPB instability strip but we  can say only about a candidate for a  coherent SPB-like variability.
Therefore we classified the star as a probable SPB pulsator. Since the frequencies varies with time, we did not look for regularities.

\subsection{KIC\,11293898}
%18
The star was classified as a hybrid SPB/$\beta$\,Cep pulsator with the frequency grouping
by \citet{2012AJ....143..101M}.

We found 388 frequency peaks and rejected 148 of them because of the $2.5\times$Rayleigh limit. Then,
%32 peaks were identified as firm independent,  148 as probably independent, 52 as combinations and 8 as harmonics.
180 peaks were identified as  independent,  52 as combinations and 8 as harmonics.
We found 10 groups of  frequencies up to $24\,\mathrm{d^{-1}}$.
At the end of the Kepler observations there is seen a kind of outburst which is reflected 
in an amplitude increase. At the same time frequencies seem to be
coherent. Despite a presence of high frequencies we identify the star as the SPB pulsator with
the frequencies shifted into higher range by the fast rotation. This classification 
is supported also by the location of the star well inside the SPB instability strip (see Fig.\,\ref{fig:kiel}).

\begin{figure}
 	\includegraphics[angle=0, width=\columnwidth]{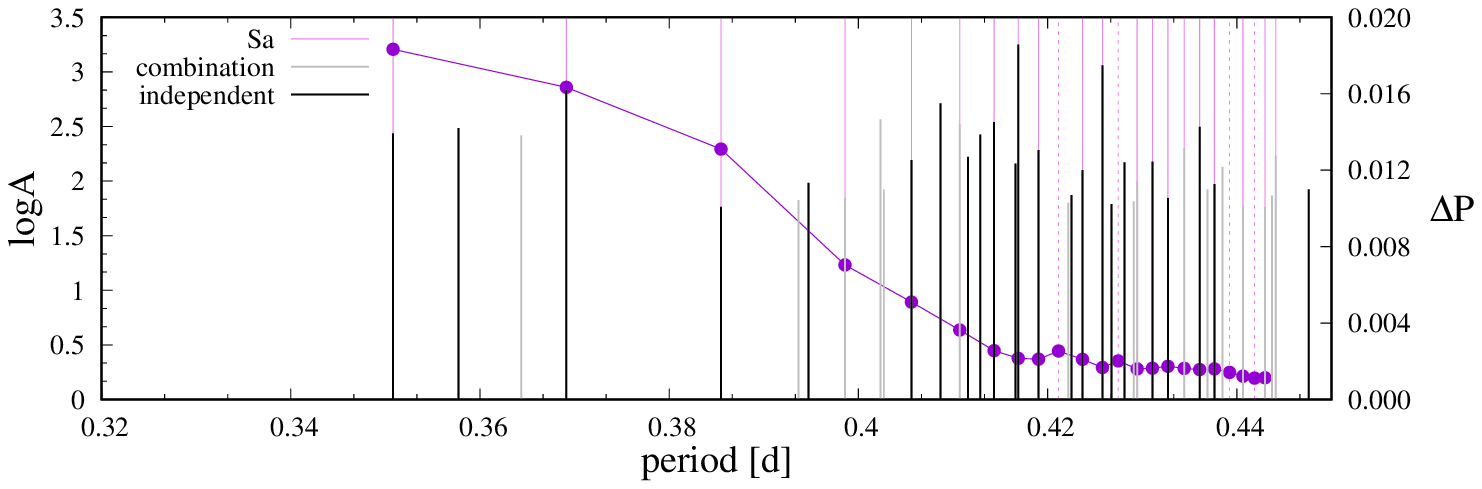}
	\includegraphics[angle=0, width=\columnwidth]{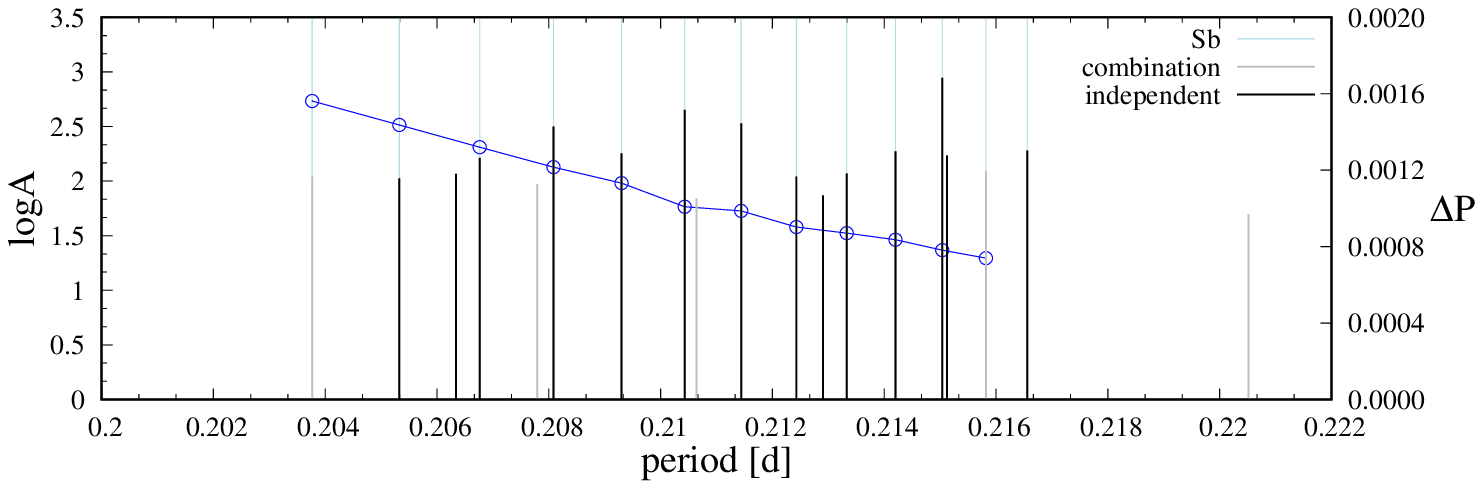}
	\includegraphics[angle=0, width=\columnwidth]{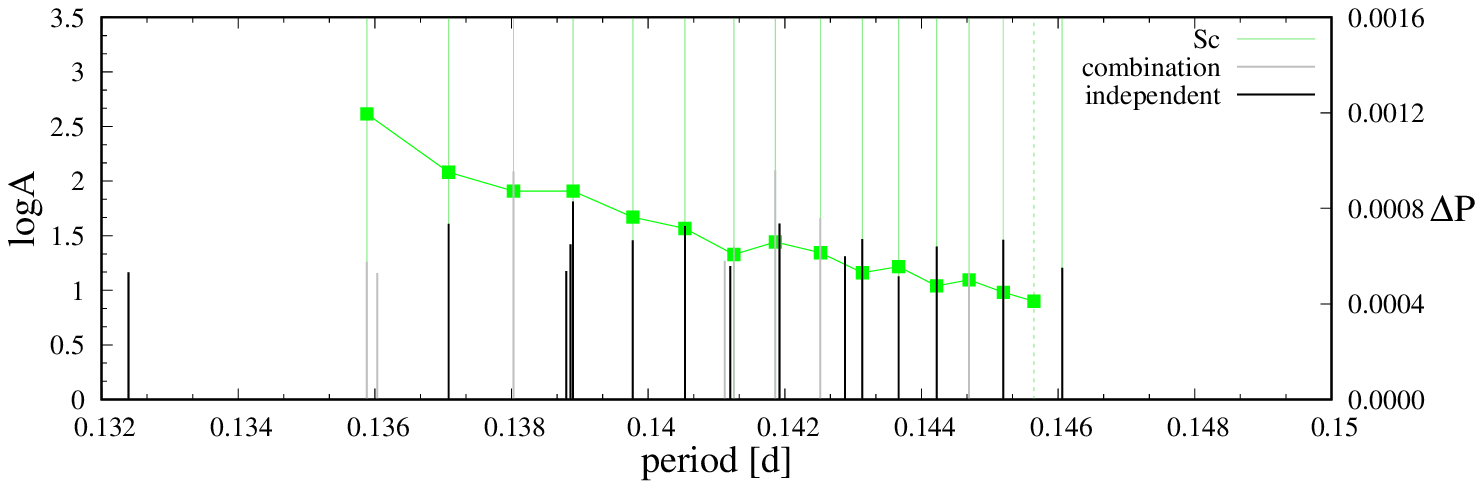} 
    \includegraphics[angle=0, width=\columnwidth]{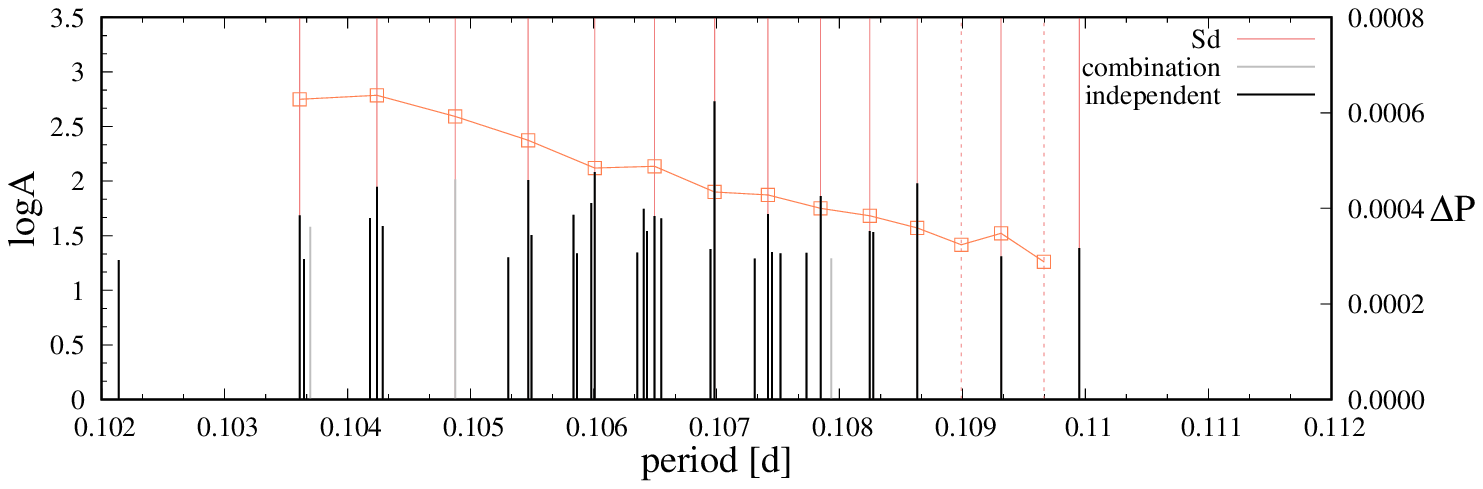}
    \includegraphics[angle=0, width=\columnwidth]{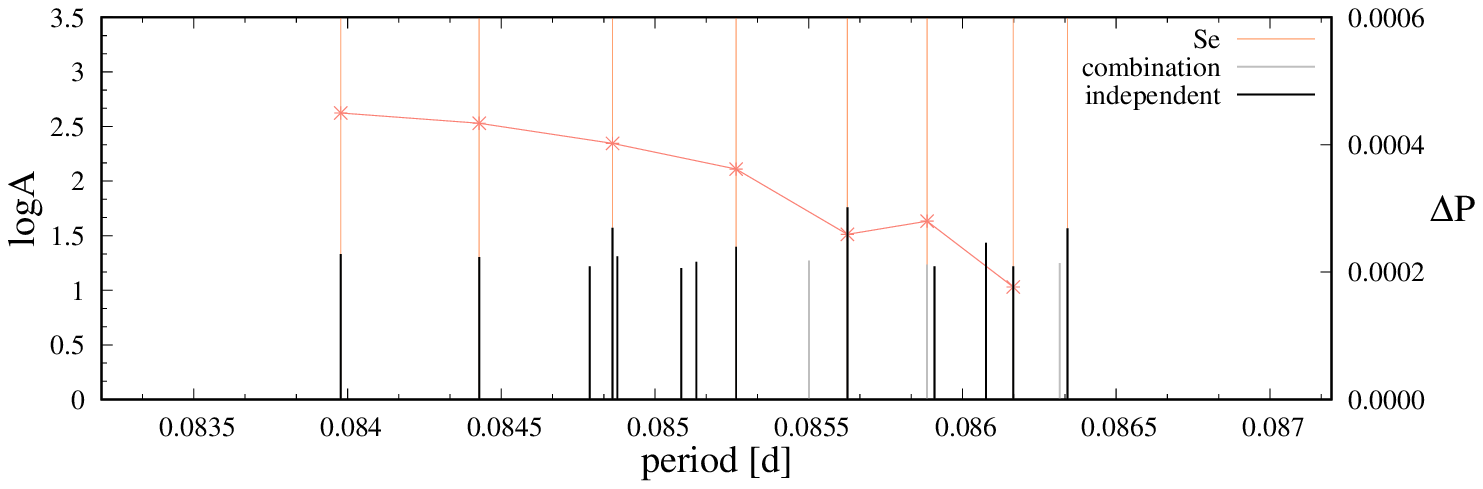} 
    \includegraphics[angle=0, width=\columnwidth]{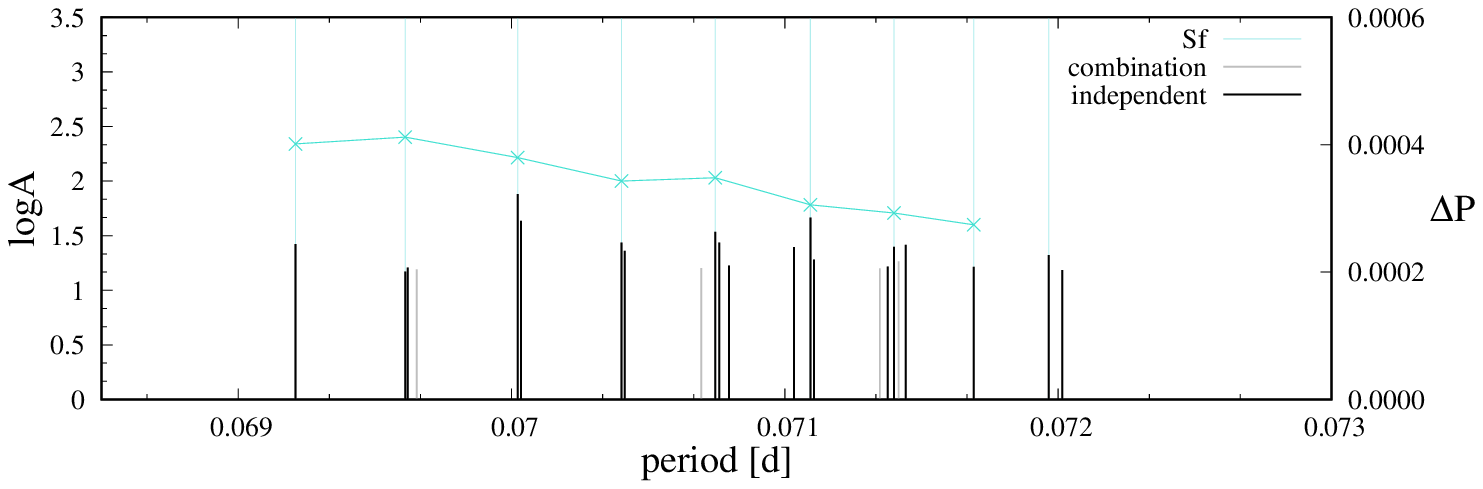} 
    \includegraphics[angle=0, width=\columnwidth]{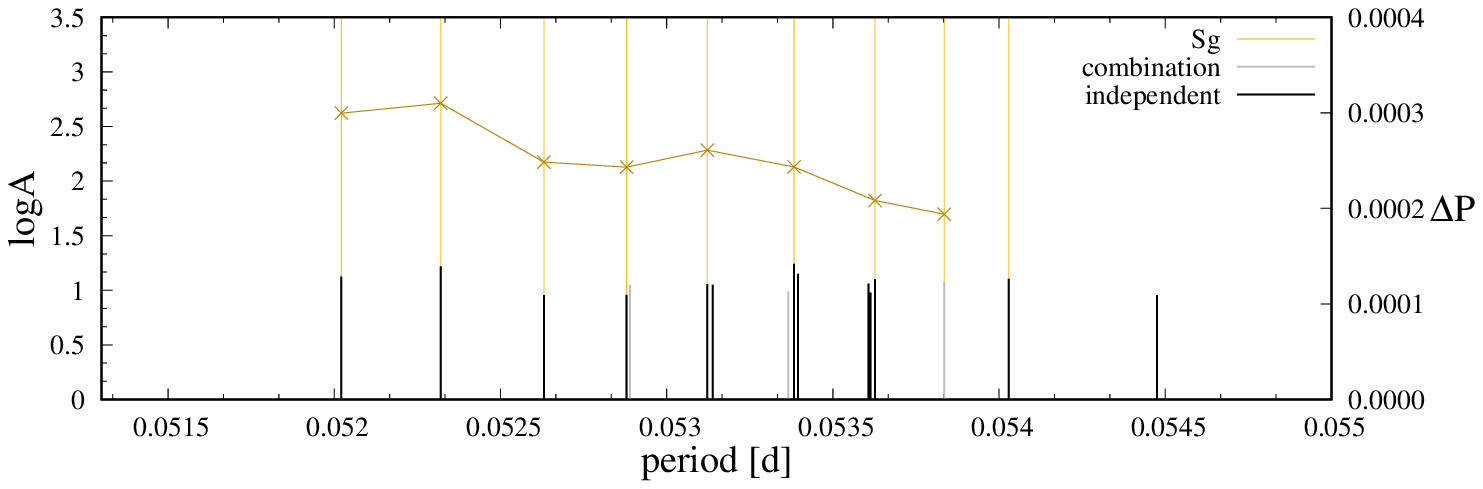}
    \vspace{-0.3cm}
 	\caption{The same as in Fig.\,\ref{fig:DP1} but for KIC\,11293898.}
 	\label{fig:DP18}
 \end{figure}

KIC\,11293898 is an example of a star with a very clear frequency grouping.
The consecutive groups may be associated with an increasing mode degrees.
The lowest frequencies can result from instrumental effects but
in the others group we have found seven frequency sequences with decreasing period spacings (see Tab.\,\ref{tab:S11293898} and Fig.\,\ref{fig:DP18}). The mean period differences are 0.00406\,d (0.02604\,d$^{-1}$) for Sa, 0.00107\,d (0.02416\,d$^{-1}$) for Sb, 0.00068\,d (0.03418\,d$^{-1}$) for Sc, 0.00045\,d (0.03975\,d$^{-1}$) for Sd, 0.00034\,d (0.04657\,d$^{-1}$) for Se, 0.00035\,d (0.06917\,d$^{-1}$) for Sf and 0.00025\,d (0.08930\,d$^{-1}$) for Sg.

Unstable dipole prograde modes with the mean value of $\Delta P$ well below 0.01\,d are predicted by our representative model (see Fig.\,\ref{fig:delty_b}\,x).
To summarize, the oscillations of KIC\,11293898 can be caused by series of consecutive dipole, quadruple and octupole modes. In-depth seismic analysis is necessary to draw 
more certain conclusions though. This work is already ongoing.

\begin{table}
	\centering
	\captionof{table}{The same as in Table\,\ref{tab:S1430353} but for  KIC\,11293898}
	\label{tab:S11293898}
	\begin{tabular}{rrrrrrr} % four columns, alignment for each
		\hline
 ID          &    $\nu$      & $P$    &$\Delta P$& $A$     & $\frac{\mathrm S}{\mathrm N}$         &   fs\\
             &    $(\mathrm{d}^{-1})$ & $(\mathrm{d})$  &$(\mathrm{d})$  & $(\mathrm{ppm})$    &               &  \\
\hline
\multicolumn{7}{|c|}{Sa}\\  
 $\nu_{36}$  & 2.25167(3) & 0.44412 & 0.001136 & 171(5) & 9.1 & c \\
 $\nu_{155}$ & 2.25744(7) & 0.44298 & 0.001114 & 58(5)  & 4.0 & c \\
 ---  & \textit{2.26313}    & \textit{0.44187} & \textit{0.001216} & ---    & ---&  ---\\
 $\nu_{145}$ & 2.26937(6) & 0.44065 & 0.001418 & 60(5)  & 4.1 & c \\
---  & \textit{2.27670}  & \textit{0.43923} & \textit{0.001598} & ---    & --- &  ---\\
 $\nu_{82}$  & 2.28501(4) & 0.43763 & 0.001559 & 94(5)  & 6.1 & i \\
 $\nu_{33}$  & 2.29318(3) & 0.43608 & 0.001631 & 313(9) & 9.6 & i \\
 $\nu_{26}$  & 2.30179(3) & 0.43444 & 0.001734 & 201(5) & 10 & c \\
 $\nu_{121}$ & 2.31102(6) & 0.43271 & 0.001632 & 70(5)  & 4.5 & i \\
 $\nu_{48}$  & 2.31977(3) & 0.43108 & 0.001604 & 151(5) & 7.7 & i \\
 $\nu_{76}$  & 2.32843(4) & 0.42947 & 0.002016 & 101(5) & 6.2 & c \\
 ---  & \textit{2.33941}    & \textit{0.42746} & \textit{0.001670} & ---    & --- & --- \\
 $\nu_{2}$   & 2.34859(7) & 0.42579 & 0.002095 & 1152(8)& 43 & i \\
 $\nu_{57}$  & 2.36020(4) & 0.42369 & 0.002538 & 126(5) & 7.2 & i \\
---  & \textit{2.37442}    & \textit{0.42116} & \textit{0.002110} &   ---  & --- & ---\\
 $\nu_{28}$  & 2.38638(3) & 0.41905 & 0.002152 & 192(5) & 9.8 & i \\
 $\nu_{1}$   & 2.398700(6)& 0.41689 & 0.002553 & 1778(9)& 49 & i \\
 $\nu_{17}$  & 2.41348(2) & 0.41434 & 0.003633 & 348(7) & 14 & i \\
 $\nu_{16}$  & 2.43483(2) & 0.41071 & 0.005095 & 331(5) & 14 & c \\
 $\nu_{40}$  & 2.46541(3) & 0.40561 & 0.007039 & 155(5) & 8.3 & i \\
 $\nu_{114}$ & 2.50895(6) & 0.39857 & 0.013100 & 70(5)  & 4.6 & c \\
 $\nu_{154}$ & 2.59421(7) & 0.38547 & 0.016338 & 58(5)  & 4.1 & i \\
 $\nu_{6}$   & 2.70903(1) & 0.36914 & 0.018324 & 685(5) & 29 & i \\
 $\nu_{19}$  & 2.85054(2) & 0.35081 & ---      & 274(5) & 18 & i \\
 
 \multicolumn{7}{|c|}{Sb}\\ 
 $\nu_{30}$ & 4.61770(3) & 0.21656 & 0.000739 & 191(5) & 12 & i \\
 $\nu_{58}$ & 4.63352(4) & 0.21582 & 0.000782 & 123(5) & 8.7 & c \\
 $\nu_{3}$  & 4.650365(9)& 0.21504 & 0.000836 & 882(6) & 47 & i \\
 $\nu_{34}$ & 4.66851(3) & 0.21420 & 0.000871 & 187(5) & 11 & i \\
 $\nu_{66}$ & 4.68756(4) & 0.21333 & 0.000903 & 117(5) & 7.6 & i \\
 $\nu_{69}$ & 4.70748(4) & 0.21243 & 0.000987 & 110(5) & 7.3 & i \\
 $\nu_{18}$ & 4.72945(2) & 0.21144 & 0.001009 & 338(6) & 16 & i \\
 $\nu_{11}$ & 4.75212(1) & 0.21043 & 0.001132 & 449(5) & 23 & i \\
 $\nu_{32}$ & 4.77782(3) & 0.20930 & 0.001216 & 179(5) & 10 & i \\
 $\nu_{14}$ & 4.80573(2) & 0.20809 & 0.001321 & 316(5) & 17 & i \\
 $\nu_{38}$ & 4.83643(3) & 0.20676 & 0.001437 & 164(5) & 9.2 & i \\
 $\nu_{71}$ & 4.87027(4) & 0.20533 & 0.001562 & 106(5) & 6.7 & i \\
 $\nu_{65}$ & 4.90760(4) & 0.20377 & ---      & 110(5) & 6.7 & c \\
 
  \multicolumn{7}{|c|}{Sc}\\
 $\nu_{313}$ & 6.8466(2)  & 0.14606 & 0.000412 & 16(5) & 4.8 & i \\
---  & \textit{6.8659}     & \textit{0.14565} & \textit{0.000449} &   --- & --- &  \\
 $\nu_{210}$ & 6.8872(1)  & 0.14520 & 0.000501 & 29(5) & 7.9 & i \\
 $\nu_{333}$ & 6.9110(3)  & 0.14470 & 0.000476 & 14(5) & 4.3 & c \\
 $\nu_{237}$ & 6.9338(2)  & 0.14422 & 0.000556 & 25(5) & 6.8 & i \\
 $\nu_{336}$ & 6.9606(3)  & 0.14367 & 0.000530 & 13(5) & 4.0 & i \\
 $\nu_{234}$ & 6.9864(2)  & 0.14314 & 0.000614 & 29(5) & 6.8 & i \\
 $\nu_{174}$ & 7.01653(8) & 0.14252 & 0.000659 & 46(5) & 11 & c \\
 $\nu_{55}$  & 7.04913(4) & 0.14186 & 0.000607 & 125(5)& 26 & c \\
 $\nu_{214}$ & 7.0794(1)  & 0.14125 & 0.000716 & 25(5) & 7.7 & c \\
 $\nu_{179}$ & 7.11550(9) & 0.14054 & 0.000763 & 39(5) & 10 & i \\
 $\nu_{221}$ & 7.1544(1)  & 0.13978 & 0.000873 & 29(5) & 7.3 & i \\
 $\nu_{133}$ & 7.19930(6) & 0.13890 & 0.000872 & 65(5) & 15 & i \\
 $\nu_{63}$  & 7.24480(4) & 0.13803 & 0.000951 & 123(5)& 26 & c \\
 $\nu_{183}$ & 7.29509(9) & 0.13708 & 0.001196 & 41(5) & 10 & i \\
 $\nu_{283}$ & 7.3593(2)  & 0.13588 & ---      & 18(5) & 5.3 & c \\

 	\end{tabular}
 	
%	\includegraphics[angle=0, width=\columnwidth]{DP_KIC11293898}
%    \captionof{figure}{The same as in Fig.\,\ref{fig:DP1} but for KIC\,11293898.}
%    \label{fig:DP18}
 	
\end{table}

 \setcounter{table}{17}
 
\begin{table}
	\centering
	\captionof{table}{Continued.}
	\label{tab:S11293898_c}
	\begin{tabular}{rrrrrrr} % four columns, alignment for each
		\hline
		ID          &    $\nu$      & $P$    &$\Delta P$& $A$     & $\frac{\mathrm S}{\mathrm N}$         &   fs\\
		&    $(\mathrm{d}^{-1})$ & $(\mathrm{d})$  &$(\mathrm{d})$  & $(\mathrm{ppm})$    &               &  \\
		\hline	
		\multicolumn{7}{|c|}{Sd}\\
		$\nu_{240}$ & 9.0950(2)  & 0.10995 & 0.000288 & 24(5)  & 5.2 & i \\
		---  & \textit{9.1189}     & \textit{0.10966} & \textit{0.000348} & ---    & --- &  ---\\
		$\nu_{265}$ & 9.1479(2)  & 0.10931 & 0.000324 & 21(5)  & 4.5 & i \\
		---  & \textit{9.1751}     & \textit{0.10899} & \textit{0.000359} & ---    & --- &  ---\\
		$\nu_{90}$  & 9.20543(5) & 0.10863 & 0.000385 & 95(5)  & 13 & i \\
		$\nu_{205}$ & 9.2382(1)  & 0.10825 & 0.000400 & 35(5)  & 6.0 & i \\
		$\nu_{194}$ & 9.2724(1)  & 0.10785 & 0.000428 & 73(9)  & 6.4 & i \\
		$\nu_{175}$ & 9.30941(8) & 0.10742 & 0.000435 & 50(5)  & 7.3 & i \\
		$\nu_{8}$   & 9.34721(1) & 0.10698 & 0.000488 & 537(5) & 59 & i \\
		$\nu_{176}$ & 9.39005(8) & 0.10650 & 0.000485 & 48(5)  & 7.2 & i \\
		$\nu_{56}$  & 9.43297(4) & 0.10601 & 0.000543 & 121(5) & 17 & i \\
		$\nu_{89}$  & 9.48152(5) & 0.10547 & 0.000593 & 102(8) & 13 & i \\
		$\nu_{147}$ & 9.5351(1)  & 0.10488 & 0.000637 & 103(8) & 8.9 & c \\
		$\nu_{99}$  & 9.59333(5) & 0.10424 & 0.000628 & 89(5)  & 12 & i \\
		$\nu_{171}$ & 9.65150(8) & 0.10361 & ---      & 49(5)  & 7.7 & i \\

		\multicolumn{7}{|c|}{Se}\\
		$\nu_{235}$ & 11.5819(2) & 0.08634 & 0.000177 & 37(6) & 5.7 & i \\
		$\nu_{307}$ & 11.6056(2) & 0.08617 & 0.000280 & 17(5) & 4.0 & i \\
		$\nu_{288}$ & 11.6435(2) & 0.08589 & 0.000259 & 17(5) & 4.3 & c \\
		$\nu_{227}$ & 11.6787(1) & 0.08563 & 0.000362 & 58(9) & 6.0 & i \\
		$\nu_{232}$ & 11.7283(1) & 0.08526 & 0.000402 & 25(5) & 6.0 & i \\
		$\nu_{190}$ & 11.7839(1) & 0.08486 & 0.000434 & 37(5) & 8.6 & i \\
		$\nu_{267}$ & 11.8444(2) & 0.08443 & 0.000450 & 20(5) & 5.0 & i \\
		$\nu_{257}$ & 11.9079(2) & 0.08398 & ---      & 22(5) & 5.4 & i  \\
		
		\multicolumn{7}{|c|}{Sf}\\
		$\nu_{275}$ & 13.8955(2)  & 0.07197 & 0.000274 & 21(5) & 4.7 & i \\
		$\nu_{309}$ & 13.9486(2)  & 0.07169 & 0.000293 & 16(5) & 4.2 & i \\
		$\nu_{250}$ & 14.0058(2)  & 0.07140 & 0.000306 & 25(5) & 5.6 & i \\
		$\nu_{177}$ & 14.06600(8) & 0.07109 & 0.000348 & 46(5) & 10 & i \\
		$\nu_{201}$ & 14.1352(1)  & 0.07075 & 0.000343 & 34(5) & 8.1 & i \\
		$\nu_{223}$ & 14.2041(1)  & 0.07040 & 0.000380 & 27(5) & 7.0 & i \\
		$\nu_{196}$ & 14.2811(1)  & 0.07002 & 0.000412 & 76(10) & 8.8 & i \\
		$\nu_{327}$ & 14.3656(3)  & 0.06961 & 0.000401 & 15(5) & 4.6 & i \\
		$\nu_{228}$ & 14.4489(1)  & 0.06921 & ---       & 27(5) & 7.8 & i \\
		
		\multicolumn{7}{|c|}{Sg}\\
		$\nu_{359}$ & 18.5086(3) & 0.05403 & 0.000194 & 13(5) & 4.6 & i \\
		$\nu_{343}$ & 18.5752(3) & 0.05384 & 0.000208 & 12(5) & 5.0 & c \\
		$\nu_{338}$ & 18.6474(3) & 0.05363 & 0.000243 & 13(5) & 5.3 & i \\
		$\nu_{272}$ & 18.7324(2) & 0.05338 & 0.000261 & 18(5) & 7.9 & i \\
		$\nu_{353}$ & 18.8244(3) & 0.05312 & 0.000243 & 11(5) & 4.8 & i \\
		$\nu_{373}$ & 18.9110(4) & 0.05288 & 0.000248 & 9(5)  & 4.2 & i \\
		$\nu_{372}$ & 19.0003(4) & 0.05263 & 0.000310 & 9(5)  & 4.3 & i \\
		$\nu_{364}$ & 19.1129(4) & 0.05232 & 0.000300 & 17(5) & 4.8 & i \\
		$\nu_{358}$ & 19.2230(3) & 0.05202 & ---      & 13(5) & 5.3 & i \\
		
	\end{tabular}
	
	%	\includegraphics[angle=0, width=\columnwidth]{DP_KIC11293898}
	%    \captionof{figure}{The same as in Fig.\,\ref{fig:DP1} but for KIC\,11293898.}
	%    \label{fig:DP18}
	
\end{table}

%\begin{figure}
%	\includegraphics[angle=0, width=\columnwidth]{DP_KIC11293898}
%    \caption{The same as in Fig.\,\ref{fig:DP1} but for KIC\,11293898.}
%    \label{fig:DP18}
%\end{figure} 

\subsection{KIC\,11360704}
%19
The star was classified as a frequency grouping pulsator by  \citet{2011MNRAS.413.2403B}, whereas,
\citet{2012AJ....143..101M} added the hybrid type of pulsations. Next, \citet{2015MNRAS.451.1445B} 
classified it as the SPB/ROT variable. \citet{2011MNRAS.413.2403B} derived the following atmospheric parameters:
$T_\mathrm{eff} =20700\,\mathrm{K}$, $\log g=4.1$ (from spectroscopy),
$T_\mathrm{eff} =17644\,\mathrm{K}$, $\log g=3.89$ (from Str\"omgren photometry)
and $T_\mathrm{eff} =18200\pm5400\,\mathrm{K}$ (from SED fitting).

In its Kepler light curve, one can see a kind of outbursts. We extracted 146 frequency peaks, of which 54 were rejected 
because of the $2.5\times$Rayleigh limit.
%We were left with 22 firm independent frequencies, 30 probably independent, 36 combinations and 4 harmonics.
We were left with 52  independent frequencies, 36 combinations and 4 harmonics.
In the oscillation spectrum we identified
frequency grouping with independent frequencies up to $\sim 15\,\mathrm{d}^{-1}$.
The star is another fast rotator located inside the SPB instability strip,
therefore we classified it as the SPB pulsator. According to the TDP analysis, the frequencies 
seem to be coherent but their amplitudes are variable.

KIC\,11360704  is a very fast rotating star with the minimum rotational velocity of 303 km\,s$^{-1}$.
It exhibits characteristic for rapid rotators frequency grouping. There are also three clear regular patterns 
in its frequency spectrum associated with asymptotic properties of high-order g modes (see Tab\,.\ref{tab:S11360704} and Fig.\,\ref{fig:DP19}).
The Sa series can be associated with dipole retrograde modes. The Sb and Sc sequences seems to be quadrupole axisymmetric and prograde modes, respectively. The mean period differences are 0.7492\,d (0.0145\,d$^{-1}$) for Sa, 0.0072\,d (0.0363\,d$^{-1}$) for Sb and 0.0078\,d (0.1516\,d$^{-1}$) for Sc.
\begin{table}
	\centering
	\captionof{table}{The same as in Table\,\ref{tab:S1430353} but for  KIC\,11360704}
	\label{tab:S11360704}
	\begin{tabular}{rrrrrrr} % four columns, alignment for each
		\hline
		ID          &    $\nu$      & $P$    &$\Delta P$& $A$     & $\frac{\mathrm S}{\mathrm N}$         &   fs\\
		&    $(\mathrm{d}^{-1})$ & $(\mathrm{d})$  &$(\mathrm{d})$  & $(\mathrm{ppm})$    &               &  \\
		\hline
		\multicolumn{7}{|c|}{Sa}\\
$\nu_{43}$ & 0.11910(6) & 8.39662 & 0.88955 & 438(22) & 4.4 & i \\
$\nu_{47}$ & 0.13321(6) & 7.50707 & 0.76234 & 366(22) & 4.2 & i \\
$\nu_{35}$ & 0.14826(5) & 6.74473 & 0.59570 & 579(27) & 5.0 & i \\
$\nu_{31}$ & 0.16263(5) & 6.14902 & ---     & 472(22) & 5.3 & i \\

		\multicolumn{7}{|c|}{Sb}\\
		$\nu_{15}$ & 2.02780(3) & 0.49315 & 0.00162 & 870(22) & 14 & c \\
		--- & 2.03448    & 0.49153 & 0.00213 &  ---    & ---& --- \\
		$\nu_{70}$ & 2.04333(7) & 0.48940 & 0.00280 & 574(40) & 5.2 & i \\
		$\nu_{26}$ & 2.05509(4) & 0.48660 & 0.00377 & 782(38) & 8.8 & i \\
		$\nu_{7}$  & 2.07115(2) & 0.48282 & 0.00434 & 1377(23)& 21 & i \\
		$\nu_{69}$ & 2.08996(7) & 0.47848 & 0.00460 & 292(22) & 5.2 & c \\
		$\nu_{74}$ & 2.11024(8) & 0.47388 & 0.00791 & 363(25) & 5.0 & i \\
		--- & 2.14603    & 0.46598 & 0.00941 &     --- & --- & ---\\
		$\nu_{39}$ & 2.19027(5) & 0.45657 & 0.00891 & 564(32) & 7.2 &  i\\
		$\nu_{95}$ & 2.23388(9) & 0.44765 & 0.01267 & 221(21) & 4.4 & i \\
		--- & 2.29897    & 0.43498 & 0.01419 &     --- & --- & --- \\
		$\nu_{89}$ & 2.37648(9) & 0.42079 & 0.01482 & 231(21) & 5.0 & i \\
		$\nu_{1}$  & 2.46321(2) & 0.40598 & ---     & 2071(22)& 35 &  i\\
		
    	\multicolumn{7}{|c|}{Sc}\\
		$\nu_{64}$ & 3.77265(7) & 0.26507 & 0.00674 & 308(21) & 6.5 & c \\
		--- & 3.87100    & 0.25833 & 0.00700 &  ---    & --- & --- \\
		$\nu_{29}$ & 3.97878(5) & 0.25133 & 0.00734 & 614(22) & 7.4 & i \\
		$\nu_{2}$  & 4.09840(2) & 0.24400 & 0.00793 & 1703(22)& 20 & i \\
		$\nu_{37}$ & 4.23610(5) & 0.23607 & 0.00734 & 423(21) & 5.4 & i \\
		$\nu_{4}$  & 4.37198(2) & 0.22873 & 0.00835 & 1526(21)& 17 & i \\
		$\nu_{33}$ & 4.53751(5) & 0.22039 & 0.00909 & 487(22) & 6.1 & c \\
		--- & 4.73275    & 0.21129 & 0.00831 &     --- & --- & --- \\
		$\nu_{11}$ & 4.92639(2) & 0.20299 & 0.00832 & 1092(23)& 23 &i  \\
		$\nu_{103}$& 5.1369(1)  & 0.19467 & ---     & 149(22) & 5.2 & i \\
		
	\end{tabular}
	\includegraphics[angle=0, width=\columnwidth]{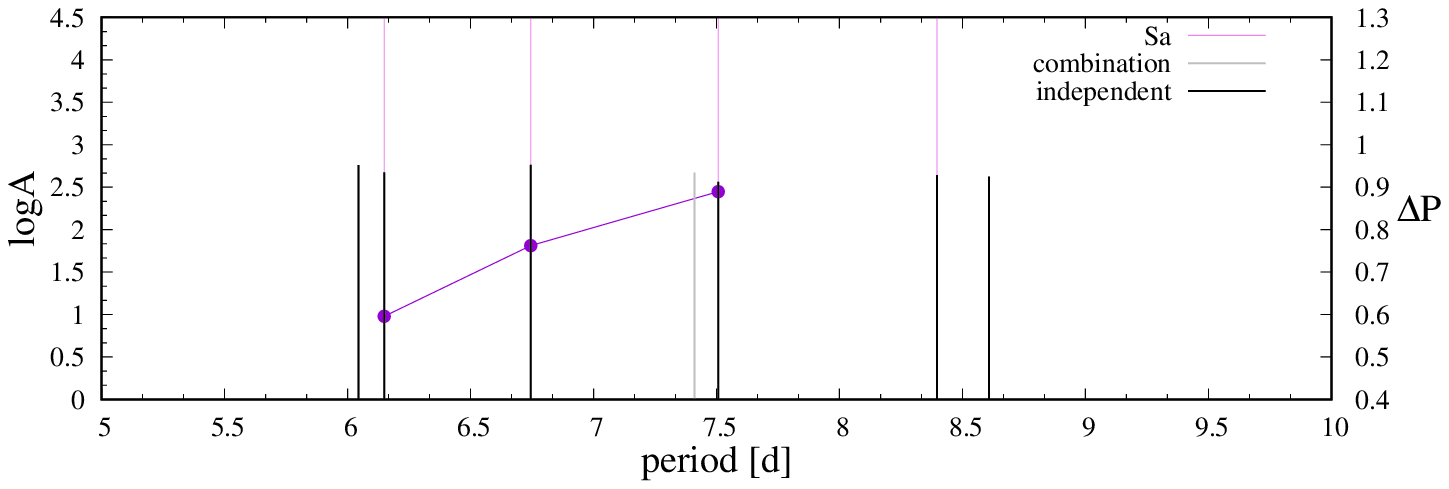}
	\includegraphics[angle=0, width=\columnwidth]{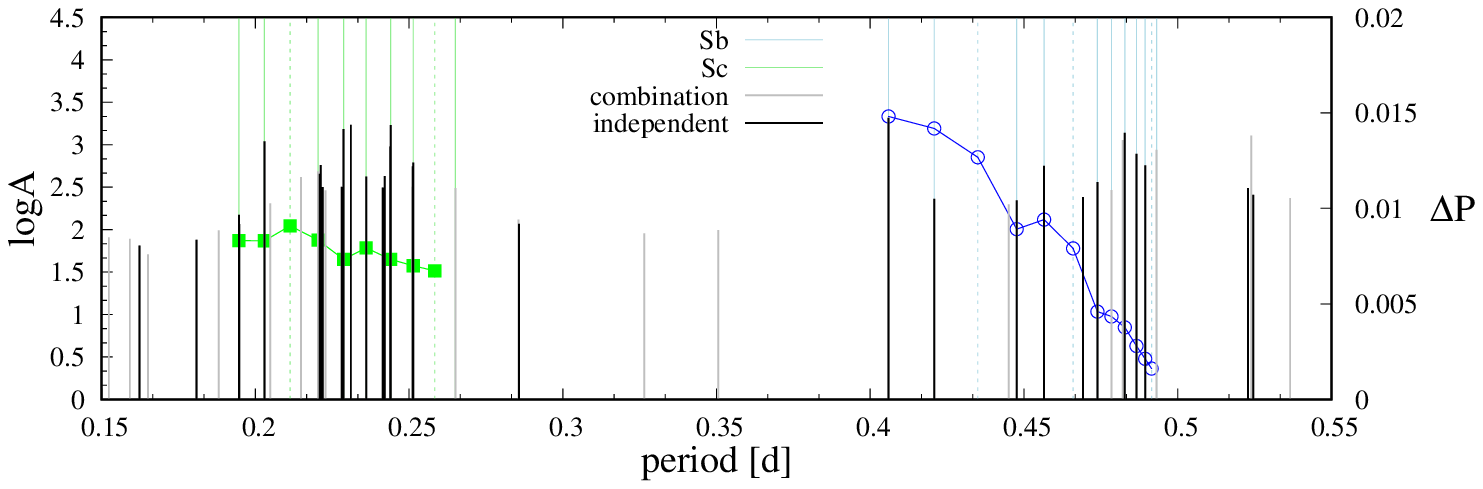}
	\vspace{-0.3cm}
	\captionof{figure}{The same as in Fig.\,\ref{fig:DP1} but for KIC\,11360704.}
	\label{fig:DP19}
	
\end{table}

\subsection{KIC\,11671923}
%20
The star was classified as the SPB variable  by \citet{2012AJ....143..101M, 2015MNRAS.451.1445B}.
From its Kepler light curve we derived 38 frequency peaks and  rejected 9 because of the $2.5\times$Rayleigh limit.
%We identified 13 firm independent frequencies,  7 probably independent, 8 combinations and 1 harmonic.
We identified 20 independent frequencies,   8 combinations and 1 harmonic. 
All frequencies are below $3.3\,\mathrm{d}^{-1}$.
Taking into account the observed frequency range and the star's location in the
Kiel diagram (see Fig.\,\ref{fig:kiel}), we classified the star as the SPB pulsator.
Frequencies seem to be coherent apart from the lowest ones $(\sim 0.1\,\mathrm{d^{-1}})$ but the
amplitudes are variable.

Since the number of detected frequencies is rather small, we could not find regular patterns.

\FloatBarrier

%%%%%%%%%%%%%%%%%%%%%%%%%%%%%%%%%%%%
%%%%%%%%%%%%%%%%%%%%%%%%%%%%%%%%%%%%%%
%%%%%%%%%%%%%%%%%%%%%%%%%%%%%%%%%%%

\section{Red noise}
\label{red_noise}
\begin{figure*}
	\includegraphics[angle=270, width=1.95\columnwidth]{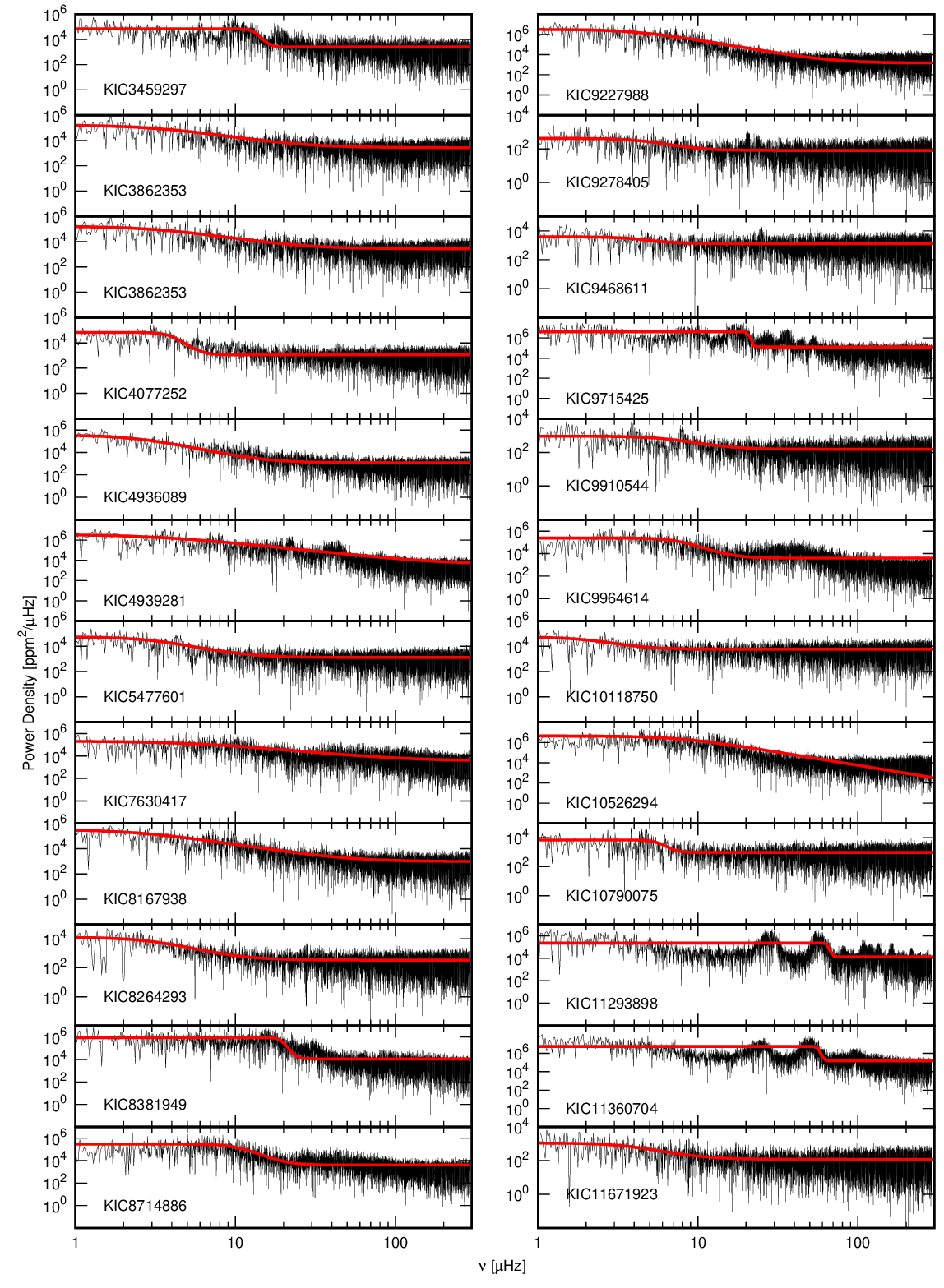}
    \caption{Fourier transforms of the data pre-whitened with all significant frequencies (black lines) and profile
    from Eq.\,\ref{eq:Lorentz}
    fitted to these fourier transforms (red lines).}
    \label{fig:red_noise}
\end{figure*}

\begin{figure*}
	\includegraphics[angle=270, width=2\columnwidth]{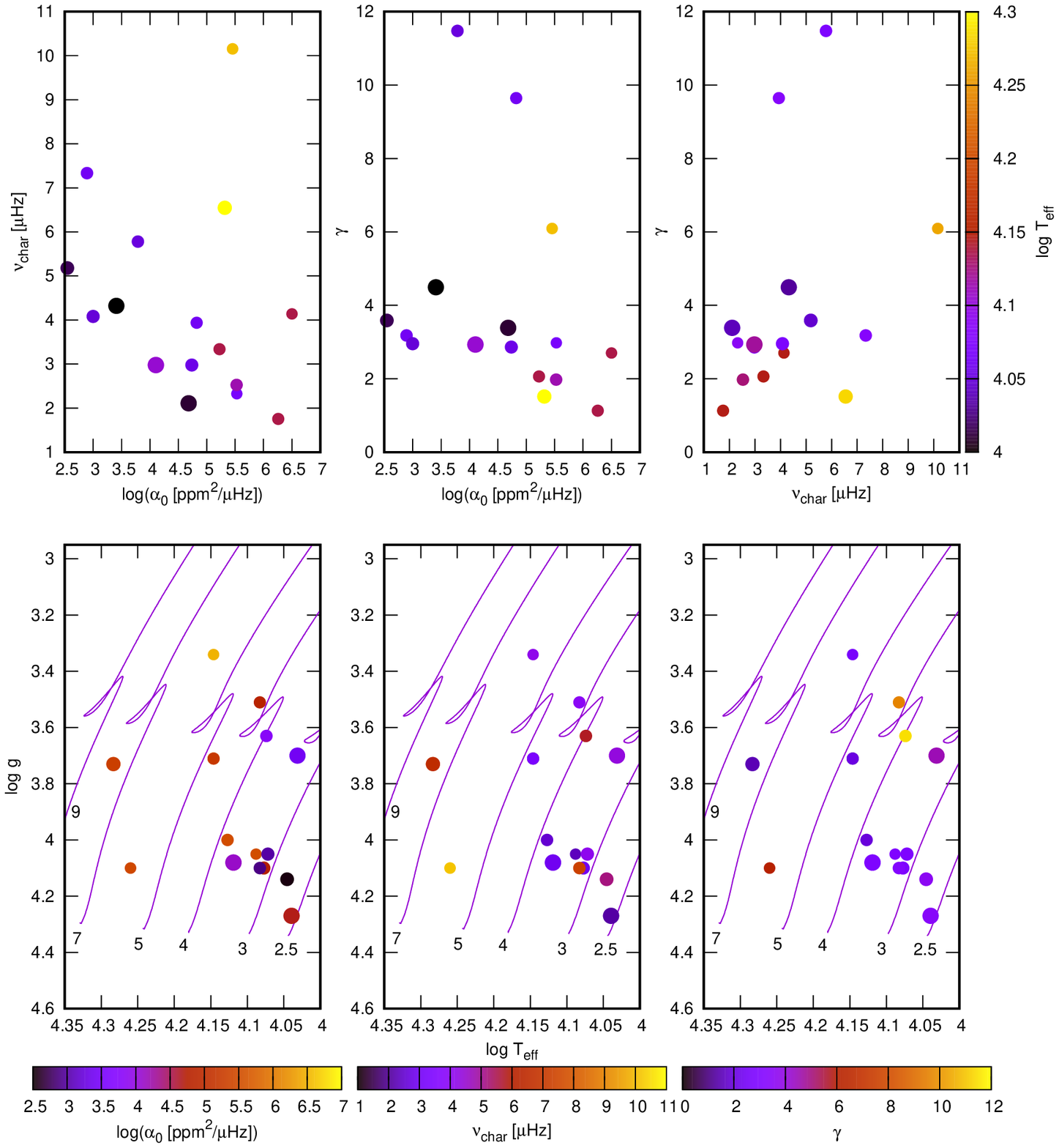}
    \caption{Fitted parameters $a_0$, $\nu_\mathrm{char}$ and $\gamma$ of profile from Eq.\,\ref{eq:Lorentz}. In the top 
    panels are shown relationships
    between parameters, $\nu_\mathrm{char}$, $\alpha_0$ and $\gamma$.
    Effective temperatures are colour coded, whereas point sizes are proportional to the rotation velocities.
    In the bottom panels the positions of the star on the Kiel diagram are shown with fitted parameters colour coded.}
    \label{fig:red_noise2}
\end{figure*}

Recently \citet{2019A&A...621A.135B} interpreted the  so-called red noise in power spectra
of a sample of O, B, A and F stars observed by {\it CoRoT}
as a manifestation of the internal  gravity waves (IGW).
The authors fitted to the power spectra of their sample stars, the profile in a form: 
\begin{equation}
\label{eq:Lorentz}
\alpha \left( \nu \right)=\frac{\alpha_0}{1+ \left( \frac{\nu}{\nu_\mathrm{char}} \right)^\gamma}+P_\mathrm{W}
\end{equation}
where $\alpha_0$ is a scaling factor,
$\gamma$ is the gradient of the linear part of the profile in a log-log graph,
$\nu_\mathrm{char}$ is the characteristic frequency and
$P_\mathrm{W}$ is white noise term.

 Then \citet{2019NatAs...3..760B} and \citet{2020A&A...640A..36B} studied red noise in a sample
of OB stars observed by K2 and TESS missions. Compared to the previous studies the authors found correlations between stochastic phtotometric variability and spectroscopic parameters. However,
this is most evident for most massive stars ($M > 20\,\mathrm M_{\sun}$)

We repeated calculations of \citet{2019A&A...621A.135B} and fitted equation \ref{eq:Lorentz}
to the power spectra of our stars, prewithened with all significant frequencies.
Similarly to the cited authors, we used a Markov Chain Monte Carlo procedure.
The results are shown in Fig.\,\ref{fig:red_noise} where fitted profiles are overplotted on the  prewithened power spectra.
We show 24 stars out of 25 from our sample because in the case of KIC\,1430353 our  MCMC procedure did not converge at all.
Apparently, the considered model is not adequate.
Furthermore, the quick inspection of Fig.\,\ref{fig:red_noise} shows that in the case of 
KIC\,3459297,
%KIC\,3839930, 
KIC\,4939281,
KIC\,8381949,
KIC\,9715425,
KIC\,9964614,
KIC\,10526294,
KIC\,11293898 and
KIC\,11360704
the adopted model is not correct for the observational data. These results were omitted in the further analysis.

The values of the fitted parameters, $\alpha_0$,
$\gamma$, $\nu_\mathrm{char}$ and $P_\mathrm{W}$ for the remaining stars are shown in Fig.\,\ref{fig:red_noise2}.
%{\bf In general parameters for our stars are in a good agreement with those found by 
%\citet{2019A&A...621A.135B, 2019NatAs...3..760B, 2020A&A...640A..36B}.
%However, in the case of our sample, we observe smaller spread in $\nu_\mathrm{char}$ and, in %general, we obtained smaller values of this parameter. In addition for two stars
%(KIC\,4077252 and KIC\,10790075) we found relatively high velue of $\gamma$ parameter.}
In general parameters for our stars are in a good agreement with those found by 
\citet{2019A&A...621A.135B, 2019NatAs...3..760B, 2020A&A...640A..36B}.
However, we observe smaller spread in $\nu_\mathrm{char}$. Also, we obtained smaller values of this parameter. In addition, for two stars
(KIC\,4077252 and KIC\,10790075) we found relatively high value of $\gamma$ parameter of the order of 10. Such high values seem not to be very common.
%Our $\alpha_0$ and $\gamma$ have similar distribution as in the case of the stars from the cited paper. 
% Our $\alpha_0$ and
%$\gamma$ have similar distributions, though.
%Only $\gamma$ for KIC\,4077252
%and KIC\,10790075 has slightly higher value than in Bowman's sample.
We also did not find any correlations between parameters from Eq.\,\ref{eq:Lorentz} and stellar parameters such as effective temperature, surface gravity, rotation  or luminosity.
Since our sample contain stars well below 20\,$\mathrm{M}_{\sun}$ this is in line with the findings od \citet{2020A&A...640A..36B}.

Finally, we can use the same arguments as  \citet{2019A&A...621A.135B, 2019NatAs...3..760B, 2020A&A...640A..36B} to explain the origin of the observed red noise.
First, the values of $\nu_\mathrm{char}$ for our stars are below a line for the granulation signal given in Fig.\,8 
of \citet{2019A&A...621A.135B}. Since we do not expect
 granulation for such massive stars, this result is consistent with the standard theory.
Second, all our stars have masses approximately below 10\,$\mathrm{M}_{\sun}$ so we also do not expect that stellar winds are responsible for the 
red noise phenomenon. Third, although we are not able to exclude instrumental origin, diversity of red noise profiles (see Fig.\,\ref{fig:red_noise})  may suggest that at least in some stars signal is of astrophysical origin.

%On the one hand, based on our analysis, we can not exclude or confirm that IGW are the source of the observed red noise.
%On the other hand, \citet{2019ApJ...886L..15L} concluded that  low-frequency variability presented by \citet{2019A&A...621A.135B}
%is not due to linearly propagating waves from the core. Their arguments seem to apply to our results as well.

Based on our analysis, we can not exclude
nor  confirm that IGW are the source of the observed red noise.
On
the one hand, \citet{2019ApJ...886L..15L}
concluded that low-frequency
variability presented by \citet{2019A&A...621A.135B} is not due to linearly
propagating waves from the core.  But on the other hand, the interpretation can be quite different when nonlinear wave propagation is considered as proposed by \citet{2020MNRAS.497.4231R}.
It was recently shown by \citet{2020A&A...641A..18H} that fully compressible nonlinear simulations of waves caused by core
convection are in full agreement with the detected red noise features.

\section{Conclusions}
\label{conclusions}
We analyzed the sample of 25 B-type stars observed during the Kepler mission,
for which \citet{2019AJ....157..129H} determined stellar parameters. All available observational data were used.
The studied stars exhibit rich and complex photometric variability.
We detect both coherent as well as incoherent variability.
The time scales of these variations are also very diverse.
Therefore the origin of the light changes can have
a various background, e.g. stellar pulsations, variability in the disk around the star, rotation or binarity.

We would also like to emphasize that distinguishing
between combinations and independent frequencies in the dense oscillation spectra is an extremely difficult task.
%Our division into firm independent, probably independent and combination frequencies should be treated with caution.

For many cases more or less regular structures were found.
%However, often they are formed from low amplitude signal, whereas high amplitude peaks remain unexplained. \textcolor{red}{Czy to dalej jest prawda? W nowej analizie staralem sie wybierac mody o najwiekszej amplitudzie}
The striking example is KIC\,8264293 for which we found clear structure 
that we associated with prograde dipole modes. In the cases where regular period spacings are not clear, i.e.,
they are created only from low signal frequencies, one should be careful since they can be of random origin.
Therefore, independent mode identification would be invaluable.
Another promising test-bed for investigating stellar structure is rare eclipsing binary  with  SPB-type
component, KIC\,8167938. 

Furthermore, our sample contains stars with different rotational velocities that range from moderate rotators to very fast rotators.
On the one hand, this raises a  hope for constraints on rotationally induced mixing.
Mixing efficiency should increase with rotation rate. Therefore for the fastest rotators it should
have the biggest impact on the oscillation spectrum and be easiest to detect. 
%Moreover, successful asteroseismology for enough large sample of stars could allow calibrate mixing efficiency with
%the rotation speed. On the other hand, successful asteroseismic
%analysis can be inhibited by a lack of unambiguous mode identification. 
%Nevertheless, we have prepared a base for detailed seismic analysis of individual
%stars with oscillation spectra showing certain regularities.
 Moreover,
successful asteroseismology for enough large sample of stars could allow calibrate mixing efficiency with
the rotation speed. And we have prepared a base for detailed seismic analysis of individual
stars with oscillation spectra showing certain regularities.

Finally, we detected clear indication of a red
noise in the Fourier spectra of our sample stars.
In the case of 17 out of 25 stars from our sample it can be manifestation of IGW exicted at
the interface between convective core and radiative envelope. But we note that other sources of variability may
contribute to the observed red noise as well.
However, detailed analysis of this phenomenon and its origin is beyond the scope of this paper.

\section*{Acknowledgements}

This work was supported financially by the Polish National Science Centre grant 2018/29/B/ST9/01940.
Calculations were carried out using resources provided by Wroclaw Centre for
Networking and Supercomputing (http://wcss.pl),  grant no.  265.  Funding for  the  Kepler
mission  is  provided  by  the  NASA  Science  Mission  directorate.
Some of the data presented in this paper were
obtained from the Multimission Archive at the Space Telescope Science Institute (MAST).
STScI is operated by the Association of Universities for Research in Astronomy, Inc.,
under NASA contract NAS5-26555. Support for MAST for non-HST data is provided
by the NASA Office of Space Science via grant NNX09AF08G and by other grants and contracts.
The publication was partially founded by Excellence Initiative – Research University grant.

\section*{Data availability}

The target pixel files were downloaded from the public data archive at
MAST\footnote{\url{https://archive.stsci.edu/kepler/data_search/search.php}}.
The lightcurves 
will be shared on reasonable request to the corresponding author.
The full lists of frequencies are available as Supplementary material to this paper.

%%%%%%%%%%%%%%%%%%%%%%%%%%%%%%%%%%%%%%%%%%%%%%%%%%



\bibliographystyle{mnras}
\bibliography{szewczuk} % if your bibtex file is called example.bib

%\begin{thebibliography}{99}
%\end{thebibliography}

%%%%%%%%%%%%%%%%%%%%%%%%%%%%%%%%%%%%%%%%%%%%%%%%%%

%%%%%%%%%%%%%%%%% APPENDICES %%%%%%%%%%%%%%%%%%%%%

\appendix

\section{Some extra material}
\label{appendix:A}
\begin{figure*}
	\includegraphics[angle=270, width=2\columnwidth]{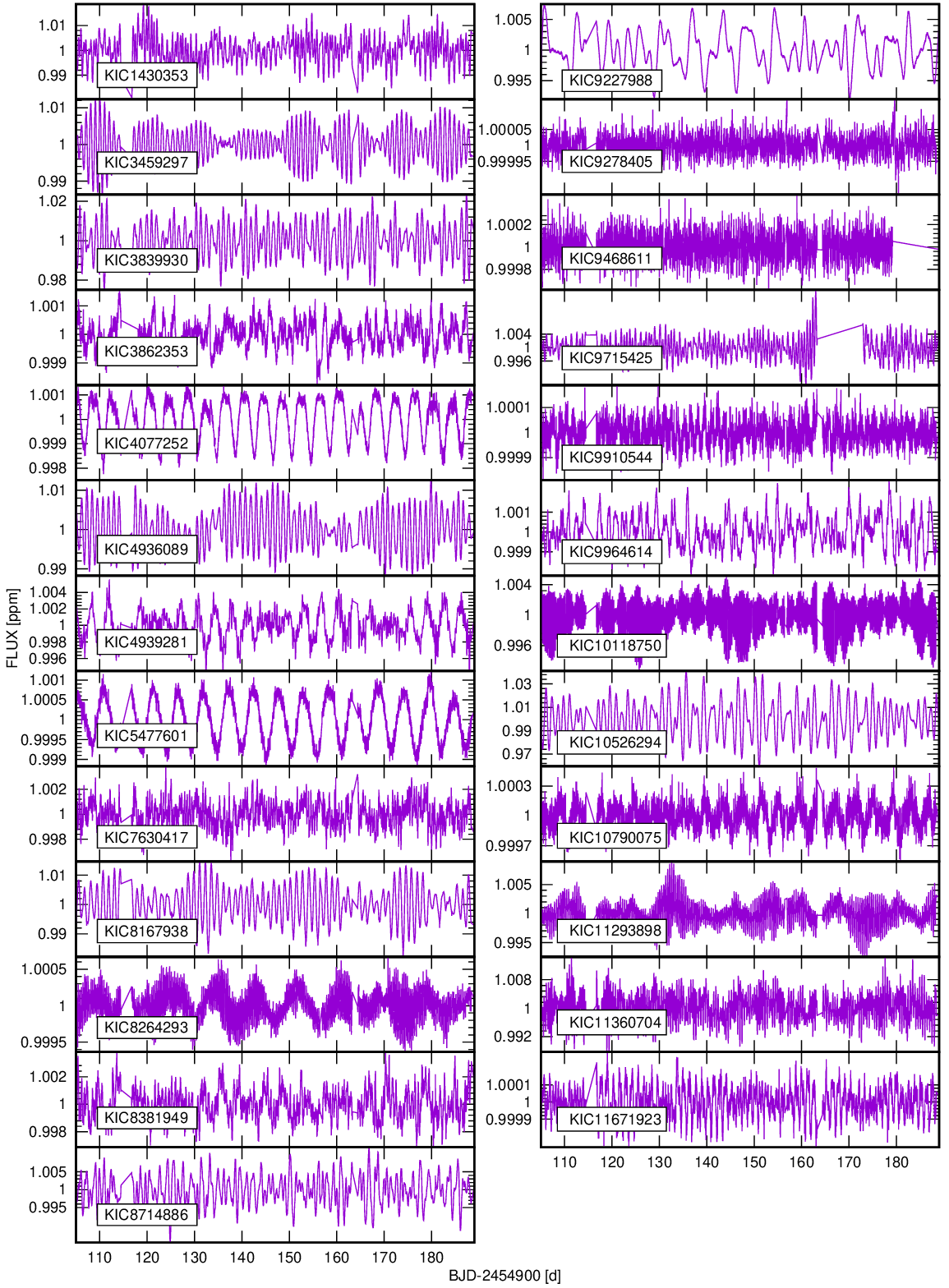}
    \caption{One quarter ($\sim 90$ days) cuts of  {\it Kepler} light curves for all program stars.}
    \label{fig:lc_all}
\end{figure*}

\begin{figure}
	\includegraphics[angle=270, width=\columnwidth]{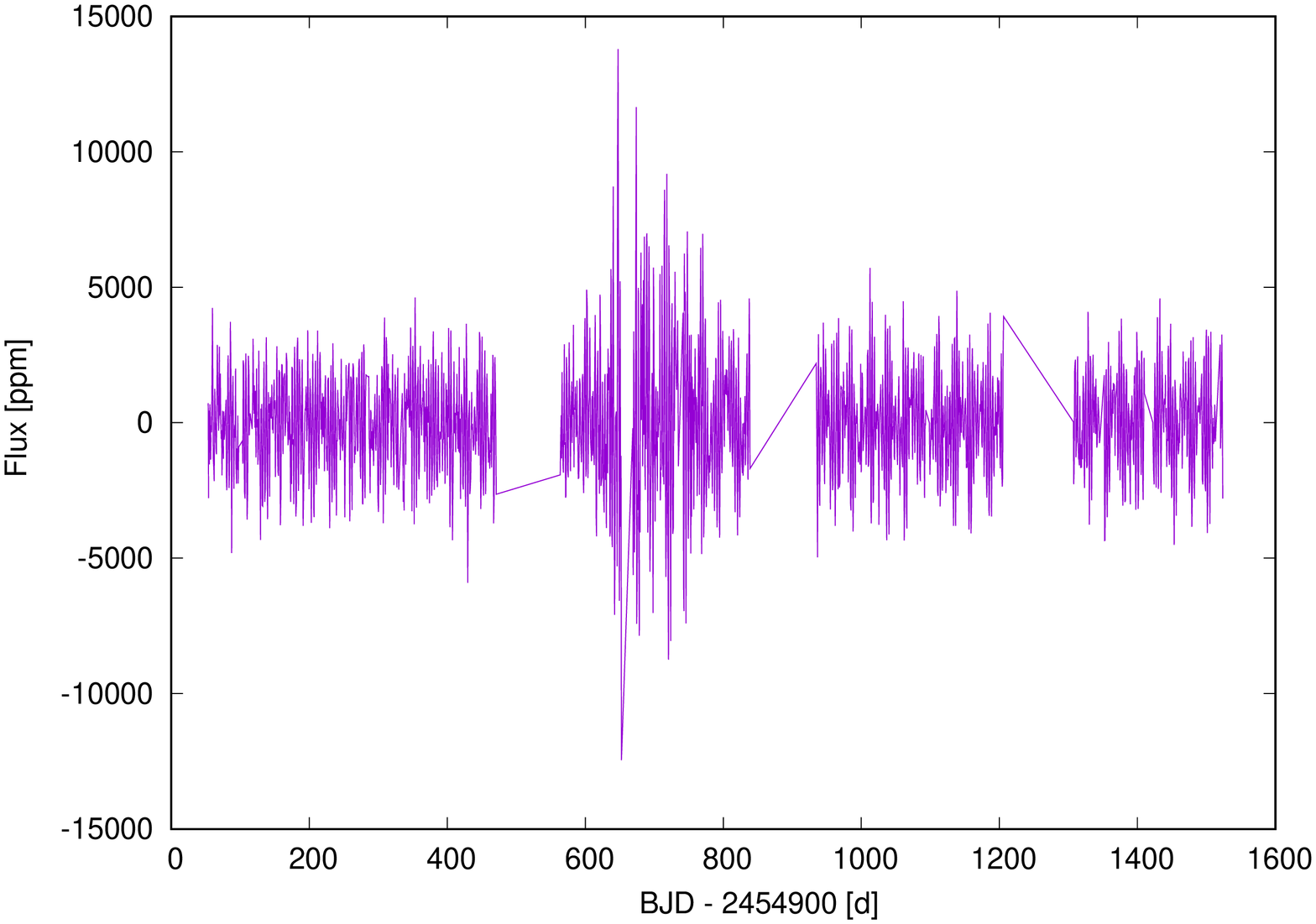}
    \caption{The full {\it Kepler} light curve of KIC\,4939281.}
    \label{fig:KIC4939281lc}
\end{figure}

\begin{figure}
	\includegraphics[angle=270, width=\columnwidth]{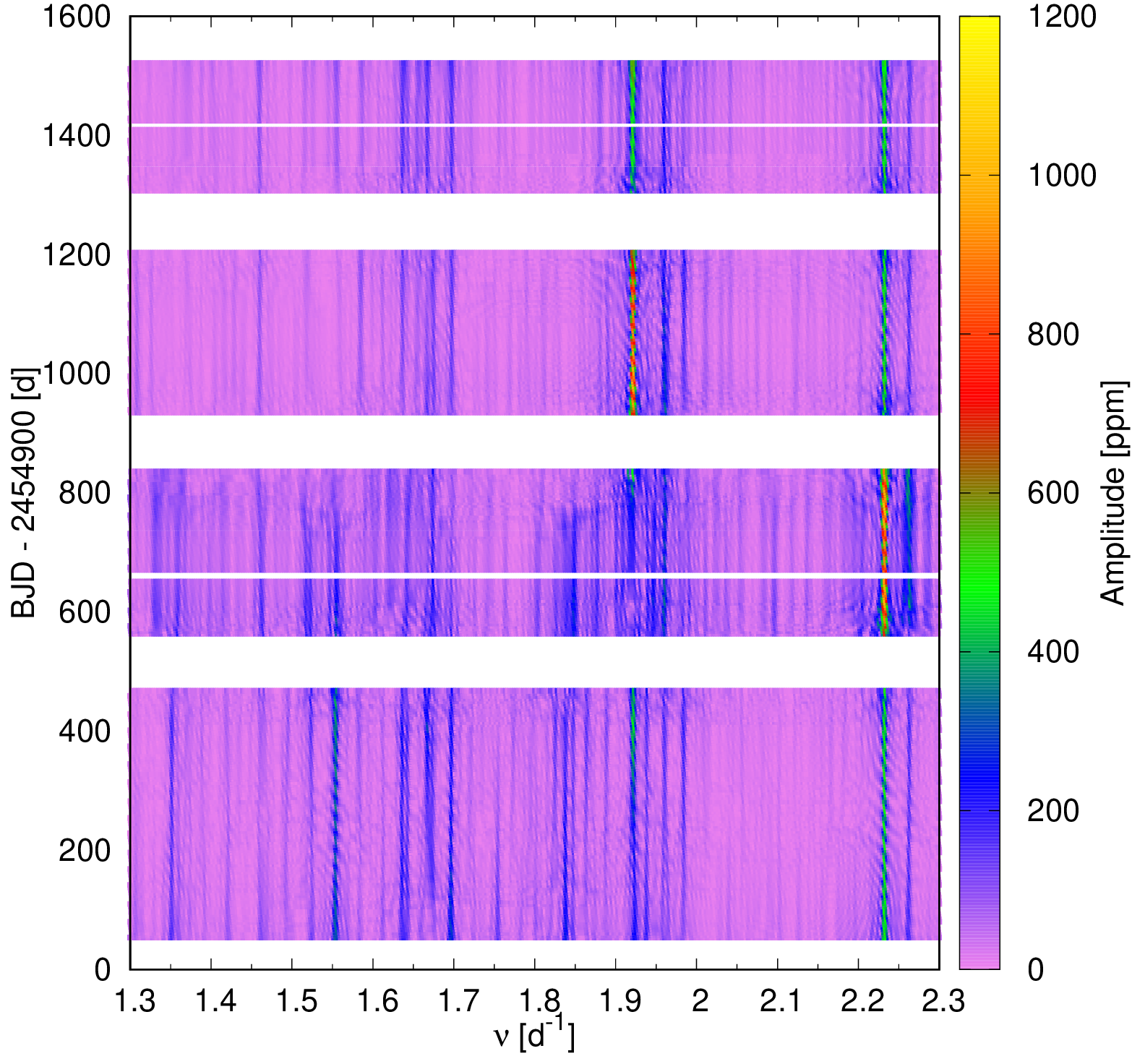}
    \caption{Time dependent periodogram of KIC\,4939281.}
    \label{fig:fourier_time_depend_lc0002KIC4939281}
\end{figure}

\begin{figure}
	\includegraphics[angle=270, width=\columnwidth]{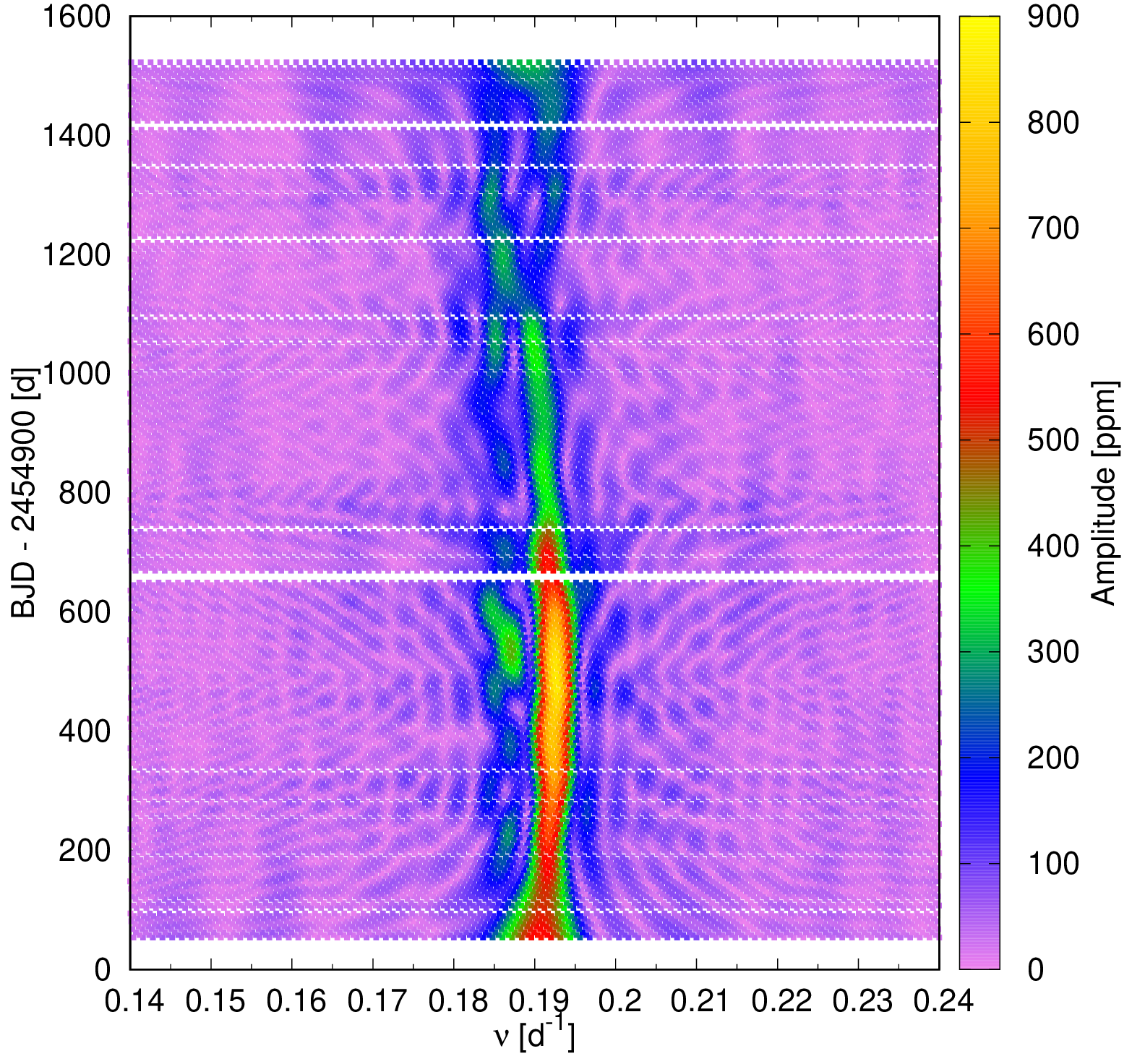}
    \caption{The same as in Fig.\,\ref{fig:fourier_time_depend_lc0002KIC4939281} but for KIC\,5477601.}
    \label{fig:fourier_time_depend_KIC5477601}
\end{figure}

\begin{figure}
	\includegraphics[angle=270, width=\columnwidth]{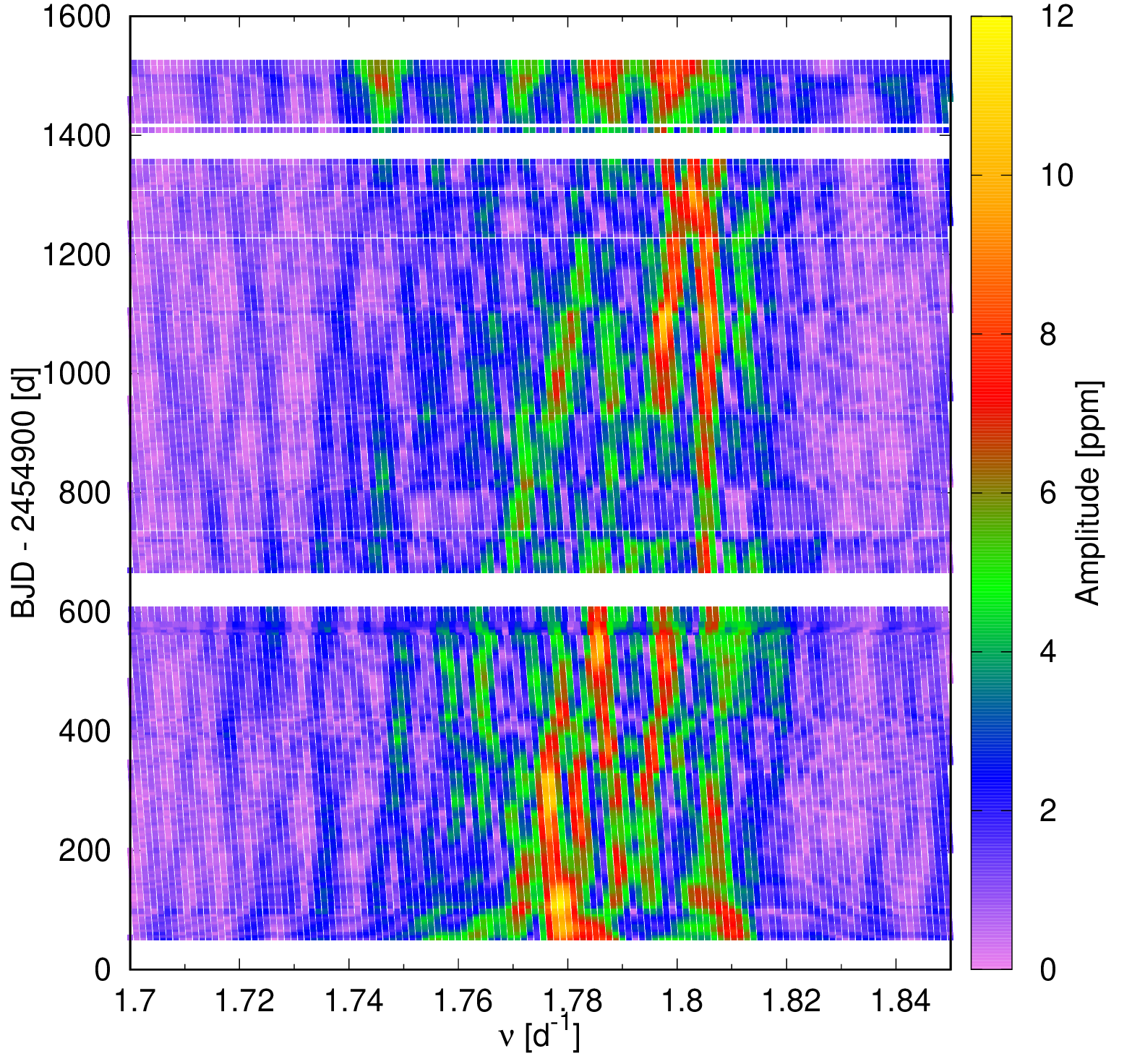}
    \caption{The same as in Fig.\,\ref{fig:fourier_time_depend_lc0002KIC4939281} but for KIC\,9278405.}
    \label{fig:fourier_time_depend_KIC9278405}
\end{figure}

\begin{figure}
	\includegraphics[angle=270, width=\columnwidth]{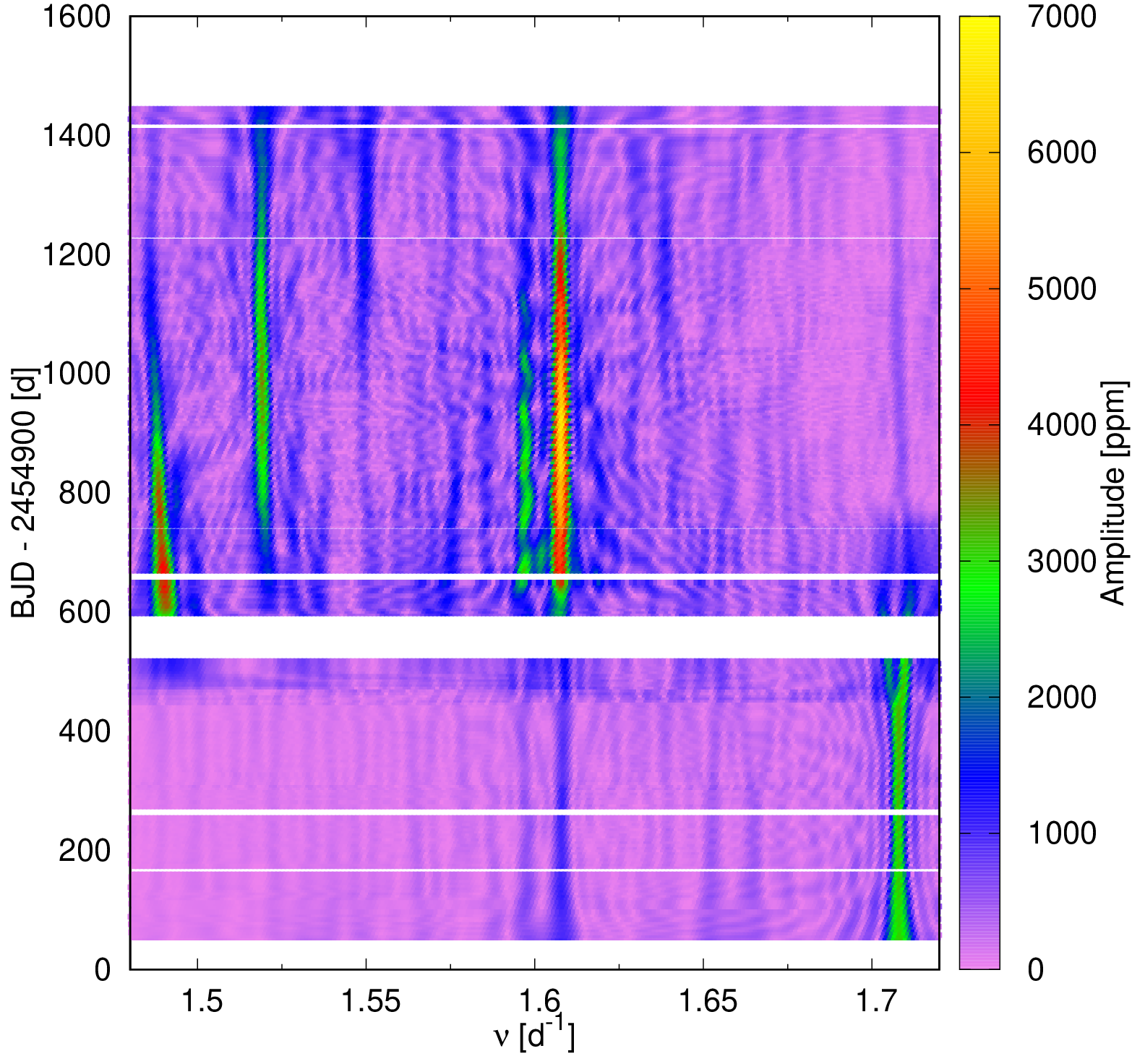}
    \caption{The same as in Fig.\,\ref{fig:fourier_time_depend_lc0002KIC4939281} but for KIC\,9715425.}
    \label{fig:fourier_time_depend_KI9715425}
\end{figure}

\begin{figure}
	\includegraphics[angle=270, width=\columnwidth]{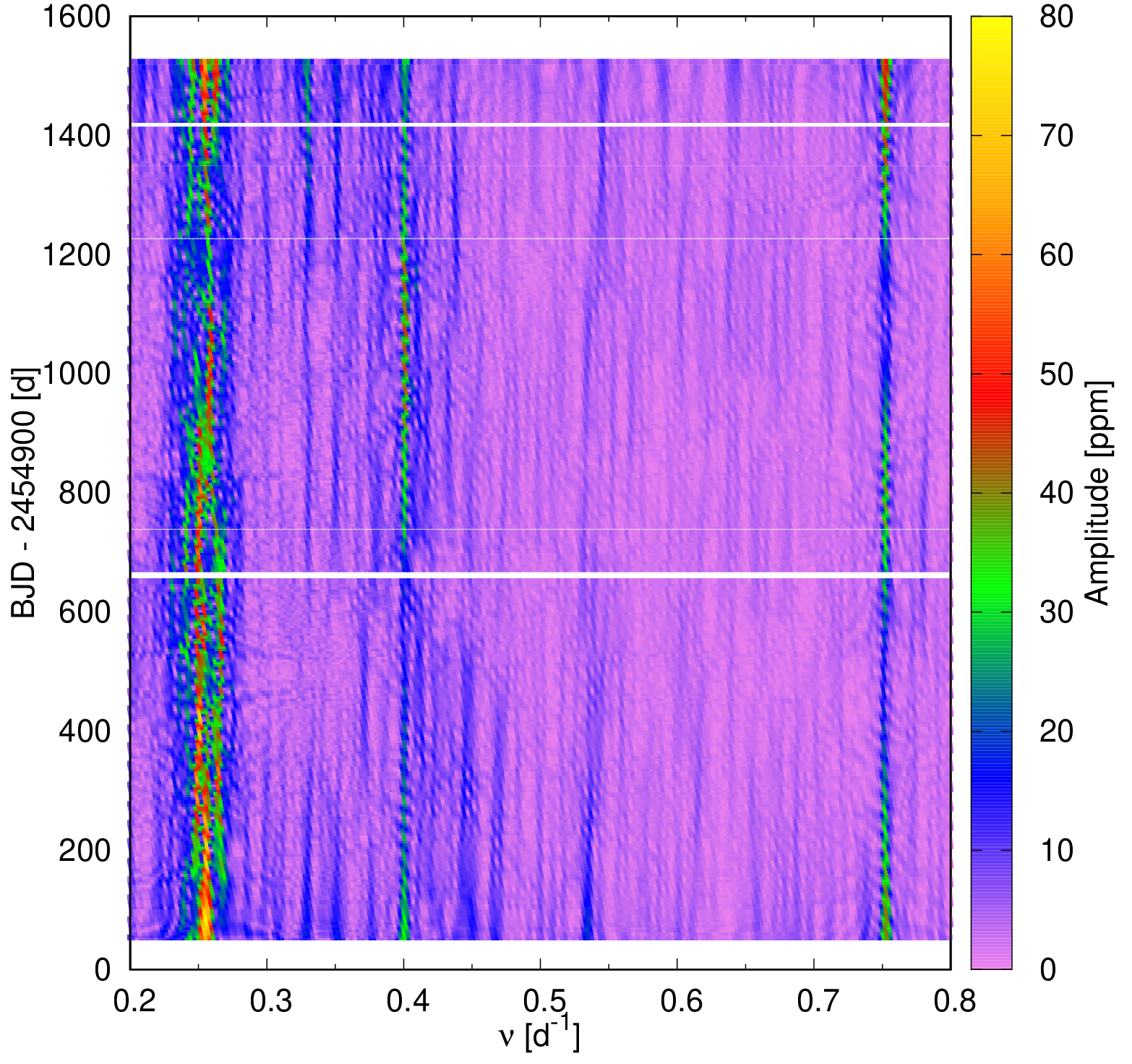}
    \caption{The same as in Fig.\,\ref{fig:fourier_time_depend_lc0002KIC4939281} but for KIC\,10790075.}
    \label{fig:fourier_time_depend_KIC10790075}
\end{figure}

Here we present one quarter cuts of the Kepler light curves for all stars from our set
(Fig.\,\ref{fig:lc_all}).
In addition there is  shown an example of the full light curve with outburst (Fig.\,\ref{fig:KIC4939281lc}).
Finally, examples of TD periodograms for stars with incoherent signal are plotted in Figs.\,\ref{fig:fourier_time_depend_lc0002KIC4939281}-\ref{fig:fourier_time_depend_KIC10790075}.

%%%%%%%%%%%%%%%%%%%%%%%%%%%%%%%%%%%%%%%%%%%%%%%%%%

\hypersetup{draft}

% Don't change these lines
\bsp	% typesetting comment
\label{lastpage}
\end{document}